\documentclass[aps,pra,twocolumn,superscriptaddress,showpacs]{revtex4-1}

\usepackage{graphicx}
\usepackage{mathptmx}      
\usepackage{subfigure}
\usepackage{color}

\newif\ifSHOWCOMMENTS
\newif\ifACCEPTCOMMENTS

\SHOWCOMMENTSfalse     
\SHOWCOMMENTStrue

\ACCEPTCOMMENTStrue    
\ACCEPTCOMMENTSfalse   %

\DeclareRobustCommand\openone{\leavevmode\hbox{\small1\normalsize\kern-.33em1}}

\begin{document}


\title{Quantum Decoherence and Thermalization at Finite Temperature \\
within the Canonical Thermal State Ensemble}

\author{M.A.\ Novotny}
\affiliation{Department of Physics and Astronomy, Mississippi State University,
Mississippi State, MS 39762-5167, USA}
\affiliation{HPC$^2$ Center for Computational Sciences, Mississippi State University,
Mississippi State, MS 39762-5167, USA}
\author{F. Jin}
\affiliation{Institute for Advanced Simulation, J\"ulich Supercomputing Centre,\\
Forschungszentrum J\"ulich, D-52425 J\"ulich, Germany}
\author{S. Yuan}
\affiliation{Institute for Molecules and Materials, Radboud University,
Heyendaalseweg 135, NL-6525AJ Nijmegen, The Netherlands}
\author{S. Miyashita}
\affiliation{
Department of Physics, Graduate School of Science,\\
University of Tokyo, Bunkyo-ku, Tokyo 113-0033, Japan
}
\affiliation{
CREST, JST, 4-1-8 Honcho Kawaguchi, Saitama, 332-0012, Japan}
\author{H. De Raedt}
\affiliation{Zernike Institute for Advanced Materials,
University of Groningen, Nijenborgh 4, NL-9747AG Groningen, The Netherlands}
\author{K. Michielsen}
\affiliation{Institute for Advanced Simulation, J\"ulich Supercomputing Centre,\\
Forschungszentrum J\"ulich, D-52425 J\"ulich, Germany}
\affiliation{%
RWTH Aachen University, D-52056 Aachen,
Germany
}%

\date{\today}

\begin{abstract}
We study measures of decoherence and thermalization of a quantum system $S$ 
in the presence of a
quantum environment (bath) $E$. The entirety $S$$+$$E$ is prepared
in a canonical thermal state at a finite temperature, 
that is the entirety is in a steady state.
Both our numerical results and theoretical predictions 
show that measures of the decoherence and the thermalization of 
$S$ are generally finite, even in the thermodynamic limit,
when the entirety $S$$+$$E$ is at finite temperature.
Notably, applying perturbation theory with respect to the system-environment 
coupling strength, we find that under
common Hamiltonian symmetries, up to first order in the coupling strength it is 
sufficient to consider $S$ uncoupled from $E$, but entangled with $E$, 
to predict decoherence and thermalization measures of $S$.
This decoupling allows closed form expressions for perturbative expansions for the
measures of decoherence and thermalization in terms of the free energies of 
$S$ and of $E$.
Large-scale numerical results for both coupled and uncoupled entireties with up to 
40 quantum spins support these findings.
\end{abstract}


\maketitle

\section{Introduction}
Decoherence and thermalization are two basic concepts
in quantum statistical physics~\cite{KUBO85}. Decoherence renders
a quantum system classical due to the loss of phase coherence of the
components of a system in a quantum superposition via
interaction with an environment (or bath). Thermalization drives the system
to a stationary state, the (micro) canonical ensemble via energy exchange
with a thermal bath. As the evolution of a quantum system is governed by the
time-dependent Schr{\"o}dinger equation, it is natural to
raise the question how the canonical ensemble could emerge from a pure quantum state.

Various theoretical and numerical studies have been performed,
trying to answer this fundamental question,
\textit{e.g.}, the microcanonical thermalization of an isolated quantum
system~\cite{YUKA11,NEUM29,PERE84,DEUT91}, canonical thermalization of a
system coupled to a (much) larger 
environment~\cite{TASA98,GOLD06,POPE06,REIM07,YUAN09,YUKA11,LIND09,
SHOR11,REIM10,GENW10,GENW13,GEMM06a}, and of two
identical quantum systems at different temperatures~\cite{PONO2011,PONO12}.
Textbooks on statistical mechanics, 
for example see \cite{LAND1980,HUANG1987,PATH2011,SWEND2012}, 
develop quantum statistical mechanics from various initial viewpoints and apply 
various assumptions and approximations.  
The standard approach to quantum statistical mechanics is to 
consider a quantum system $S$ coupled to a 
quantum environment $E$, with the time evolution of 
the entirety $S$$+$$E$ governed by the laws of 
quantum mechanics.  

There are many quantum technologies where a physical understanding and the ability to 
make quantitative predictions of quantum 
decoherence and thermalization is critical to the design and to the functioning of a 
device.  A few such technologies include 
gate-based quantum computers \cite{RIEF2011,MERM2007},
adiabatic quantum computers \cite{BOIX2014,MCG2014,ALBA15},
electron transport through nanodevices \cite{DATT2005,HANS2008}, 
and quantum dots \cite{TART2012,HANS2007}.
The ability to make finite temperature quantitative predictions  
based on quantum statistical mechanics 
is also critical to experiments in fields such as 
cold atoms \cite{BLOC2012,GARD2014,GARD2015}, 
quantum optics \cite{FOX2006},
and atom/cavity systems \cite{RAIM2001}.   
Equally important technologically is to understand when the quantum world 
allows adequate approximation in terms of classical statistical mechanics, 
with applications ranging from physical chemistry \cite{MCQU1997}
to electrical engineering and materials science \cite{SENT2004}.  

Both here and in our earlier work~\cite{JIN13a} we measure the decoherence 
of the system S in terms of $\sigma$, defined below in terms of 
the off-diagonal components of the reduced density matrix which 
describes the state of the system S. If $\sigma =0$, then the system is in 
a state of full decoherence. The difference between the diagonal elements 
of the reduced density matrix and the canonical or Gibbs distribution is 
expressed by our measure of thermalization $\delta$. Hence, for the system S 
being in its canonical distribution it is expected that its measures of 
decoherence and thermalization are zero.

In our earlier work~\cite{JIN13a} we analyzed the decoherence 
and thermalization for the quantum system $S$ being part of the quantum 
entirety $S+E$, of which the time evolution is governed by the time-dependent 
Schr{\"o}dinger equation.  
We focused on closed entireties $S+E$ with a Hilbert space of size $D=D_SD_E$ 
with $D_S$ ($D_E$) being the size of the Hilbert space of $S$ ($E$). 
We found analytically that at infinite temperature ($T = +\infty$) 
the degree of decoherence of $S$ scales with $1/\sqrt{D_E}$ 
if $D_E \gg 1 \gg D_S^{-1}$ and if the final (steady) state of the time evolution 
of the entirety $S+E$ corresponds to a state that can be picked uniformly 
at random from the unit sphere in the Hilbert space of $S+E$.
We showed that in the thermodynamic limit 
$D_E\longrightarrow +\infty$ the system $S$ decoheres thoroughly.
We demonstrated by numerically solving the time-dependent Schr\"{o}dinger 
equation (TDSE) for spin-$1/2$ ring systems that this scaling holds 
as long as the dynamics drives the initial state of $S+E$ to a state 
which has similar properties as such a random state. However, we have also shown 
that for $T$$=$$\infty$ there exist exceptions, namely entireties and initial states 
for which the  dynamics  cannot  drive  the system to decoherence. 

In this paper, we study measures of decoherence and thermalization 
of a system $S$ which is part of an entirety $S + E$ that is at a finite temperature $T$.
We mainly focus on the case that the entirety $S$$+$$E$ is in a canonical thermal state,
a pure state at finite temperature $T$~\cite{HAMS00,JIN10x,SUGI13}.
This canonical thermal state could be the resulting steady state of 
a thermalization process of the entirety $S$$+$$E$ coupled to 
a large quantum bath, a bath which we do not consider 
any further, as it has been decoupled from the entirety for a long time 
before we begin our measurements on $S$. 

The research is twofold.  
First, we perform simulations for the entireties $S$$+$$E$ being spin-$1/2$ ring systems. 
In our simulation work we first study the thermalization and decoherence process 
by solving the TDSE for an entirety at finite temperature 
starting in a canonical thermal 
state and in a product state. For both cases, the final state after some time evolution 
is a steady state which is or is close to the canonical thermal state of the entirety. 
From our infinite temperature simulations~\cite{JIN13a} we know that 
there may exist exceptions 
to this dynamical behavior. We do not consider these exceptions in this paper. 
Therefore for the remainder of our numerical simulations we assume that the entirety 
simply is in a canonical thermal state for calculating 
the measures of decoherence and thermalization.  
The Hamiltonian $H$ of the entirety includes,
besides a Hamiltonian $H_S$ and Hamiltonian $H_E$ describing the 
system and environment, respectively,
a Hamiltonian $\lambda H_{SE}$ describing the coupling of 
$S$ to $E$, with $\lambda$ the overall coupling strength.  
Our simulation results demonstrate that both 
$\sigma$ and $\delta$ are generally finite when $\lambda H_{SE}$ is not negligible.  
The finite value does not scale with $D_E$ and therefore our simulations suggest that 
this lack of complete decoherence remains even if the environment size 
goes to infinity.  
The simulation results suggest that if we want complete decoherence, 
either the entirety must be at infinite temperature 
or the entirety must be in the weak interaction regime where $\lambda H_{SE}$ 
goes to zero in the thermodynamic limit.
Our numerical results are by necessity for a particular system 
with less than forty spin-$1/2$ particles (see Fig.~\ref{figS00}). Our results can 
nevertheless be viewed as the normal behavior for any quantum entirety $S$$+$$E$. 
This statement is bolstered by the second part of our work.  

Second, we present analytical work based on perturbation theory 
for any entirety with a finite size $D$ of its Hilbert space. 
Our perturbation theory shows that the 
conclusions and inferences drawn from our large-scale simulation data 
on specific Hamiltonians $H$ for the entirety are applicable in general, 
{\it i.e.\/} applicable for any entirety.  
Furthermore, our perturbation theory provides quantitative predictions not inferred 
from our simulation data. Therefore, we performed 
additional large-scale simulations of spin-$1/2$ Hamiltonians in order to both 
test and illustrate these predictions (without any adjustable parameters).  
We perform perturbation theory for small $\langle\lambda H_{SE}\rangle$, and 
show that under symmetry transformations that leave the
Hamiltonians of $H_S$ and $H_E$ invariant but
reverse the sign of the interaction Hamiltonian $H_{SE}$,
conditions which are usually satisfied for example in
quantum spin systems, the first-order term of the perturbation expansion of
$\sigma^2$ in terms of the interaction between $S$ and $E$ is exactly zero.
Therefore, up to first order in our perturbation theory, it is sufficient to study 
the case when $\lambda H_{SE}$$=$$0$. 
Even if the first-order term in the expansion of $\lambda H_{SE}$ did not vanish, 
the leading contribution is still the zero-th order term.  
Because the entirety $S+E$ is in a pure state 
from the ensemble of all canonical thermal states, the state for the 
case $\lambda H_{SE}$$=$$0$ is not a direct product of states from $S$ and $E$. 
Hence, even the zero-th order term for the perturbation theory in 
$\lambda H_{SE}$ is not simple to calculate. 
A canonical thermal state is given by an imaginary-time projection 
$\exp\left(-\beta H/2\right)$  applied to a state drawn uniformly from the 
Hilbert space of the entirety (together with a normalization of this pure state).  
The probability that a particular state is drawn uniformly 
from the Hilbert space of the entirety is $D^{-1}$.  
These facts allow us to perform a Taylor expansion in the expectation value 
as a difference from the average of $D^{-1}$, and we calculate this expansion to 
second order.  
By combining the perturbation theory for small $\lambda H_{SE}$ with the 
Taylor expansion about the expectation values $D^{-1}$ of a random state drawn from the 
Hilbert space of the entirety, we demonstrate that 
the leading term in the expressions for $\sigma^2$ and $\delta^2$
is a product of factors of the free energy of $E$ and the free energy of $S$.
Hence, these expressions for $\sigma^2$ and $\delta^2$ allow one to study the influence
of the environment on the decoherence and thermalization of $S$ 
starting from a canonical thermal state.  In other words, 
only knowing the free energy of $S$ and of $E$ is sufficient to predict the 
degree of decoherence and thermalization that $S$ exhibits due to the influence of 
the environment $E$.  These perturbation predictions hold for any $H_S$ and $H_E$, 
not just for the spin Hamiltonians like we have studied numerically.  

The paper is organized as follows. 
In Sec.~II we describe the basic theory and provide definitions for $\sigma$, $\delta$, 
and the canonical thermal state ensemble.
The model spin-$1/2$ systems and simulation results are presented in Sec.~III.
Section~IV contains the results from our perturbation theory.  The 
perturbation derivations are very lengthy, and hence are relegated to Appendix~B.  
Further discussion of our results and additional conclusions are given in Sec.~V.

\section{Theory and Definitions}

The time evolution of a closed quantum system is governed by the TDSE~\cite{NEUM55,BALL03}.
If the initial density matrix of an isolated quantum system is non-diagonal then,
according to the time evolution dictated by the TDSE, it remains non-diagonal.
Therefore, in order to decohere the system $S$, it is necessary to have the
system $S$ interact with an
environment $E$, also called a heat bath or quantum bath, or called a spin bath 
if the environment is composed of spins.
Thus, the Hamiltonian of the entirety $S+E$ can be expressed as
\begin{equation}
\label{Eq:H}
H=H_S + H_E + \lambda H_{SE}
\>,
\end{equation}
where  $H_S$ and $H_E$ are the system and environment Hamiltonian, respectively 
and $H_{SE}$ describes the interaction between the system $S$ and the environment $E$. 
Here $\lambda$ denotes the global system-environment coupling strength. 
We focus only on Hamiltonians $H_S$, $H_E$ and $H_{SE}$ 
for the closed quantum system that are time-independent.  

The state of the quantum system $S$ is described by the reduced density matrix
\begin{equation}
\hat{\rho}(t)\equiv\mathbf{Tr}_{E}\rho \left( t\right)
\>,
\label{eq1}
\end{equation}
where 
$\rho \left( t\right) =\left|\Psi(t)\right\rangle \left \langle \Psi(t)\right |$ 
is the density matrix of the entirety $S$$+$$E$ at time $t$
and $\mathbf{Tr}_{E}$ denotes the trace over the degrees of freedom of the environment.
The state $\left|\Psi(t)\right\rangle$ of the entirety $S$$+$$E$ evolves in time 
according to (in units of $\hbar=1$)
\begin{eqnarray}
\label{Eq:TimeEv}
\left|\Psi(t)\right\rangle
&=&
e^{-itH}\left|\Psi(0)\right\rangle=
\sum_{i=1}^{D_S} \sum_{p=1}^{D_E} c(i,p,t)\left|i,p\right\rangle
\>,
\end{eqnarray}%
where the set of states $\{ |i,p\rangle \}$ denotes
a complete set of orthonormal states in some chosen basis.  
We assume that $D_S$ and $D_E$ are both finite. 
Although $\left|\Psi(t)\right\rangle$ can be decomposed in any basis, 
we find it often beneficial to use a basis that is a direct product of the states 
$|i\rangle$ of $S$ and states $|p\rangle$ of $E$, even though these states are 
not eigenstates of the entirety Hamiltonian 
in Eq.~(\ref{Eq:H}) if $\lambda\ne 0$.  
In terms of the expansion coefficients $c(i,p,t)$, the matrix element $(i,j)$
of the reduced density matrix reads
\begin{eqnarray}
\label{Eq:rhohat}
\hat\rho_{ij}(t) &=&
\mathbf{Tr}_{E} \sum_{p=1}^{D_E}\sum_{q=1}^{D_E}
c^\ast(i,q,t)c(j,p,t)\left|j,p\right\rangle\left\langle i,q\right|
\nonumber \\
&=&\sum_{p=1}^{D_E} c^\ast(i,p,t)c(j,p,t) \left|j\right\rangle\left\langle i\right|
\>.
\end{eqnarray}

\subsection{Measures of decoherence and thermalization}
We characterize the degree of decoherence of the system by~\cite{YUAN09,JIN13a}
\begin{equation}
\sigma (t) =\sqrt{\sum_{i=1}^{D_S-1}\sum_{j=i+1}^{D_S}
\left\vert\widetilde\rho_{ij}(t) \right\vert ^{2}} \>,
\label{eqsigma}
\end{equation}%
where $\widetilde\rho_{ij}(t)$ is the matrix element $(i,j)$
of the reduced density matrix $\hat\rho$ in the basis
that diagonalizes $H_S$.
It is important to emphasize that in order to study the classic canonical ensemble 
one has to study $\widetilde\rho$, wherein we have picked the basis in 
Eq.~(\ref{Eq:rhohat}) to be the eigenbasis of $H_S$ of the system $S$.  
We do not study a general ${\hat \rho}$ of Eq.~(\ref{Eq:rhohat}) 
which could be in any basis that spans $S$.  
Clearly, $\sigma(t)$ is a global measure for the size of the
off-diagonal terms of $\widetilde{\rho}$. 
If $\sigma(t)=0$ the system is in a state of full decoherence
(relative to the representation that diagonalizes $H_S$).
We define a quantity measuring the
difference between the diagonal elements of $\widetilde{\rho }$ and
the canonical distribution as~\cite{YUAN09}
\begin{equation}
\delta(t) =\sqrt{\sum_{i=1}^{D_{S}}\left( \widetilde{\rho }_{ii}(t)-\left. {%
e^{-b(t)E_{i}^{(S)}}}\right/ {\sum_{i'=1}^{D_{S}}e^{-b(t) E_{i'}^{(S)}}}%
\right) ^{2}},  
\label{eqdelta}
\end{equation}%
where $\{E_i^{(S)}\}$ denote the eigenvalues of $H_S$ and $b(t)$ is a fitting parameter 
which is given by
\begin{equation}
\label{eqbt}
b(t)=\frac{\sum_{i<j,E_{i}^{(S)}\neq E_{j}^{(S)}}
\left[\ln \widetilde\rho_{ii}(t) -
\ln \widetilde\rho _{jj}(t)\right]/({E_{j}^{(S)}-E_{i}^{(S)}})}
{\sum_{i'<j',E_{i'}^{(S)}\neq E_{j'}^{(S)}}1}.
\end{equation}
For excellent fits to the classic canonical ensemble the fitting 
parameter $b(t)$ should approach the inverse temperature $\beta=1/T$ 
(in units $k_B=1$) at large times.  
The quantities $\sigma(t) $ and $\delta(t) $ are respectively general 
measures for the decoherence and the thermalization of $S$.
The values of $\sigma(t)$ and $\delta(t)$ are generally time dependent.
If the pure state of the entirety $S+E$ is drawn from the ensemble of 
canonical thermal states at a particular temperature then these quantities are 
constant in time, except small quantum or thermal fluctuations.  
Moreover, as seen below (see Fig.~\ref{FigRelax}) for most, if not all, 
initial pure states both $\sigma(t)$ and 
$\delta(t)$ converge to a constant value after some time 
(neglecting small fluctuations). 
Therefore, in what follows we leave out the time index in the expressions 
for $\sigma$, $\delta$ and $b$. We here only study one measure of decoherence 
and one measure of thermalization, 
namely $\sigma(t)$ from Eq.~(\ref{eqsigma}) and $\delta(t)$ from Eq.~(\ref{eqdelta}).  
Any other measurement of the degree of decoherence or the degree of thermalization 
would of necessity be different functions of the 
reduced density matrix ${\widetilde \rho}_{ij}(t)$.  

In our previous work for infinite temperature~\cite{JIN13a}, 
we demonstrated that $\sigma$ and $\delta$ in 
Eqs.~(\ref{eqsigma}) and (\ref{eqdelta}) scale with the 
dimension of the Hilbert space of the environment $E$, \textit{i.e.},
\begin{equation}
\label{Eq:Tinfinity}
\sigma \propto \frac{1}{\sqrt{D_E}},
\quad\quad {\rm and} \quad\quad
\delta \propto \frac{1}{\sqrt{D_ED_S}},
\end{equation}
if the state of the entirety $S$$+$$E$ is prepared in a random state.  
In this paper, we investigate the properties of $\sigma$ and $\delta$, 
measures respectively of the decoherence and the thermalization, at finite temperatures.
This allows us to compare and contrast with the infinite-temperature results of 
\cite{JIN13a}.  

\subsection{Random state for the entirety}
A random (\textsl{i.e.} infinite-temperature) state of the entirety 
$S$$+$$E$ reads, 
\begin{equation}
\label{randomstate}
\left | \Psi_0\right \rangle \>=\> 
\sum_{i=1}^{D_S} \sum_{p=1}^{D_E} d_{i,p} \left | i,p\right \rangle
\> ,
\end{equation}
where the coefficients $\{d_{i,p}\}$ are complex Gaussian random numbers.
Note that the wave function $|\Psi_0\rangle$ must be normalized, so 
\begin{equation}
\sum_{i=1}^{D_S} \sum_{p=1}^{D_E} \left| d_{i,p} \right|^2 
\>=\> 1 
\>.  
\label{randomstatenormalize}
\end{equation}
A pure state $\left | \Psi_0\right \rangle$ 
is a state drawn uniformly at random from the unit hypersphere 
of all states of the Hilbert space of the entirety $S+E$.  
Appendix~B describes the algorithm used to 
calculate $\left | \Psi_0\right \rangle$ numerically.  
The pure state $\left | \Psi_0\right \rangle$ 
corresponds to an equilibrium state at infinite temperature for the entirety 
Hamiltonian $H$.  
The time evolution of a state is given by Eq.~(\ref{Eq:TimeEv}).   Hence 
both mathematically and physically (since at infinite temperature all states 
are equally probable) 
the time evolution of a particular state $\left | \Psi_0\right \rangle$ 
gives another pure state, one which had the same probability of being 
drawn from the ensemble.  Therefore at infinite temperature as long as one 
starts in any state $\left | \Psi_0\right \rangle$  one gets the same values for 
$\sigma$ and $\delta$ whether or not the state is evolved in time, 
except for small fluctuations \cite{JIN13a}.  

\subsection{Canonical thermal state}
A canonical thermal state is a pure state at a finite inverse
temperature $\beta $ defined by 
(the imaginary-time projection)~\cite{HAMS00, JIN10x, SUGI13}
\begin{equation}
\label{psi_ft}
\left | \Psi_\beta\right \rangle=\frac{e^{-\beta H/2} \left | \Psi_0\right \rangle}
{\left \langle \Psi_0\right | e^{-\beta H}\left | \Psi_0\right \rangle^{1/2}},
\end{equation}
where $\left | \Psi_0\right \rangle$ is a random state defined in 
Eq.~(\ref{randomstate}).
The justification of this definition can be seen from the fact that for any 
quantum observables of the entirety $S$$+$$E$~\cite{HAMS00,SUGI13},
one has
\begin{equation}
\left \langle \Psi_\beta\right | A\left | \Psi_\beta\right \rangle\approx 
\mathbf{Tr} A e^{-\beta H}/\mathbf{Tr} e^{-\beta H}.
\end{equation}
The error in the approximation is of the order of the inverse square root of the Hilbert
space size of the entirety $S$$+$$E$~\cite{HAMS00}, 
and therefore the approximation improves for increasing $D$.
One may consider the state $\left\vert \Psi_\beta
\right\rangle $ as a \textquotedblleft typical" canonical thermal state~\cite{SUGI13}, 
in the sense that if one measures observables their expectation
values agree with those obtained from the canonical distribution at the
inverse temperature $\beta $.

The time evolution of a state, Eq.~(\ref{Eq:TimeEv}), is given by 
acting on the state with the operator $e^{-i t H}$.  The imaginary time 
projection for $\left | \Psi_\beta\right \rangle$ in Eq.~(\ref{psi_ft}) uses the operator 
$e^{-\beta H/2}$.  The Hamiltonian $H$ of the entirety commutes with 
itself.  Consequently, the time evolution of a pure state 
$\left | \Psi_\beta\right \rangle$ drawn from the canonical thermal ensemble 
gives a state with the same probability of being drawn from the 
canonical thermal ensemble.  
Therefore just as at infinite temperature, at finite temperature as long as one 
starts in any state $\left | \Psi_\beta\right \rangle$ one gets the same values for 
$\sigma$ and $\delta$ whether or not the state is evolved in time, 
except for small fluctuations (for an example, see Fig.~\ref{MANdiffEtP}). 

\section{Numerical simulation}
\begin{figure}[t]
\includegraphics[width=8cm]{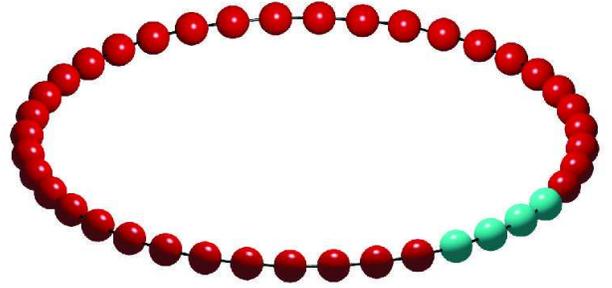}
\caption{
\label{figS00}
(Color online).  
Sketch of one of the largest entireties studied numerically.   
The environment has $N_E=36$ spins (red), and the system 
has $N_S=4$ spins (light green).  
The dimension of a vector in the Hilbert space of the entirety is 
$2^{40}=1,099,511,627,776\approx 1.1\times 10^{12}$.  
}
\end{figure}

We performed large-scale numerical simulations of a spin-$1/2$ entirety divided into a 
system $S$ and an environment $E$ 
in order to investigate the measures of decoherence $\sigma$ and 
thermalization $\delta$ of $S$. 
The geometry of one of the largest systems we have studied is shown in Fig.~\ref{figS00}.  

Most of our calculations used imaginary time projections 
to obtain a canonical thermal state (see Eq.~(\ref{psi_ft})).  
Only for the results presented in Figs.~\ref{FigRelax} and \ref{MANdiffEtP} 
we solved the TDSE for the entirety starting from the initial states 
given by Eq.~(\ref{psi_ft}) or a product state defined later,
which evolves in time according to Eq.~(\ref{Eq:TimeEv}).  

\subsection{Model and method}
We consider a quantum spin-$1/2$ model defined by the Hamiltonian
of Eq.~(\ref{Eq:H})
where
\begin{eqnarray}
\label{hamiltonian}
H_{S} &=&-\sum_{i=1}^{N_{S}-1}\sum_{j=i+1}^{N_{S}}\sum_{\alpha
=x.y,z}J_{i,j}^{\alpha }S_{i}^{\alpha }S_{j}^{\alpha }, \label{HAMS} \\ 
H_{E} &=&-\sum_{i=1}^{N_E-1}\sum_{j=i+1}^{N_E}\sum_{\alpha =x,y,z}\Omega
_{i,j}^{\alpha }I_{i}^{\alpha }I_{j}^{\alpha },  \label{HAME}\\ 
H_{SE} &=&-\sum_{i=1}^{N_{S}}\sum_{j=1}^{N_E}\sum_{\alpha =x,y,z}\Delta
_{i,j}^{\alpha }S_{i}^{\alpha }I_{j}^{\alpha }.  \label{HAMSE}
\end{eqnarray}%
Here, $S_i^\alpha$ and $I_i^\alpha$ denote the spin-$1/2$ operators of the 
spins at site $i$ of the system
$S$ and the environment $E$, respectively. The number of spins in $S$ and $E$
are denoted by $N_S$ and $N_E$, respectively. The total number of spins in
the entirety is $N=N_S+N_E$.
The parameters $J_{i,j}^\alpha$ and $\Omega_{i,j}^\alpha$ 
denote the spin-spin interactions of
the system $S$ and environment $E$, respectively, while $\Delta_{i,j}^\alpha$ denotes
the local coupling interactions between the spins of $S$ and the spins of $E$.
The dimensions of the Hilbert spaces of the system and environment are 
$D_S=2^{N_S}$ and $D_E=2^{N_E}$, respectively.

In our simulations we use the spin-up -- spin-down basis and use units such
that $\hbar =1$ and $k_{\rm B}=1$ (hence, all quantities are dimensionless).
Numerically, the imaginary- and real-time
propagations by $\exp(-\beta H)$ and $\exp(-iHt)$, respectively
are carried out by means of exact diagonalization or by using 
the Chebyshev polynomial algorithm~\cite{TALE84,LEFO91,IITA97,DOBR03,RAED06}.  
These algorithms yield results that are very
accurate (close to machine precision).
The simulations use out of necessity specific values for 
$J_{i,j}^\alpha$, $\Omega_{i,j}^\alpha$, and $\Delta_{i,j}^\alpha$.  
However, as we show in Sec.~IV the simulation results are 
representative for any quantum system $S$ coupled by any Hamiltonian 
$H_{SE}$ to any quantum bath $E$.  

\subsection{Simulation results}
We performed numerical simulations of the spin-$1/2$ Hamiltonian for the 
entirety given by Eq.~(\ref{Eq:H}), with the Hamiltonians written explicitly 
in Eqs.~(\ref{HAMS}-\ref{HAMSE}).
All simulations are carried out for a system $S$ consisting of a chain of 
$N_S=4,6,8,10$ spins coupled to an 
environment $E$ being a chain of spins with $14\le N_E\le 36$.
Two interaction bonds connect the ends of the system and the environment, 
making the entirety a ring.  The ring entireties are the same as 
some of the entireties studied at infinite temperature \cite{JIN13a}.  
The interaction strengths $J_{i,i+1}^\alpha$ with 
$1\le i\le N_S-1$ are set to $J=-1$, and all non-zero 
$\Omega_{i,j}^\alpha$ and $\Delta_{i,j}^\alpha$ are randomly 
generated from the range $[-4/3,4/3]$. 
Here we present only simulation results 
for the decoherence measure $\sigma$, 
as the thermalization measure $\delta$ behaves similarly.  
We have included the graphs for $\delta$ and 
$b$ only in Appendix~A.

\subsubsection{Different initial states}
\begin{figure}[t]
\includegraphics[width=8cm]{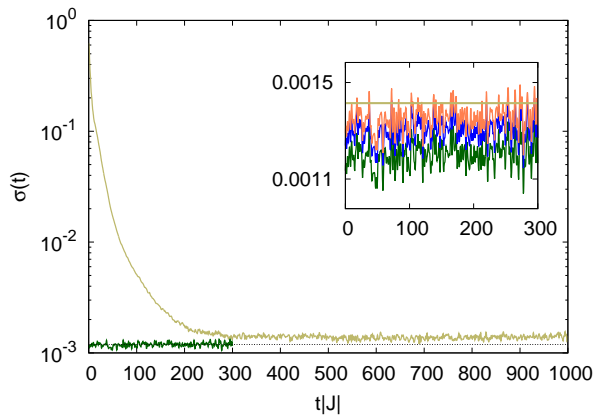}
\caption{
\label{FigRelax}
(Color online).  
Simulation results for $\sigma(t)$ for a coupled ring entirety with $N_S=4$, $N_E=22$ 
and $\lambda=1$ for two different initial states \textsl{X} (flat curve, green) 
and \textsl{UDUDY} (decay curve, dark khaki) with $\beta |J|=0.900$.  
The dotted (green) horizontal line is a guide for the eyes.
The inset shows the time average for long times for the \textsl{UDUDY} initial state 
as a horizontal line.  The bottom curve (green), the middle curve (blue) 
and the top curve (red) are for an initial state X 
with $\beta |J| =0.900$, $0.930$, $0.945$ respectively.  
}
\end{figure}

We first study the decoherence process by solving the TDSE for an entirety 
at finite temperature starting in two different initial states:
\begin{enumerate}
	\item ``\textsl{X}". The initial state of the entirety $S$$+$$E$ is in a 
        canonical thermal state defined by Eq.~(\ref{psi_ft}).
        The real-time dynamics will not play a significant role in measurements of 
        $\sigma(t)$ and $\delta(t)$ for such an initial 
        state, except for some small fluctuations 
        due to quantum and/or thermal effects.
        However, for other quantities, for example expectation values for 
        time-displaced expectation values such as 
        $\left\langle S_1^z(0) S_1^z(t)\right\rangle$, 
        the time dependence can be significant.   
	\item ``\textsl{UDUDY}". For $N_S=4$, the initial state of the entirety is a 
        product state of the system and environment.
        The first four spins (those in $S$) are in the up, 
        down, up, down state, 
        and the remaining spins (those in $E$) are in a 
        canonical thermal state ``\textsl{Y}".
\end{enumerate}
The quantum dynamics may drive the entirety with arbitrary initial state, 
including the UDUDY state, into a state which is indistinguishable from a state 
drawn from the ensemble of canonical thermal states of the entirety.  
The state observed after sufficiently long times may be expected to 
resemble a canonical state \textsl{X}. 
For an initial state \textsl{UDUDY}, the initial temperature of 
$E$ used to calculate the canonical thermal state \textsl{Y}  
will be different from the temperature of the corresponding long-time value 
of the entirety canonical thermal state \textsl{X}.  

Figure~\ref{FigRelax} presents the time evolution of $\sigma(t)$ 
for a spin entirety with 
$N_S$$=$$4$ and $N_E$$=$$22$ prepared in these two different initial states.
For both initial states the inverse temperature is set to $\beta |J|$$=$$0.900$. 
From Fig.~\ref{FigRelax}, one sees that for the entirety prepared in
the product state \textsl{UDUDY} $\sigma(t)$
evolves closely to the value obtained for the entirety prepared 
in the canonical thermal state \textsl{X}. 
Of course the fitting parameter $b$ from Eq.~(\ref{eqbt}) 
calculated for the initial state \textsl{UDUDY} is larger than 
the initial $\beta$ for the canonical state \textsl{X} because 
the initial state of the system is closer to the ground state energy.  

The bottom (green) curve (in both the main figure and the 
inset of Fig.~\ref{FigRelax}) depicts $\sigma(t)$ for 
an initial state drawn from \textsl{X} at inverse temperature $\beta |J|=0.900$, 
and has an average fitting parameter $b |J|=0.895$.  
The inset shows the time average for long times for $\sigma(t)$ for 
the \textsl{UDUDY} initial state with $\beta |J|=0.900$ (dark khaki curve).  
The standard deviation of the time average for $t>300/|J|$ of $\sigma(t)$ 
for the \textsl{UDUDY} 
initial state is $6$$\times$$10^{-5}$, while the fit to the parameter $b$ from 
Eq.~(\ref{eqbt}) gives the average $b |J|=0.926$.   
The green bottom curve in the inset is the same curve 
as shown in the main figure, for 
the initial state \textsl{X} with $\beta |J|=0.900$.  
As seen from the inset the initial states \textsl{X} (green curve) 
and \textsl{UDUDY} (dark khaki curve) lead to
different average values for $\sigma(t)$.  The final state obtained 
for the simulation with the \textsl{UDUDY} initial state 
is expected to correspond closely to an \textsl{X} 
state at a different temperature.  Therefore, in the inset we show two other 
curves for \textsl{X} states with different values of $\beta |J|$.  
The middle curve (blue) is for an initial state \textsl{X} with $\beta|J|=0.930$ 
(giving an average fitting parameter $b|J|=0.924$).  
The top curve (red) is for an initial state \textsl{X} with $\beta|J|=0.945$ 
(yielding an average fitting parameter $b|J|=939$).  
Thus for sufficiently long times, the value of $\sigma (t)$ obtained 
for the entirety being in the initial \textsl{UDUDY} state at a given 
temperature is well approximated by its value obtained for the entirety being in 
a state \textsl{X} at a different temperature.

As seen from Fig.~\ref{FigRelax} the time needed to reach a stationary value 
for $\sigma(t)$ (with small fluctuations) is quite long 
for the entirety starting in the \textsl{UDUDY} state. 
For the ring geometry of the entirety used in Fig.~\ref{FigRelax} there 
are only two terms in the interaction Hamiltonian $H_{SE}$.  
If more terms were added in $H_{SE}$ the relaxation time could be reduced 
dramatically, as was observed at infinite temperature~\cite{JIN13a}.
There are also cases in which the entirety cannot be driven into a 
state which is close to the state obtained for the entirety being 
initially in a canonical thermal state. 
For example, at infinite temperature this was observed when  
conserved quantities other than the total energy or when 
particular geometric structures were involved~\cite{JIN13a}.
Such exceptional cases will not be considered in the present paper.  

\begin{figure}[t]
\includegraphics[width=8cm]{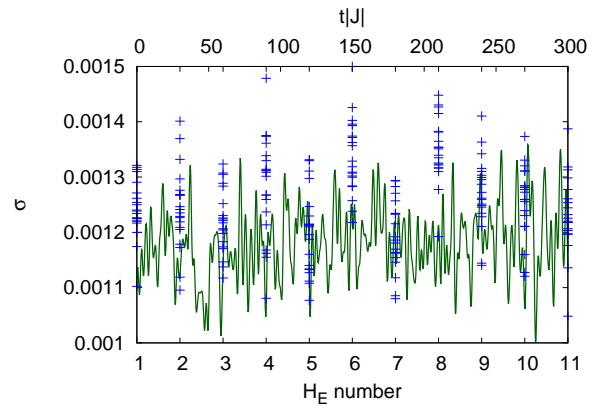}
\caption{
\label{MANdiffEtP}
(Color online).  
Simulation results for $\sigma$ for a coupled ring entirety with $N_S$$=$$4$, 
$N_E=22$ and $\lambda=1$ starting from different initial states \textsl{X}
with $\beta |J|=0.90$.  
Results for eleven different realizations of the environment Hamiltonian $H_E$ 
are shown ($x$-axis label at the bottom), 
each with different initial states drawn from the ensemble that gives an \textsl{X}
state (blue pluses).  
The time dependence of $\sigma$ for the first realization of $H_E$ 
and one of the initial states \textsl{X} 
is shown by the solid (green) curve ($x$-axis label on top) which is the 
same curve (green) as depicted in Fig.~\protect\ref{FigRelax}.  
}
\end{figure}

In principle, high statistics for our measure of decoherence $\sigma$ 
for a particular $H_S$ could be obtained from performing four different averages.  
As seen in Fig.~\ref{FigRelax}, an average over time starting 
from a particular initial \textsl{X} state could be performed.  
Another average would be an average over a large number of 
different initial states, each drawn from the ensemble that gives an \textsl{X} state.  
In addition to the time average and ensemble average over \textsl{X} states 
for a fixed environment Hamiltonian $H_E$, one could 
also average over different $H_E$.  
For each $H_E$ the coupling coefficients $\Omega_{i,j}^\alpha$ are randomly generated.
One could also average over different Hamiltonians $H_{SE}$ that couple $S$ to $E$.  
There is only one realization for $H_E$ used for the results 
shown in Fig.~\ref{FigRelax}.
In order to demonstrate that different realizations of $H_E$ do not significantly 
affect the values of $\sigma$ and $\delta$, 
we present simulation results for $\sigma$
with different $H_E$ in Fig.~\ref{MANdiffEtP}.
For each realization of $H_E$, a number of different initial states drawn from 
the ensemble that gives an \textsl{X} state at $\beta |J|=0.90$ are shown.
The average and standard deviation of $\sigma$, obtained from all 
(blue pluses) data points 
in Fig.~\ref{MANdiffEtP}, are $1.25\times 10^{-3}$ 
and $6.62\times 10^{-5}$, respectively.  
Figure~\ref{MANdiffEtP} demonstrates that the value of $\sigma$ does not differ 
significantly for different $H_E$ or for different initial \textsl{X} states.
For comparison, Fig.~\ref{MANdiffEtP} also shows the time
dependence of $\sigma$ for the first realization of $H_E$ and 
one of the initial states \textsl{X} by the green curve 
which is the same as the one in Fig.~\ref{FigRelax}.  
A high precision calculation for an average value of $\sigma$ 
would require taking into account 
a time average, an ensemble average over initial states \textsl{X}, 
and an average over different Hamiltonians $H_E$ and $\lambda H_{SE}$ 
(with fixed $D_E$ and $D_S$).  
In this paper we are interested in how $\sigma$ and $\delta$ vary with different 
values of $D_E$, $D_S$, $\beta$, and $\lambda$.  The trends we focus on do not require 
extremely high precision measurements.  
Therefore, we conclude that for our 
investigation of $\sigma$ and $\delta$ it is sufficient to consider only
one realization of $H_E$ and $H_{SE}$, one realization of the initial 
\textsl{X} state, and averaging over time is not necessarily required.  

In the remainder of the paper we focus only on the initial state of 
the entirety $S$$+$$E$ being an \textsl{X} state.
In addition, we will
omit the time index $t$ for the measures of 
decoherence $\sigma$ and thermalization $\delta$.
For entireties of size $N=N_S+N_E<32$ the
values of $\sigma$ ($\delta$) are taken either from the time averages 
or the last time step of $\sigma(t)$.  
For large system sizes ($N>32$),  
it is not necessary to perform real-time 
simulations as the fluctuations are very small (data not shown).  

\subsubsection{Coupled spin entirety}
\begin{figure}[t]
\includegraphics[width=8cm]{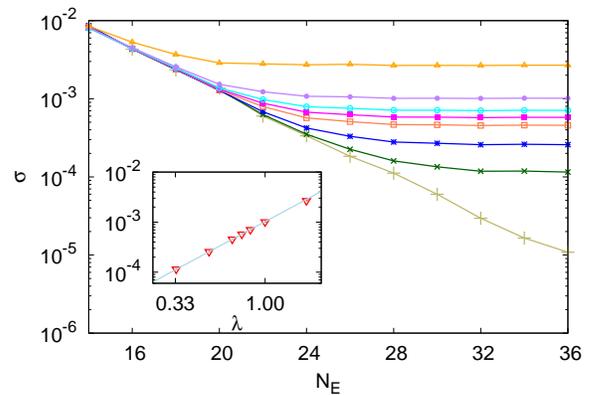}
\caption{
\label{SigmaNeLambda}
(Color online).  
Simulation results for $\sigma$ for a coupled ring entirety with $N_S$$=$$4$ 
and $N_E=14,\ldots,36$ for different global interaction strengths $\lambda$. 
The entirety is in a canonical thermal state with $\beta|J|$$=$$0.90$.  
Curves from bottom to top correspond to 
$\lambda=$ $0.00$, $0.33$, $0.50$, $0.67$, $0.75$, $0.83$, $1.00$, $1.67$.
Inset: $\sigma$ as a function of $\lambda$ for $N_E$$=$$36$.
The (light blue) solid line is a fitting curve for non-zero $\lambda$, and gives 
$\sigma\sim 0.001\lambda^2$.
}
\end{figure}

We consider the coupled ring entirety with $\lambda\neq 0$, and investigate 
how $\sigma$ behaves with changing global interaction strength $\lambda$ 
and inverse temperature $\beta$.
In all cases we start with an entirety prepared in the canonical thermal 
state \textsl{X} and measure $\sigma$. 
The strengths for the two interaction bonds in the Hamiltonian $H_{SE}$ 
are randomly generated, and are kept the same for all considered entireties. 
Note that $H_E$ is totally different for each 
realization of the environment with size $N_E$.

Figure~\ref{SigmaNeLambda} presents simulation results for $\sigma$ for 
a fixed system size $N_S=4$ and different environment sizes $N_E$. 
The initial state is prepared at inverse temperature $\beta|J|=0.90$.
From Fig.~\ref{SigmaNeLambda} two regimes with different behaviors of $\sigma$ 
as a function of $N_E$ can be observed.   
The two regimes are separated by a given environment size that 
depends on the global interaction strength $\lambda$ and is denoted by $L(\lambda)$.
For $N_E<L(\lambda)$, $\sigma$ decreases approximately exponentially 
with increasing $N_E$. 
For $N_E>L(\lambda)$, $\sigma$ converges to a finite value that also depends on 
$\lambda$. 
The smaller $\lambda$ is, the larger $L(\lambda)$ and the smaller 
the value to which $\sigma$ converges are.
We infer from this that $\sigma$ may not go to zero once $H_{SE}$ is present, 
that is once the system and environment are coupled.
This would imply that $S$
does not decohere thoroughly even when the size of the environment 
reaches the thermodynamic limit ($N_E=+\infty$). 
The inset in Fig.~\ref{SigmaNeLambda} shows 
$\sigma$ as a function of $\lambda$ for $N_E=36$.
It is seen that $\sigma \sim 0.001 \lambda^2$.
This implies that complete decoherence for $S$ 
requires both $N_E\rightarrow +\infty$ and $\lambda\rightarrow 0$. However, 
numerically we cannot rule out a slow decrease of $\sigma$ with $N_E$ for finite 
$\lambda$.  

\begin{figure}[t]
\includegraphics[width=8cm]{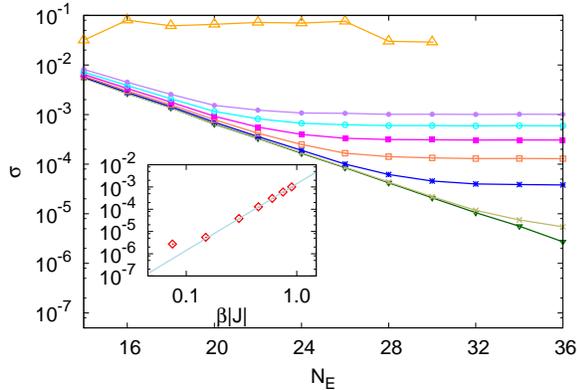}
\caption{
\label{SigmaNeBeta}
(Color online).  
Simulation results for $\sigma$ for a coupled ring entirety with $N_S=4$, 
$N_E=14,\ldots,36$ and $\lambda=1$ for different inverse temperatures $\beta$.   
Curves from bottom to top correspond to 
$\beta|J|=0.075$, $0.15$, $0.30$, $0.45$, $0.60$, $0.75$, $0.90$, $+\infty$.
Inset: $\sigma$ as a function of $\beta |J|$ for $N_E=36$.
The (light blue) solid line is a fitting curve and gives 
$\sigma \sim 0.0014\>\left|J\right|^3\beta^3$ for $\beta |J|\ge 0.15$.  
}
\end{figure}

Figure~\ref{SigmaNeBeta} 
presents simulation results for $\sigma$ for the coupled ring entirety 
for different temperatures $\beta$. 
In this case $\lambda=1$.
We observe the same features as for the results shown in Fig.~\ref{SigmaNeLambda}
for varying $\lambda$. 
In Fig.~\ref{SigmaNeBeta}, $\sigma$ first decreases approximately exponentially 
for small $N_E$,  
and then gradually converges to a finite value for large $N_E$.  
The  point of crossover shifts to larger $N_E$ for smaller values of $\beta$.
Although Fig.~\ref{SigmaNeBeta} presents only results for finite $\beta |J|<1$, we 
observe the same type of curves for finite $\beta |J|\ge 1$ (not shown).  

In Fig.~\ref{SigmaNeBeta} we also present results for the entirety 
being in the ground state ($\beta = +\infty$). 
We used the Lanczos algorithm to obtain the ground state of the entirety $S$$+$$E$.
The fluctuations of $\sigma$ for different $N_E$ are large compared to the 
fluctuations in the results for $\sigma$ at finite temperature.
One cause of this is the unavoidable error made in finding the exact ground state,  
leading to a different effective inverse temperature $\beta$ for different $N_E$. 
Another cause is that for every value of  $N_E$ the bath is 
completely different, and for each value of $N_E$ we 
performed the Lanczos calculations for only one particular bath 
described by the Hamiltonian $H_E$.  
Different baths (different values of the $\Omega_{i,j}^\alpha$ in Eq.~(\ref{HAME})) 
for the same value of $N_E$ may be expected to give very different values for $\sigma$, 
which should be more pronounced for large value of $N_E$ at low temperature.  
Due to limited computer resources, it was not possible to 
run the Lanczos for even larger systems. 
Within the calculational accuracy and with these caveats, 
we speculate that $\sigma$ is flat and converges to a large value 
at the ground state.

The insets of Figs.~\ref{SigmaNeLambda} 
and~\ref{SigmaNeBeta} present the results for $\sigma$ as a 
function of $\lambda$ and $\beta$, respectively for $N_E=36$.  
At relatively large values of $\lambda$ and $\beta$, $\sigma$ already approaches its 
plateau value for $N_E=36$.  
The only outlier point is for $\beta|J|=0.075$ 
in the inset of Fig.~\ref{SigmaNeBeta}. 
We ignored this point in the fit because from Fig.~\ref{SigmaNeBeta} the 
asymptotic value for large $N_E$ had not yet been reached for $N=40$ spins.  
From these insets we find that the plateau values for $\sigma$ for large $N_E$ 
can be fitted well by functions of $\lambda^2$ and $\beta^3$ 
for $\lambda<1$ and $\beta |J| <1$.  

We have previously shown that $\sigma$ goes to zero in the thermodynamic limit if 
$\beta=0$~\cite{JIN13a} [see Eq.~(\ref{Eq:Tinfinity})].
From Figs.~\ref{SigmaNeLambda} and~\ref{SigmaNeBeta}, 
it can be concluded that for large sizes of the environment,
$\sigma$ converges to a value $(\beta\lambda)^2(c_2+c_3\beta)$ for 
$0.1<\beta |J|<1$ and $0.33<\lambda<1$, 
where the coefficients $c_2$ and $c_3$ depend on the specific 
form of the interaction Hamiltonian $H_{SE}$, even in the thermodynamic limit.
The presence of finite interactions between the system and the environment results in 
the system not decohering thoroughly ($\sigma$ remains finite) 
even when the size of the environment goes to infinity ($D_E\rightarrow+\infty$).
In order to retrieve $\sigma \rightarrow 0$ in the 
thermodynamic limit ($D_E\rightarrow+\infty$), 
one might have to go simultaneously to the weak interaction region. 
Hence complete decoherence of the system with fixed $N_S$ 
at finite temperature may require a limiting procedure in which 
$N_E \lambda$ is kept fixed as $N_E\rightarrow +\infty$ and $\lambda\rightarrow 0$. 

\begin{figure}[t]
\includegraphics[width=8cm]{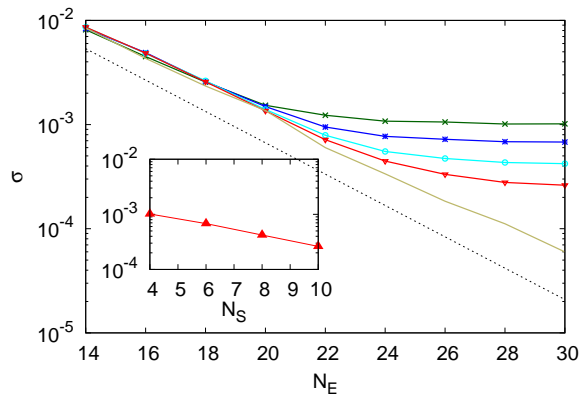}
\caption{\label{fig5}
(Color online).  
Simulation results for $\sigma$ for a coupled ring entirety with $N_S=4, 6, 8, 10$ 
(symbols, top to bottom), $N_E=14,\ldots,30$ and $\lambda=1$ for 
$\beta|J|=0.90$. 
The solid (dark khaki) line depicts the simulation results for the uncoupled entirety 
($\lambda=0$) with $\beta|J|=0.90$. 
The dotted line depicts the analytical results for infinite temperature
\protect\cite{JIN13a}. 
Inset: $\sigma$ as a function of $N_S$ for $N_E=30$.  
}
\end{figure}

All the results shown in Fig.~\ref{SigmaNeLambda} and 
\ref{SigmaNeBeta} are for system size $N_S=4$.
In Fig.~\ref{fig5}, we present results for different system sizes 
$N_S=4, 6, 8, 10$.  
It is seen that the values of $\sigma$ converge to a different finite value
for different $N_S$, and this value decreases as $N_S$ increases.
Therefore, $\sigma$ might go to zero if 
$N_S\rightarrow +\infty$ and $N_E \rightarrow+\infty$.
Effectively in this limit one enters the weak interaction regime 
for a ring geometry because $\lambda$ is fixed while both $N_E$ and $N_S$ 
approach infinity.  

\begin{figure}[t]
\includegraphics[width=8cm]{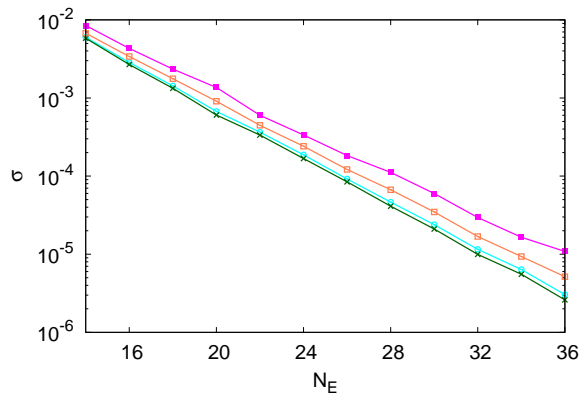}
\caption{
\label{FigUncoupled}
(Color online).  
Simulation results for $\sigma$ for an uncoupled entirety ($\lambda=0$) with $N_S=4$ and 
$N_E=14,\ldots,36$ for different inverse temperatures.
Curves from bottom to top correspond to $\beta|J|=0.075$, $0.30$, $0.60$, $0.90$.
}
\end{figure}

\subsubsection{Uncoupled spin entirety}
As shown in the previous section, one may have $\sigma=0$ in the thermodynamic limit 
if $\lambda$ goes to zero (see Fig.~\ref{SigmaNeLambda}).
The uncoupled case ($\lambda=0$) is a special case which we explore further 
in this section. 
Even though $\lambda H_{SE}=0$ the states of the entirety which are 
drawn from the ensemble of 
canonical thermal states (see Eq.~(\ref{psi_ft})) are not direct product states.  
In other words, the states of $S$ and $E$ are 
entangled even if $\lambda=0$, because the entirety is prepared in a 
canonical thermal state.  
Figure~\ref{FigUncoupled} shows the simulation results of $\sigma$ for an 
uncoupled entirety as a 
function of the size of the environment $N_E$ 
for a number of values for the inverse temperature $\beta$.
The value of $\sigma$ decreases approximately exponentially 
with the size of the environment.  

In Fig.~\ref{FigUncoupled} the absolute value of the slope decreases slightly as 
$\beta|J|$ increases.  When $\beta \rightarrow +\infty$, the slope of $\sigma$ becomes 
zero and the curve is a horizontal line.  
The entirety stays in the ground state as $\beta \rightarrow +\infty$.
If the ground state of $S$ is 
non-degenerate then $\sigma=0$, and if the ground state of $S$ is 
degenerate then $\sigma$ is generally finite for $\beta \rightarrow +\infty$. 

\subsection{Summary of simulation results}
Unlike what we found in our previous work for $\beta=0$ \cite{JIN13a}, 
at finite $\beta$ the behavior of our measure $\sigma$ for the decoherence of $S$ 
is quite different.  
For any finite values of $\beta$ and $\lambda$, 
$\sigma$ decreases approximately exponentially with $N_E$ if $N_E$ is smaller than 
a certain threshold, and converges to a finite value for large $N_E$.
This implies that $S$ will not totally decohere even if $N_E\rightarrow +\infty$.
The numerical results suggest that $\sigma \approx (\beta\lambda)^2(c_2+c_3\beta)$ for 
certain ranges of $\lambda$ and $\beta$ in the thermodynamic ($N_E\rightarrow +\infty$) 
limit.  In order to have $\sigma=0$ in the thermodynamic limit, 
either $\beta$ goes to zero (our previous results~\cite{JIN13a}), 
or $\lambda$ goes to zero, which is an uncoupled entirety.
We emphasize that the uncoupled entirety must be understood as a limiting 
case of $\lambda\rightarrow 0$, since the states of $S$ and $E$ are entangled 
in a canonical thermal state \textsl{X}.
If one instead directly starts with the initial entirety state being an uncoupled 
direct product state, 
then the dynamics always will remain a direct product state.  

We stress that the calculations presented in this section were extremely expensive 
to perform in terms of computer resources.  
Computer memory and CPU time put limitations on the size of the quantum system that 
can be simulated. The required CPU time is mainly determined by the number of operations 
to be performed and does not currently put a hard limit on the simulation.  
However, the memory of the computer does put on a hard limitation. We have studied 
sizes of the entirety $S$$+$$E$ ranging from $N=18$ to $N=40$. 
The largest and most costly 
simulations were the computations of the decoherence for a $N=40$ spin-$1/2$ system at 
various temperatures $\beta$ and global interaction strengths $\lambda$. 
It took about $1.6$ million core hours to complete the eight data points for 
$N_E$$=$$36$ ($N$$=$$40$) in Fig.~\ref{SigmaNeLambda} on 
131,072 processors of JUQUEEN, an IBM 
Blue~Gene/Q located at the J{\"u}lich Supercomputer Centre in 
J{\"u}lich Germany \cite{JUQUEEN15}.  The $N=40$ points require using $64$~TB (Tera bytes) of memory 
(SDRAM-DDR3) just to store the four required wave vectors.  However some additional 
memory is required to store other quantities, necessitating to run with an 
allocation of $128$~TB spread over the 131,072 processors.  

\section{Perturbation theory}
Most of the interesting numerical results in Sec.~III are based on an initial state 
of the type ``\textsl{X}", 
which means that the entirety is in a canonical thermal state.
As seen in Figs.~\ref{FigRelax} and \ref{MANdiffEtP}, except for small fluctuations 
the quantum dynamics does not play a significant role for our decoherence measure 
$\sigma(t)$ [nor does it play a significant role for $\delta(t)$]. 
Therefore, we again leave the time index $t$ from our expressions for $\sigma$
and $\delta$. 
This allows us to perform certain analytical calculations 
dealing only with the imaginary-time propagation $\exp(-\beta H/2)$
of Eq.~(\ref{psi_ft}), which we do here.
The derivations are long, and hence only the sketch of the calculations and the 
final results are presented in the main text.  The long derivations are relegated to 
Appendix~B.  
Especially for the uncoupled entirety $S$$+$$E$ ($\lambda=0$), 
we are able to derive closed 
forms for the measures of decoherence and thermalization, namely $\sigma$ and $\delta$.  
It is important to remember that even when $\lambda=0$ 
the state of the entirety is not a direct 
product state of states of $S$ and $E$. 
These closed forms for $\sigma$ and $\delta$ 
may be useful for understanding and making predictions of physical 
systems in certain circumstances. 
For the coupled case, we derive the first-order perturbation term in the global 
interaction strength $\lambda$, and show that the first order term 
is exactly zero if the system obeys a certain common symmetry introduced below.
The vanishing of the first order term in $\lambda$ means that the results of the 
closed expressions for the uncoupled entirety fit extremely well results for 
the coupled entirety at small values of $\lambda\beta$.  

Hereafter, we investigate the properties of the decoherence measure $\sigma$ 
of a quantum system $S$ when the entirety $S$$+$$E$ is in the canonical thermal state 
[see Eq.~(\ref{psi_ft})]. In essence, our calculations average over the 
entire ensemble of canonical thermal states \textsl{X} 
for a fixed $\beta$ for any entirety Hamiltonian $H$.  

\subsection{Canonical thermal state}
In the eigenenergy basis $\{\left | E_k \right \rangle\}$ of the Hamiltonian $H$ 
of the entirety, the state of Eq.~(\ref{psi_ft}) is given by
\begin{equation}
\left | \Psi_\beta\right \rangle \>=\>
\sum_{k=1}^D \frac{d_k e^{-\beta E_k/2}}{\sqrt{\sum_{k'=1}^D|d_{k'}|^2 
e^{-\beta E_{k'}}}}\left | E_k \right \rangle
=\sum_{k=1}^D a_k \left | E_k \right \rangle,
\end{equation}
where $a_k$ is given by 
\begin{eqnarray}
a_k&=& \frac{d_k p_k^{1/2}}{\sqrt{\sum_{k'=1}^D |d_{k'}|^2 p_{k'}}}, \\
p_k&=& \frac{e^{-\beta E_k}}{\sum_{k'=1}^D e^{-\beta E_{k'}}}.
\end{eqnarray}
Note that, in general, the probability density of the coefficient $a_k$ is not 
Gaussian any more as it was at infinite temperature.
The $\{a_k\}$ satisfy the required normalization condition, 
$\sum_{k=1}^D \left|a_k\right|^2 \>=\>1.$
For sufficiently large $D$ (the dimension of the entirety), 
we have~\cite{JIN10x}
\begin{equation}
\sum_{k=1}^D |d_k|^2 p_k \approx \frac{1}{D} \>.
\label{norm1}
\end{equation}
Eq.~(\ref{norm1}) is a good approximation for all values of $\lambda$ and $\beta$
(see Fig.~\ref{fig0} in Appendix~B), 
in fact Eq.~(\ref{norm1}) is exact both for $\beta=0$ and $\beta=\infty$.  
Therefore, the canonical thermal state can be written to a good approximation as
\begin{equation}
\left | \Psi_\beta\right \rangle 
\>=\> 
D^{1/2} \sum_{k=1}^D d_k p_k^{1/2} \left | E_k \right \rangle.
\label{psi_ft2}
\end{equation}

\subsection{Uncoupled entirety with Eq.~(\ref{psi_ft2}) approximation}
\label{sec_ucsA}
First we consider an uncoupled entirety with $H_{SE}=0$ or $\lambda=0$.
There exist simple relations for the eigenvalues $E_{k}$ (eigenstates $%
|E_{k}\rangle $) of the entirety Hamiltonian $H$ in terms of the eigenvalues $%
E_{i}^{(S)}$, $E_{p}^{(E)}$ (eigenstates $|E_{i}^{(S)}\rangle $, 
$|E_{p}^{(E)}\rangle $) of the system Hamiltonian $H_{S}$ and environment
Hamiltonian $H_{E}$, respectively, \textit{i.e.}, 
$E_{k}=E_{i}^{(S)}+E_{p}^{(E)}$ and 
$\left\vert E_{k}\right\rangle =\left\vert E_{i}^{(S)}\right\rangle \left\vert
E_{p}^{(E)}\right\rangle $.
The canonical thermal state reads (from the Eq.~(\ref{psi_ft2}) approximation)
\begin{equation}
\left | \Psi_\beta\right >=\>
D^{1/2} \sum_{i=1}^{D_S}\sum_{p=1}^{D_E} d_{i,p} p_{i,p}^{1/2}
\left | E_i^{(S)} \right \rangle
\left | E_p^{(E)} \right \rangle.
\label{psi_ft3}
\end{equation}
The matrix element ($i,j$) of the reduced density matrix of $S$, 
in the basis that diagonalizes $H_S$, is given by
\begin{equation}
\widetilde{\rho}_{ij}\>=\>
\mathbf{Tr}_E \left | \Psi_\beta\right \rangle \left \langle \Psi_\beta\right |
\>=\>
D \sum_{p=1}^{D_E} d_{i,p}^*p_{i,p}^{1/2} d_{j,p} p_{j,p}^{1/2}.
\end{equation}
The expectation value of the off-diagonal matrix elements ($i\neq j$) 
with respect to the probability distribution of 
the random variables $d_{i,p}$ is given by~\cite{HAMS00,JIN13a}
\begin{eqnarray}
\label{ar1}
{\cal E}\left(2\sigma^2\right)&=&
   {\cal E}\left(\sum_{i\neq j}^{D_S} \left | D
   \sum_{p=1}^{D_E} d_{i,p}^*p_{i,p}^{1/2} d_{j,p} p_{j,p}^{1/2}  \right |^2 \right ) \cr
&=& D^2 \sum_{i\neq j}^{D_S}
   \sum_{p=1,p^\prime=1}^{D_E} 
   {\cal E}\left(  d_{i,p}^* d_{j,p}d_{i,p^\prime}d_{j,p^\prime}^* \right)
   p_{i,p}^{1/2} p_{j,p}^{1/2} p_{i,p^\prime}^{1/2} p_{j,p^\prime}^{1/2} \cr
&=& D^2 \sum_{i\neq j}^{D_S}\sum_{p=1}^{D_E} 
    {\cal E}\left(  |d_{i,p}|^2 |d_{j,p}|^2\right)p_{i,p} p_{j,p} \cr
&=& D^2 E\left(  |d_{i,p}|^2 |d_{j,p}|^2\right) \left (1- 
   \frac{Z_S(2\beta)}{Z_S^2(\beta)} \right )\frac{Z_E(2\beta)}{Z_E^2(\beta)},
\end{eqnarray}
where $Z_{S}(n\beta)=\sum_{i} e^{-n\beta E_i^{(S)}}$ and 
$Z_{E}(n\beta)=\sum_{p} e^{-n\beta E_p^{(E)}}$ denote the 
partition functions of the system $S$ and the environment $E$ at inverse temperature $n\beta$, 
respectively.  
Here and in the following ${\cal E}(\cdot)$ denotes
the expectation value with respect to the probability distribution
of the random numbers $\{d_{i,p}\}$.
We change from the partition function to the free energy
\begin{equation}
Z(n\beta)=\sum_{k} e^{-n\beta E_k}=e^{-n\beta F(n\beta)},
\end{equation}
where $F(n\beta)=-(n\beta)^{-1}\ln Z(n\beta)$, 
for either the entirety (no subscript), the system with subscript $S$, or 
the environment with subscript $E$.  
We have
\begin{eqnarray}
{\cal E}\left(\sigma^2\right)&=&
  \frac{D^2}{2} {\cal E}\left(  |d_{i,p}|^2 |d_{j,p}|^2\right)  \cr
  &&\times\left (1-e^{-2\beta \left(F_S(2\beta)-F_S(\beta)\right)} \right )
    e^{-2\beta \left(F_E(2\beta)-F_E(\beta)\right)} \cr
&=&\frac{D}{2(D+1)} \left (1-e^{-2\beta \left(F_S(2\beta)-F_S(\beta)\right)} \right ) \cr
&& \times e^{-2\beta \left(F_E(2\beta)-F_E(\beta)\right)},
\label{sigma1}
\end{eqnarray}
where ${\cal E}\left(  |d_{i,p}|^2 |d_{j,p}|^2\right)={1}/{D(D+1)}$~\cite{HAMS00}.
From Eq.~(\ref{sigma1}), we see that $\sigma$ scales with the size of the environment 
for the uncoupled entirety because 
the free energy $F_E$ scales with the size of the environment.
Hence, $\sigma$ goes to zero in the thermodynamic limit ($N_E\rightarrow +\infty$) 
for this uncoupled case.  

For $\delta$, we obtain the following expression
\begin{eqnarray}
&& {\cal E}\left(\delta^2\right) = \cr
&&\frac{D}{D+1} e^{-2\beta\left(F_S(2\beta)-F_S(\beta)\right)}
  \left(e^{-2\beta\left(F_E(2\beta)-F_E(\beta)\right)}-\frac{1}{D}\right)
\label{delta1}
\end{eqnarray}
from a similar analysis.  

\subsection{Uncoupled entirety with full $\left|\Psi_\beta\right\rangle$}
\label{sec_ucsF}
These expressions Eq.~(\ref{sigma1}) and (\ref{delta1}) only work for very 
high or very low temperatures where the approximation in Eq.~(\ref{psi_ft2}) is valid.
The reason is that the derivation of Eqs.~(\ref{sigma1}) and~(\ref{delta1}) 
is based on an approximate expression of the 
canonical thermal state [see Eq.~(\ref{psi_ft3})] by using Eq.~(\ref{norm1}).
In order to improve the above results, we have to perform calculations which start 
from the canonical thermal state in Eq.~(\ref{psi_ft}).
We perform a Taylor series expansion of $\sigma^2$ up to second order in $|d|^2$ 
about the value $1/D$, and then
calculate the expectation value of $\sigma^2$. A very lengthy calculation, 
relegated to Appendix~B, gives
\begin{eqnarray}
&&{\cal E}\left(\sigma^2\right)= \frac{1}{2}e^{-2\beta 
  \left(F_E(2\beta)-F_E(\beta)\right)}
  \left(1-e^{-2\beta \left(F_S(2\beta)-F_S(\beta)\right)}\right)  \cr
&& - \frac{2D}{D+1}e^{-3\beta \left(F_E(3\beta)-F_E(\beta)\right)}  \cr
&&\times \left(e^{-2\beta \left(F_S(2\beta)-F_S(\beta)\right)}
  -e^{-3\beta \left(F_S(3\beta)-F_S(\beta)\right)}  \right)   \cr
&& + \frac{3D}{2(D+1)} e^{-4\beta \left(F_E(2\beta)-F_E(\beta)\right)}
   e^{-2\beta \left(F_S(2\beta)-F_S(\beta)\right)}		\cr
&&\times \left(1-e^{-2\beta \left(F_S(2\beta)-F_S(\beta)\right)} \right)
\>.
\label{sigma2}
\end{eqnarray}
Obviously, in most cases the first term will dominate, which 
approaches Eq.~(\ref{sigma1}) for $D$ large.

Two special cases are of interest. If $\beta =0$, we recover the previous result 
${\cal E}\left(\sigma ^{2}\right)=\frac{D_{S}-1}{2(D+1)}$~\cite{JIN13a}. 
In the vicinity of $\beta =0$, the first-order term of the Taylor expansion of
Eq.~(\ref{sigma2}) vanishes. Hence in the high temperature limit,
${\cal E}\left(\sigma^{2}\right) =
\frac{D_{S}-1}{2(D+1)}+\mathcal{O}\left(\beta ^{2}\right)$.

If the temperature approaches zero, Eq.~(\ref{sigma2}) becomes
\begin{equation}
\label{Eq:gg}
\lim_{\beta \rightarrow +\infty }E(\sigma ^{2})=\frac{g_{S}-1}{2g_{S}g_{E}}%
\left( 1-\frac{D_SD_E}{\left(D_S D_E+1\right)g_{S}g_{E}}\right) ,  \label{sigmalt}
\end{equation}%
where $g_{S}$ and $g_E$ refer to the degeneracy of the ground state of the system $S$ 
and environment $E$, respectively. This expression yields zero if the
ground state of the system is non-degenerate.  
For a system with a highly degenerate
ground state ($g_{S}\gg 1$) the expression goes to $1/2g_{E}$.
For a system with known $g_S>1$ and a large environment $D_E \gg 1$, at small $\lambda$
and at low temperature, if one measures
${\cal E}\left(\sigma^2\right)$, one can determine the degeneracy $g_E$ of the 
ground state
of the environment.  This is a new, strong prediction.   The ground state degeneracy 
$g_E$ of the environment can be obtained 
by only measuring quantities in the system $S$.  

Similarly, we can make the Taylor expansion for $\delta^2$ up to second order with 
respect to both $|d|^2$ and $b$ about the values $1/D$ and $\beta$, respectively. 
The full derivation is in Appendix~B.  
The expectation value of $\delta^2$ is given by
\begin{eqnarray}
&& {\cal E}\left(\delta^2\right)=\frac{D}{D+1} 
   e^{-2\beta \left(F_E(2\beta)-F_E(\beta)\right)}
   \left(e^{-2\beta \left(F_S(2\beta)-F_S(\beta)\right)} \right . \cr 
&& \left. -2e^{-3\beta \left(F_S(3\beta)-F_S(\beta)\right)}+ 
   e^{-4\beta \left(F_S(2\beta)-F_S(\beta)\right)}\right) \cr 
&&+e^{-2\beta \left(F_S(2\beta)-F_S(\beta)\right)} 
  \left[ \left(C_S(2\beta)/(4\beta^2)\right) \right. \cr 
&& \left. + \left( U_S(2\beta) -U_S(\beta) \right)^2)\right] \left(\Delta b\right)^2,
\label{delta}
\end{eqnarray}
where $\Delta b=b-\beta$, 
$C_S(n\beta)$ and $U_S(n\beta)$ are, respectively, the specific heat and average energy 
of the system $S$ at inverse temperature $n\beta$. It is obvious that for the
uncoupled entirety $b=\beta$. For the coupled entirety, as we find below, $b$ is not
necessarily equal to $\beta$, but should usually be close to the value of $\beta$.

\subsection{Coupled entirety}
For a generic entirety, a system $S$ is coupled to an environment $E$.
To solve such a coupled entirety analytically, we have to resort to a perturbation 
theory. 
Up to first order in the global system-environment coupling strength $\lambda$, 
we have~\cite{WILC67}
\begin{equation}
e^{-\beta {H}} \approx \left(1 - \left\{ \int_0^1 d\xi e^{-\beta\xi%
{H}_0} {H}_{SE} e^{\beta\xi{H}_0}\>\> \right\} \beta
\lambda \right)e^{-\beta {H}_0},
\label{expbH}
\end{equation}
where ${H}_0=H_S+H_E$ denotes the Hamiltonian of the uncoupled system and environment.

The first-order perturbation comes from both the denominator and numerator 
of Eq.~(\ref{psi_ft}).
First let us deal with the denominator.
Up to the first order, we have
\begin{eqnarray}
&&D \left \langle \Psi(0)\right | e^{-\beta H}\left | \Psi(0)\right \rangle \cr
&&\approx
{\bf Tr} e^{-\beta H_0} - \beta \lambda \int_0^1 d\xi 
{\bf Tr} e^{-\beta\xi H_0} H_{SE}
e^{-\beta(1-\xi) H_0}.
\label{denominator}
\end{eqnarray}

Hereafter, we introduce a kind of symmetry which makes the first-order term in 
Eq.~(\ref{denominator}) be zero, 
and restrict ourselves to a system which obeys such a symmetry.
The symmetry is a kind of unitary transformation such that if we reverse the 
components in the system $S$ or in the environment $E$, 
the sign of the interaction Hamilton $H_{SE}$ is reversed while
the Hamiltonians $H_S$ and $H_E$ are unchanged.
Let $Z_0$ be the partition function of the unperturbed system 
(the uncoupled entirety where $H_{SE}$$=$$0$).
The complete symmetry requirement can easily be seen by performing the integration 
over $\xi$ in Eq.~(\ref{denominator}) to give 
\begin{equation}
D \left \langle \Psi(0)\right | e^{-\beta H}\left | \Psi(0)\right \rangle 
\approx
Z_0
- \beta \lambda {\bf Tr}_{S,E} \left( H_{SE}
e^{-\beta H_E} e^{-\beta H_S} \right)
\> ,
\label{denominatorInt}
\end{equation}
and asking when the trace that multiplies $\beta\lambda$ vanishes.  
With such a symmetry involved, it is clear that the first-order term in 
Eq.~(\ref{denominator}) has to be zero. 
Then the first-order perturbation term can only come from the numerator of 
Eq.~(\ref{psi_ft}).

Consequently up to the first order, we have
\begin{equation}
\left \langle \Psi(0)\right | e^{-\beta H}\left | \Psi(0)\right \rangle\approx
{\bf Tr} e^{-\beta H_0}/D=Z_0/D
\> .
\end{equation}
The wave function is thus given approximately by
\begin{eqnarray}
&&\left|\Psi_\beta\right\rangle
\>\approx \>  
\sqrt{\frac{D}{Z_0}}\> 
e^{-\beta H/2}\left|\Psi(0)\right\rangle   \cr
&\approx &\sqrt{\frac{D}{Z_0}}\>    \left(1 - \left\{
\int_0^1 d\xi e^{-\beta\xi H_0 /2} H_{SE}
e^{\beta\xi H_0 /2}\>\> \right\}
\beta \lambda/2  \right)  \cr
&&\times e^{-\beta H_0/2}     \left|\Psi(0)\right\rangle
\>.
\label{psi_numerator}
\end{eqnarray}
Based on the expression in Eq.~(\ref{psi_numerator}), 
we find that the first-order term of the
perturbation expansion in $\lambda$ of the expectation value of $\sigma^2$ is given by
\begin{eqnarray}
&& {\cal O}\left({\cal E}\left(\sigma^2\right)\right)_{\lambda^1} 
 =
 - \beta\lambda  \left(\frac{D}{Z_0}\right)^2\frac{D}{D+1}  \cr
&&\times   \left[
 Z_S {\bf Tr} e^{-\beta H_S}e^{-2\beta H_E} H_{SE} 
\right .
  \left . -
 {\bf Tr} e^{-2\beta (H_S+H_E)}H_{SE}
 \right ].
\label{firstorder}  
\end{eqnarray}
Applying the same symmetry transformation as discussed before results in 
${\cal O}\left({\cal E}\left(2\sigma^2\right)\right)_{\lambda^1} =0$.
In other words, the same symmetry that makes the $\beta\lambda$ term in 
Eq.~(\ref{denominatorInt}) zero will make both traces in 
Eq.~(\ref{firstorder}) zero.  
Hence, to study the decoherence of a system $S$ coupled to an environment $E$
up to first order in $\lambda$ it is 
{\it sufficient to study the uncoupled entirety\/} ($\lambda=0$) 
(see the results in Sec.~\ref{sec_ucsF}).  

Calculating the second-order perturbation term of $\sigma^2$ is much more complicated 
as the perturbation term 
comes from both the denominator and numerator of Eq.~(\ref{psi_ft}).
In terms of perturbation theory, the reduced density matrix of $S$ can be written by
\begin{eqnarray}
\widetilde{\rho} &=& \frac{\mathbf{Tr}_E e^{-\beta H/2} \left | \Psi(0)\right \rangle 
    \left \langle \Psi(0)\right | e^{-\beta H/2}}
    {\left \langle \Psi(0)\right | e^{-\beta H}\left | \Psi(0)\right \rangle}  \cr
&=& \widetilde{\rho}_0+\beta\lambda \widetilde{\rho}_1 
    + (\beta\lambda)^2 \widetilde{\rho}_2+\cdots \>\>,
\end{eqnarray}
where $\widetilde{\rho}_0$ is the zeroth-order term which represents the reduced 
density matrix of the uncoupled entirety ($\lambda=0$), and $\widetilde{\rho}_1$ 
and $\widetilde{\rho}_2$ are matrices representing the first- and second-order 
perturbation terms.
We have shown that $\widetilde{\rho}_1=0$ if the Hamiltonian of the entirety has 
the previously discussed symmetry.
If $\widetilde{\rho}_2$ or higher-oder terms are non-zero, 
then $\sigma$ will be finite at finite $\lambda$.
If $\beta\lambda\ll 1$, we can safely use the results 
obtained from the uncoupled entirety 
for the measures of decoherence and thermalization.  It is important to remember 
that the initial state of uncoupled entirety ($\lambda=0$) is not a direct product 
state of states of $S$ and $E$.  

\subsection{Verification by spin Hamiltonians}
From Eqs.~(\ref{HAMS}-\ref{HAMSE}) it is seen that the Hamiltonian of the spin entirety 
obeys the symmetry property required to make the first-order term $\lambda^1$ of the 
perturbation expansion
of the expectation value of $\sigma^2$ [see Eq.~(\ref{firstorder})] exactly zero.
Namely, reversing all spin components of the system or of the environment spins 
does not change $H_S$ or $H_E$, 
but the sign of $H_{SE}$ changes. Note that such a symmetry is also obeyed in the 
case that there is no interaction
between the environment spins, \textit{e.g.} for an environment Hamiltonian
$H_{E}=-\sum_{i=1}^{N_{E}}\sum_{\alpha =x,y,z}
h_{i}^{\alpha }I_{i}^{\alpha }$~\cite{NOVO12b,NOVO12a}.
In this particular case, it is only required that $H_S$ is an even function 
and $H_{SE}$ an odd function under reversal of all spin components of the system spins.

\begin{figure}[t]
\includegraphics[width=8cm]{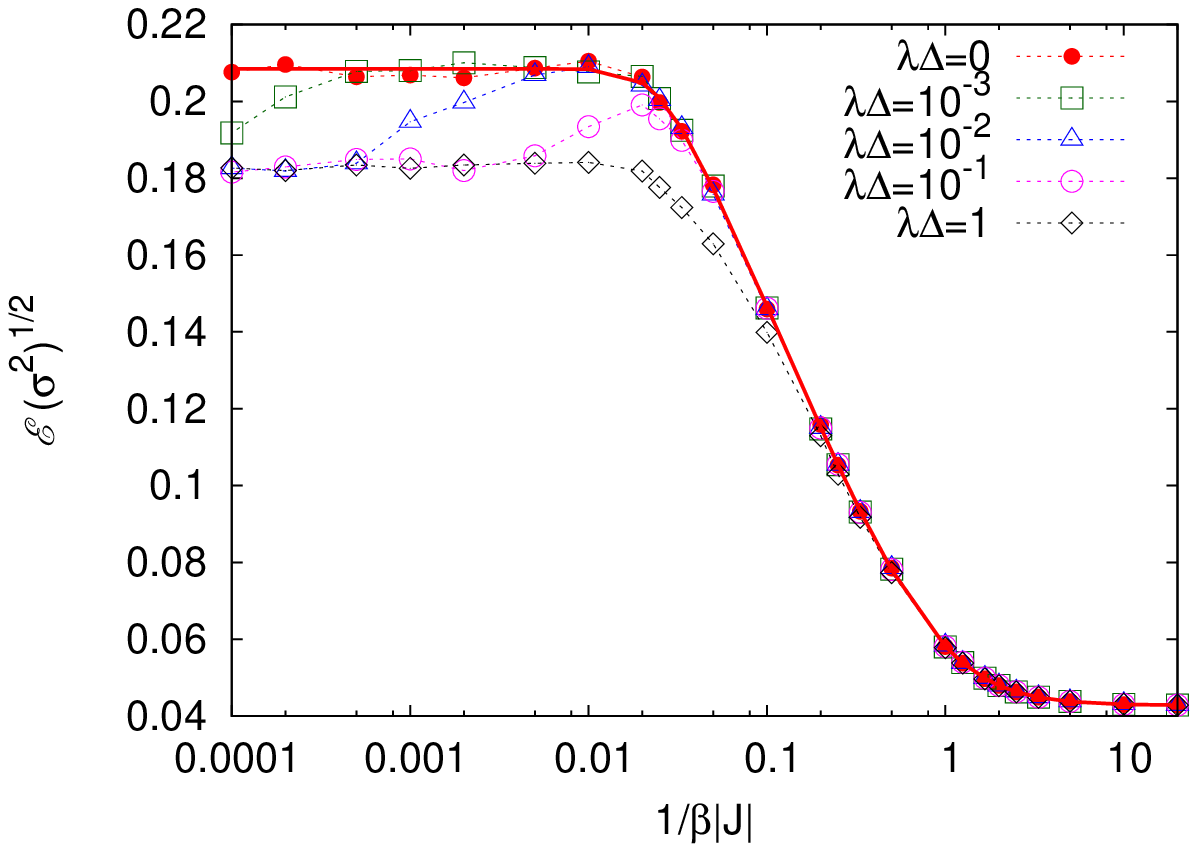} %
\caption{
\label{FigExactD}
(Color online).  
Simulation results for $\sqrt{{\cal E}\left(\sigma ^{2}\right)}$ 
for ferromagnetic spin-$1/2$ chains with $N_{S}=4$,
$N_{E}=8$, $J=\Omega=1$, and various interaction strengths $\lambda\Delta$
as a function of the temperature $T/J=1/(\beta J)$. 
The solid line (red) is obtained from Eq.~(\protect\ref{sigma2}) 
by using numerical values for the free energies $F_S(n\beta)$ and $F_E(n\beta)$.
The dotted lines are guides to the eye.}
\end{figure}

For a small size of the system such as $N\leq 12$, 
we can diagonalize the system exactly, 
find all the eigenvalues and eigenstates of the Hamiltonians $H_S$ and $H_E$,
and directly calculate the values of $\sigma$ and $\delta$ according to the 
analytical expression of Eqs.~(\ref{sigma2}) and~(\ref{delta}), respectively.

Figure~\ref{FigExactD} shows the simulation results for $\sqrt{{\cal E}(\sigma ^{2})}$
obtained by exact diagonalization
for the entirety $S$$+$$E$ being a spin chain with $N_{S}=4$ and $N_{E}=8$. The
system $S$ and environment $E$ consist of two ferromagnetic spin chains with
isotropic spin-spin interaction strengths 
$J_{i,j}^\alpha=J=\Omega_{i,j}^\alpha=\Omega =1$.
They are connected by one of their end-spins, with an interaction strength 
$\Delta_{N_S,1}^\alpha=\Delta$.
The global system-environment coupling strength is $\lambda=1$.
The simulation results (symbols) are averages
over $1000$ simulations with different initial random state vectors drawn from 
the ensemble \textsl{X}.
Substituting the numerically obtained values for the free energy of the system and 
environment for $\lambda\Delta=0$
in the analytical expressions for ${{\cal E}(\sigma^2)} $ 
given by Eq.~(\ref{sigma2}) 
results in the solid lines depicted in Fig.~\ref{FigExactD}.
The simulation
results for the uncoupled entirety ($\lambda\Delta =0$) and for the coupled cases when
$\beta \lambda \Delta \leq 1$
agree with the analytical results for the whole range of temperatures.
As the temperature decreases the state of the entirety $S$$+$$E$ 
approaches the ground state, and
${\cal E}\left(\sigma^2\right)$ becomes constant with 
its numerical value being given by Eq.~(\ref{Eq:gg}).
For the case at hand, $g_S=5$, $g_E=9$, $D_S=16$ and $D_E=256$, hence 
Eq.~(\ref{Eq:gg}) yields $\sqrt{{\cal E}\left( \sigma^2\right)}=0.21$, in excellent
agreement with the numerical data.
In the coupled case and for small temperatures $1/\beta J$, 
$\sqrt{{\cal E}\left( \sigma^2\right)}$ develops a plateau
different from that of the uncoupled case.
The dependence of this plateau on $\beta$ or $\lambda\Delta$ is nontrivial,
requiring a detailed analysis of how the ground state of $S$$+$$E$ leads to the 
reduced density matrix of $S$
(in the basis that diagonalizes $H_S$).
In this respect, the $\beta$ or $\lambda\Delta$ dependence of the data shown in 
Fig.~\ref{FigExactD} are somewhat special because
the ferromagnetic ground state of the system does not depend on $\lambda\Delta$.

For the spin system under study with $\lambda\Delta\neq 0$, the first-order term of
the perturbation expansion of the expectation value of $\sigma^2$ in terms of 
$\beta \lambda\Delta$
is exactly zero. Hence, for a weakly coupled entirety ($\lambda\Delta$ small) 
deviations from the analytical
results Eq.~(\ref{sigma2}) obtained for the uncoupled entirety ($\lambda\Delta=0$),
are, as expected, seen {\it only\/} in the low temperature region.  The numerical 
results (symbols) in Fig.~\ref{FigExactD} are in
excellent agreement with the predicted results (solid line, red) as long as 
$\beta\lambda\Delta$ is small.
For a finite $\beta\lambda\Delta$, the plateaus at low temperature may or may not be 
reached, and therefore the perturbation results may no longer be applicable.
The results in Fig.~\ref{FigExactD} are in amazingly good 
agreement for all temperatures with the perturbation theory predictions of 
Eq.~(\ref{sigma2}).  
The excellent agreement is also seen for low temperatures whenever 
$\beta \lambda \Delta \leq 1$, giving agreement with the expression Eq.~(\ref{Eq:gg}) 
wherein the ground state degeneracy of the environment $E$ enters the measured 
value of $\sigma$ in the system $S$.  

In the low temperature limit for ${\cal E}(\sigma^2)$ from Eq.~(\ref{Eq:gg})
or (\ref{EqS:EsigmaLowT})
the perturbation expression gives
\begin{equation}
\label{EqS:EsigmaLowTa}
\begin{array}{lcl}
\lim_{\beta\rightarrow \infty} \> {\cal E}\left(\sigma^2\right) 
& \> \approx \> &
\frac{\left(g_{S}-1\right)\left(g_{S} g_{E} - 1 \right)}{2 g_{S}^2 g_{E}^2}
\end{array}
\end{equation}
with the approximation valid for large $D$.  
In Fig.~\ref{FigExactD} results for the approach to the low temperature 
limit for one case with $N_S=4$, $N_E=8$ and $g_{S}=5$, $g_{E}=9$. For $g_{S}>1$ 
the expression in Eq.~(\ref{EqS:EsigmaLowTa}) is finite at $T=0$.  However, when $g_{S}=1$ 
the expression in Eq.~(\ref{EqS:EsigmaLowTa}) is zero at $T=0$. 
Therefore the predicted 
curve looks much different from the curve in Fig.~\ref{FigExactD}.  

\begin{figure}[t]
\includegraphics[width=8cm]{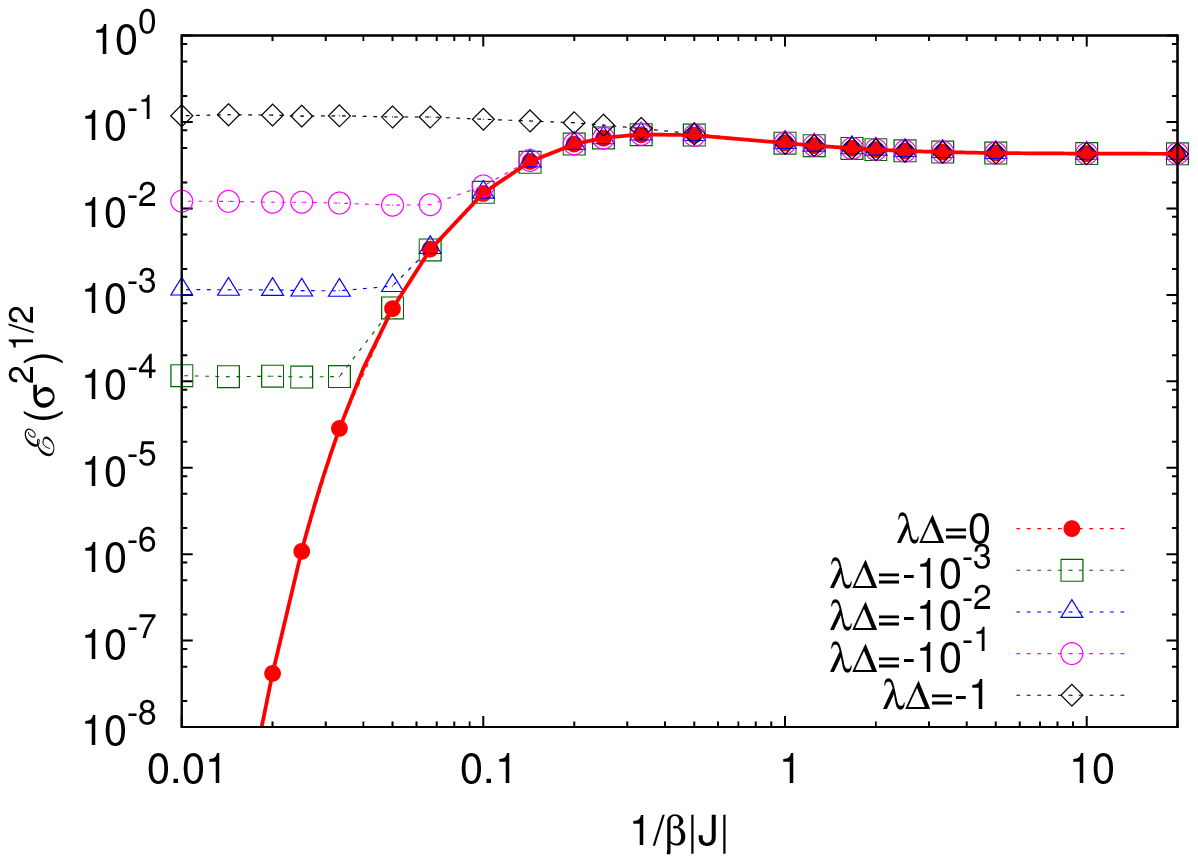}
\caption{
\label{figSIIIa}
(Color online).  
Simulation results for $\sqrt{{\cal E}\left(\sigma^2\right)}$ for spin-$1/2$ chains 
with $N_S=4$, $N_E=8$, $J=-1$, $\Omega=1$ and various interaction strengths 
$\lambda \Delta$ as a function of temperature $T/|J|=1/\beta|J|$. 
The solid line (red) is obtained from Eq.~(\protect\ref{sigma2}) 
by using numerical values for the
free energies $F_S(n\beta)$ and $F_E(n\beta)$.
The dotted lines are guides to the eyes. 
Note that this figure is for $g_{S}=1$, which looks very different compared to 
Fig.~\protect\ref{FigExactD} for $g_{S}>1$. 
}
\end{figure}

Therefore, we here present results for a case with $g_{S}=1$.  
The system is a spin chain with
$N_S=4$ and isotropic antiferromagnetic spin-spin interactions $J^\alpha=-1$ 
with $\alpha =x, y, z$, so $g_{S}=1$.  
The environment is a spin chain with $N_E=8$ and isotropic ferromagnetic
spin-spin interactions $\Omega^\alpha=1$.  
The environment and system are connected by one of their end spins to form  
the entirety $S+E$ with a chain geometry. 
The coupling interactions $\lambda \Delta^{\alpha}$ take various isotropic values.   
Figure~\ref{figSIIIa} for $g_{S}=1$ looks completely different compared to Fig.~\ref{FigExactD}
for $g_{S}>1$. Nevertheless, as the system-environment 
coupling strength $\lambda\Delta$ becomes small, the data from the calculations 
fall nicely on the theoretical curve obtained from Eq.~(\ref{sigma2}) 
(red solid line). 
Note the extremely small values for 
$\sqrt{{\cal E}\left(\sigma^2\right)}$ for low 
temperatures.  Calculating the theoretical curves (red solid lines) 
for these quantities at low temperatures required 
quadruple precision in the floating point numbers.   

\begin{figure}[t]
\includegraphics[width=8cm]{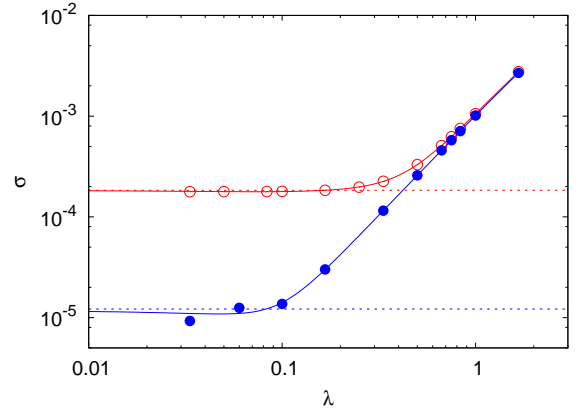}
\caption{
\label{fig8}
(Color online).  
Simulation results for $\sigma$
for rings with $N_{S}=4$, $N_{E}=26$ (open circles) 
and $N_{S}=4$, $N_{E}=36$ (solid circles)
as a function of the global interaction strength $\lambda $ for $\beta |J|=0.90$.
For the values of the interaction parameters, see text.
The solid lines are fits to the data as described in the text.
The top (bottom) horizontal dashed line represents the value obtained by simulating the 
non-interaction system, $\lambda=0$, with $30$ ($40$) spins.
}
\end{figure}

In order to study the behavior of $\sigma$ as a function of the global coupling 
interaction strength $\lambda$,
we performed further simulations
for a spin entirety configured as a ring with $N_S=4$ and $N_E=26,36$ at the 
inverse temperature $\beta |J|=0.90$.
In Fig.~\ref{fig8} we present the simulation results for $\sigma$ as a function of 
$\lambda$.  The entirety is a ring, and the system Hamiltonian 
$H_S$ is antiferromagnetic 
(the Hamiltonians and geometry have the same structure as in 
Figs.~\ref{FigRelax} through \ref{FigUncoupled}).  
Least squares fitting of the data for $\sigma^2$ to polynomials in $\lambda$,
we find that a polynomial of degree~$7$ yields the best fit, for both the 30- and 
40-spin entirety data~\cite{30spins,40spins}.
The behavior of $\delta$ is very similar to that of $\sigma$ and is again only 
shown in Appendix~A.
From Fig.~\ref{fig8} it is seen that for $\lambda\approx 1$,
$\sigma$ changes very little as the dimension of the Hilbert space of the environment
increases. This is a pronounced finite temperature effect, as
for $\beta=0$ the scaling $\sigma\sim 1/\sqrt{D_E}$ holds independent of the coupling 
$\lambda$~\cite{JIN13a}.

\section{Conclusions and Discussion}
In this paper, we investigated measures $\sigma$ for the decoherence
and $\delta$ for the thermalization of a quantum system $S$ coupled to
a quantum environment $E$ at finite temperature. The entirety $S$$+$$E$ is 
a closed quantum system of which the time evolution is governed by the
time-dependent Schr\"{o}dinger equation (TDSE).   

Today many technologies are being driven by necessity to the quantum regime, rather 
than operating in a classical or semi-classical regime.  In the quantum regime 
maintaining the coherence of the state of the system under investigation is paramount.  
Therefore an understanding and quantitative predictions of how difficult it is for 
a quantum system $S$ to decohere, and how effective a particular quantum environment 
$E$ is at decohering any system is critical to quantum 
technologies and experiments such as gate-based 
quantum computers \cite{RIEF2011,MERM2007},
adiabatic quantum computers \cite{BOIX2014,MCG2014,ALBA15},
quantum dots \cite{TART2012,HANS2007},
quantum optics \cite{FOX2006},
cold atoms \cite{BLOC2012,GARD2014,GARD2015}, 
coherent electron transport \cite{DATT2005,HANS2008} 
(including nanoelectronics \cite{NAZA2009,WONG2011} and 
quantum dragon nanodevices \cite{NOVO2014,NOVO2015}), 
and atom/cavity systems \cite{RAIM2001}.   
We have found that at finite and small $\beta \lambda$, 
where $\beta$ denotes the inverse temperature and $\lambda$ 
the global system-environment coupling strength (see Eq.~(\ref{Eq:H})),
the important quantities to answer these questions about 
decoherence are the free energy $F_S$ of the system $S$ and the free energy $F_E$ 
of the environment $E$.  
Therefore, experimentally it is important to measure or to estimate $F_S$ and $F_E$.  
The lowest order result for $\sigma$ is given in Eq.~(\ref{sigma1}), with the 
full result given in Eq.~(\ref{sigma2}).  
Similar statements hold for the measure of thermalization $\delta$, with the 
lowest order result given in Eq.~(\ref{delta1}) and the full result given in 
Eq.~(\ref{delta}) both in terms of the free energies of $S$ and $E$.  

We have investigated $\sigma$ and $\delta$ at finite temperature both 
numerically and analytically. 
Most of the numerical results can be understood within the 
framework of our analytic results.
If the entirety $S$$+$$E$ is prepared in a canonical thermal state, we showed
by means of perturbation theory that $\sigma^2$,
the degree of the decoherence of $S$, is of the order $\beta ^{2}\lambda ^{2}$. 
Similar results were found for our measure of thermalization $\delta^2$.  
Up to the first order in the system-environment interaction we found
\begin{equation}
\sigma^2 , \delta^2 \propto 
\exp\left\{-2\beta \left[F_{E}(2\beta )-F_{E}(\beta )\right]\right\}
\>.
\label{sigma}
\end{equation}
A related decoherence result, for a somewhat different context,  
was found in reference~\cite{SUGI13}.  
Note that $F_E$ is the environment free energy, and consequently is an 
extensive quantity.  
This provides a measure for how well a weakly-coupled specific finite environment can
decohere and thermalize a system at an inverse temperature $\beta$.
A measure for how difficult it is to decohere a quantum system is
given by ratios of free energies of the system, as in Eq.~(\ref{sigma2}).  

To illustrate the power of our conclusions, one could ask of any bath how effective 
it is to decohere any system.  The simplest bath, one often used in theoretical 
calculations with spin baths, is a collection of non-interacting environment 
spins ($H_E=0$). The partition function is then $Z_E=2^{N_E}$ and the free energy is 
$F_E=-N_E \rm ln(2)/\beta$.  From Eq.~(\ref{sigma}) this gives 
$\sigma\> ,\>\delta \propto 2^{- N_E}$ for any temperature $\beta$.  
Even if $H_{SE}=0$ the decoherence goes as $2^{-N_E}$, but one needs to remember that 
the thermal canonical state of the entirety is 
not a direct product of states of the system and environment.  
Other related questions can be raised.  For example for the 
case where $H_E=-\sum_{i=1}^{N_E} \sum_{\alpha=x,y,z}h_i^\alpha I_i^\alpha$
the partition function is 
$Z_E=2^{N_E} \prod_{i=1}^{N_E} \cosh\left(\beta\left|h_i\right|\right)$.
Therefore it does not matter whether or not all the environment fields point in 
the same direction or in random directions in terms of the efficiency of the 
environment to decohere and thermalize any system.  Of course for the same 
system $S$ but different $h_i$ for this type of environment the ensemble of 
canonical thermal states will be different.  

\begin{figure}[t]
\includegraphics[width=8cm]{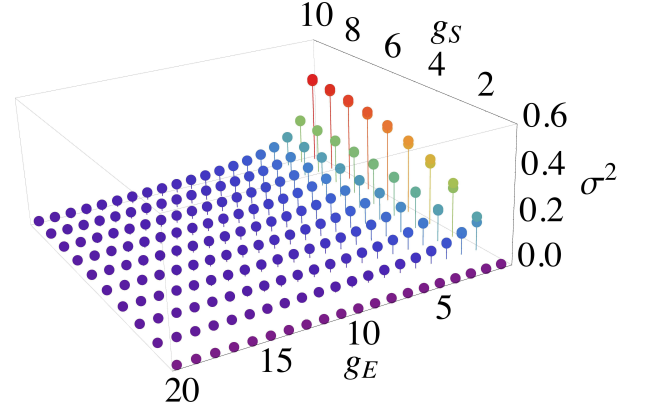}
\caption{\label{figZen}
(Color online).  
Predicted results for $\sigma^2$ at very low temperatures in terms of the 
degeneracy $g_S$ of the system and $g_E$ of the environment.  These are from 
Eq.~(\protect\ref{Eq:gg}).  Two values for the dimension $D$ of the Hilbert 
space of the entirety $S$$+$$E$ are plotted, $D$$=$$4$ and $D$$=$$2^{30}$.  
The difference between these two values of $D$ are only discernible 
in the case $g_E$$=$$1$.  
}
\end{figure}

We have obtained a very strong prediction at low temperatures for 
the decoherence, namely Eq.~(\ref{Eq:gg}).  
At very low temperatures and for large dimension of the Hilbert space 
for the entirety $S$$+$$E$ this prediction is 
\begin{equation}
\label{Eq:ggD}
{\cal E}\left(\sigma^2\right) =  
\frac{\left(g_S-1\right)\left( g_S g_E - 1 \right)}{2 g_S^2 g_E^2}
\end{equation}
with the ground state degeneracy of $S$ ($E$) given by $g_S$ ($g_E$).
Eq.~(\ref{Eq:ggD}) shows that it is possible to perform measurements 
only on the system $S$, but from that extract the ground state degeneracy of 
the environment $E$.  
The results in Fig.~\ref{FigExactD} are for $g_S>1$, and a corresponding graph is 
shown for a case with $g_S=1$ in Fig.~\ref{figSIIIa}.  
As predicted by Eq.~(\ref{Eq:ggD}) these two cases look very different in the 
low-temperature limit.  
Furthermore, at low temperatures in order for a system to not 
be able to decohere it is best to have the system $S$ have a high degeneracy while 
the environment $E$ is non-degenerate.  This is shown in Fig.~\ref{figZen}.  

We performed large-scale real- and imaginary-time simulations for $N_S$ spins in the 
system and $N_E$ spins in the environment.
A canonical thermal state (see Eq.~(\ref{psi_ft})) can be prepared 
by imaginary-time propagation based on the Chebyshev polynomial algorithm.
Starting with such a canonical thermal state, the simulation results for the 
uncoupled entirety agree very well with the analytical results 
(see in particular Figs.~\ref{FigExactD} and \ref{figSIIIa}).

Once the interaction Hamiltonian $H_{SE}$ is turned on,  
we observe that the decoherence measure $\sigma$ generally converges to a 
finite value when the environment size is above a threshold number 
which depends on the inverse temperature $\beta$ and 
the global interaction strength $\lambda$ (see Figs.~\ref{SigmaNeLambda} 
and~\ref{SigmaNeBeta}).
The smaller $\beta$ and $\lambda$ are, the larger the threshold number is.
When the system size is smaller than the threshold number, $\sigma$ (and $\delta$) 
behave as they do for an uncoupled entirety.  By an uncoupled entirety we mean that 
$\lambda H_{SE}$$=$$0$, but the initial state of the system is a 
canonical thermal state of the entirety $S$$+$$E$ and hence is not a direct 
product state of states of $S$ and $E$.  
After the system size reaches the threshold number, $\sigma$ (and $\delta$) 
quickly converges to a finite value, 
due to the high-order contributions from the interaction $H_{SE}$.
From the numerical simulations, the stationary value of $\sigma$ has the form 
$(\beta\lambda)^2(c_2+c_3\beta)$ 
for our range of simulation parameters.
 
Strictly speaking, the system $S$ completely decoheres
if there is no interaction between $S$ and $E$ and if $N_E\rightarrow \infty$. If $S$
is coupled to $E$, the $H_{SE}$ interaction is important and both 
$\sigma$ and $\delta$ are finite for a finite system $S$
even in the thermodynamic limit ($N_E\rightarrow +\infty$). 
However, if the canonical
ensemble is a good approximation for the state of the system for some 
inverse temperatures $\beta$
up to some chosen maximum energy $E_{hold}>0$ (measured from the ground
state), then it is required that $\exp(-\beta E_{hold}) \gg \sigma$.
By determining the crossover of the left- and right-side
functions, we find a threshold for the temperature above which the state of
the system is well approximated by a canonical ensemble, and below which
quantum coherence of the system is well preserved.

We emphasize that the entirety $S$$+$$E$ is initially prepared in a pure state given 
by a particular choice of a canonical thermal state \textsl{X} in Eq.~(\ref{psi_ft}).  
With such a state as the initial state for the TDSE, the real-time dynamics does not 
have much effect on our measures for decoherence ($\sigma$) or 
thermalization ($\delta$).  
If we start with a non-equilibrium state, such as a product state of $S$ and $E$, 
where $S$ is in the ground state and $E$ is in a canonical thermal state,
the real-time dynamics play an important role in both the decoherence and the 
thermalization of $S$~\cite{YUAN11, JIN10x, JIN13a}, as seen in Fig.~\ref{FigRelax}.
At infinite temperature there may exist certain geometric structures or 
conserved quantities which prevent the system from having 
complete decoherence~\cite{JIN13a}.  In contrast to the 
infinite temperature results, we have found here that at finite 
temperature the lack of complete decoherence is the normal scenario for any 
coupled entirety (finite $\lambda H_{SE}$).  

In this paper we have answered important questions about how easily a given system $S$ 
can decohere or thermalize, and how efficient a given bath is to 
decohere or thermalize any system.  We have not addressed 
the equally important question of how quickly $S$ thermalizes or decoheres.  
Nevertheless, we believe that our methodology of simulations and perturbation 
calculations with thermal canonical states can also be important to address 
the time-dependent question.  For full time dependence, the real-time version of 
Eq.~(\ref{expbH}) would need to be used, most likely leading to even more 
complicated perturbation theory calculations than are detailed in Appendix~B.

\section*{Acknowledgements}
The authors gratefully acknowledge the computing time granted by the JARA-HPC
Vergabegremium and provided on the JARA-HPC Partition part of the 
supercomputer JUQUEEN \cite{JUQUEEN15} at Forschungszentrum J{\"u}lich.
MAN is supported in part by US National Science Foundation grant DMR-1206233.


\appendix

\section{Numerical results for $\delta$}

In the main text, we only present the simulation results for $\sigma(t)$, 
a measure of the decoherence of a quantum $S$ under the influence of a 
quantum environment $E$.
The simulation results for $\delta(t)$, a measure of the thermalization of $S$, 
given by Eq.~(\ref{eqdelta}),
are shown in this appendix.
The largest entireties we were able to study contained $40$ spins, 
as it requires about $10^{12}$ 
floating-point numbers to represent a vector of the Hilbert space 
of an entirety with this size.   
A sketch of the ring geometry for $N=40$ and $N_S=4$, is given in 
Fig.~\ref{figS00}.
We will see that besides the size of the statistical fluctuations,
$\delta(t)$ (or the time-independent average $\delta$) 
behaves very similar as $\sigma(t)$ (or the time-independent average $\sigma$).
For a single run with one realization of $H_E$ and one representation of 
the canonical thermal state (see Eq.~(\ref{psi_ft})), 
it is obvious that the data for $\delta(t)$ 
may have stronger statistical
fluctuations than those for $\sigma(t)$ shown in the main text, as the number of 
diagonal elements of the reduced density matrix of the system $S$ 
are much smaller than the number of the off-diagonal elements.

\begin{figure}[t]
\includegraphics[width=8cm]{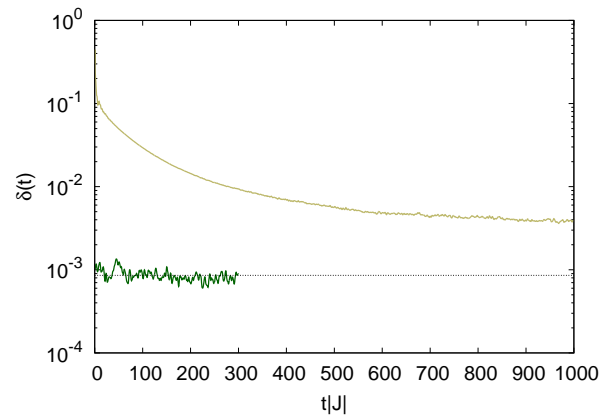}
\caption{
\label{figS1}
(Color online).  
Simulation results for $\delta(t)$ for a coupled ring entirety 
with $N_S=4$, $N_E=22$ and $\lambda=1$ 
for two different initial states $X$ (flat curve, green) and 
$UDUDY$ (decay curve, dark khaki) with $\beta |J|=0.90$.
The dotted (green) horizontal line is a guide for the eyes. 
This figure corresponds to Fig.~\protect\ref{FigRelax} in the main text.
}
\end{figure}

Figure~\ref{figS1} presents the time evolution of $\delta(t)$ 
for a spin system with $N_S=4$
and $N_E=22$ prepared in two different initial states $X$ and $UDUDY$.
From Fig.~\ref{figS1}, one sees that $\delta(t)$ obtained from $UDUDY$ evolves closely to
the value obtained from $X$, which is very similar to the behavior of $\sigma(t)$ shown in 
Fig.~\ref{FigRelax}. The difference of the values of $\delta(t)$ between these two 
initial states at long times is about $0.003$. This difference is larger than that for 
$\sigma(t)$ at long times. The reason is that the diagonal elements of 
the reduced density matrix $\widetilde\rho$ for $S$
keeps a strong memory about its initial state. The memory effects would be reduced 
for a larger system $S$.

\begin{figure}[t]
\includegraphics[width=8cm]{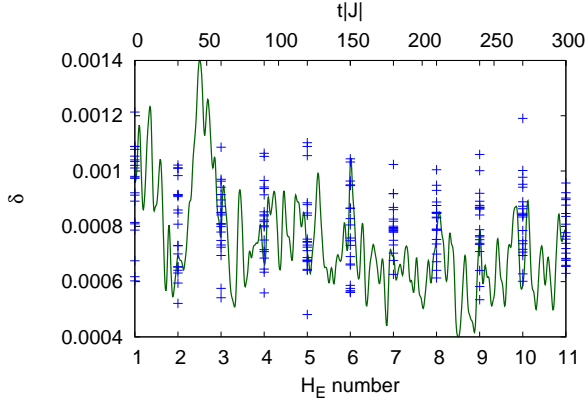}
\caption{
\label{figS2}
(Color online).  
Simulation results for $\delta$ for a coupled ring entirety 
with $N_S=4$, $N_E=22$ and $\lambda=1$ starting from different initial states $X$
with $\beta |J|=0.90$. Results for eleven different realizations of the environment
Hamiltonian $H_E$ are shown ($x$-axis label at the bottom),
each with different initial states drawn from the ensemble 
that gives an $X$ state (blue pluses).
The time dependence of $\delta$ for the first realization of $H_E$ and one of the initial
states $X$ is shown by the solid (green) curve ($x$-axis label on top) which is the same
(green) curve as depicted in Fig.~\protect\ref{figS1}.
This figure corresponds to Fig.~\protect\ref{MANdiffEtP} in the main text.
}
\end{figure}

Figure~\ref{figS2} presents the corresponding results for $\delta$ 
as in Fig.~\ref{MANdiffEtP} for $\sigma$.
The average and the standard deviation of the data points shown in Fig.~\ref{figS2}
are $8.0\times 10^{-4}$ and $1.4\times 10^{-4}$, respectively.
As is the case for $\sigma$ in the main text, the time-average for $\delta$ 
and the average over different environment Hamiltonians $H_E$ and different representations 
of the initial state $X$ all behave similarly.  

\begin{figure}[t]
\includegraphics[width=8cm]{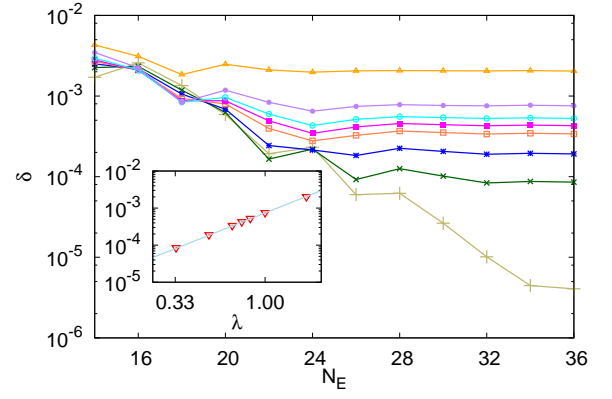}
\skip 0.1 true cm
\includegraphics[width=8cm]{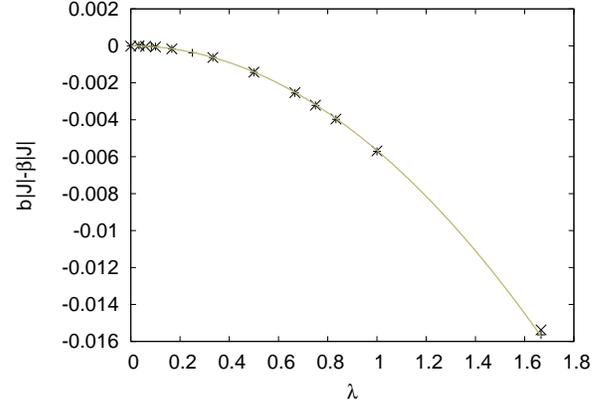}
\caption{
\label{figS3}
(Color online).  
Top: Simulation results for $\delta$ for a coupled ring entirety 
with $N_S=4$ and $N_E=14$, 
$\ldots$, $36$ for different global interaction strengths $\lambda$.
The entirety is in a thermal canonical state with $\beta |J|=0.90$. 
Curves from bottom to top correspond to
$\lambda=0.00$, $0.33$, $0.50$, $0.67$, $0.75$, $0.83$, $1.00$, $1.67$.
Inset: $\delta$ as a function of $\lambda$ for $N_E=36$. 
The (light blue) solid line is a fitting curve
for non-zero $\lambda$, and gives $\delta \approx 0.00074\lambda^2$. 
This figure corresponds to Fig.~\protect\ref{SigmaNeLambda} for $\sigma$. 
Bottom: Simulation results for the difference between the fitting temperature $b$ 
and the inverse temperature $\beta$ for entireties with $N_E=26$ (pluses) and 
$N_E=36$ (crosses). For $\lambda<1$, the data points
fit very well to the curve $b|J|-\beta|J| \approx -0.00566\lambda^2$ (solid curve).
}
\end{figure}

Figure~\ref{figS3} presents the simulation results for $\delta$ 
for scaling $H_{SE}$ by the global interaction strength $\lambda$. 
From Fig.~\ref{figS3} (top), it is obvious that we observe similar behavior for $\delta$ as 
we did for $\sigma$ shown in Fig.~\ref{SigmaNeLambda} in the main text.  
The difference is in the stronger fluctuations for the data points for $\delta$.
There are two regimes of $\delta$ separated by some threshold number of $N_E$, 
labeled as $L(\lambda)$.
If $N_E<L(\lambda)$, $\delta$ decreases approximately exponentially as $N_E$ increases.
If $N_E>L(\lambda)$, $\delta$ converges to a finite value that depends on $\lambda$.
The constant values for $\delta$ for $N_E>L(\lambda)$ is well fitted to 
$\lambda^2$ (see the inset of Fig.~\ref{figS3}).
Figure~\ref{figS3} (bottom) shows the simulation results for the fitting temperature $b$, 
see Eq.~(\ref{eqbt}), which has the inverse temperature $\beta$ subtracted,
where $\beta$ is the inverse temperature used to prepare the canonical thermal state 
of Eq.~(\ref{psi_ft}) from the initial state $X$.
The data points are well fit to $-\lambda^2$ for $\lambda <1$.
This implies that only for $\lambda\rightarrow 0$ (the uncoupled entirety), 
does one have $b=\beta$, 
which is consistent with the analysis for $\sigma$ in the main text.

\begin{figure}[t]
\includegraphics[width=8cm]{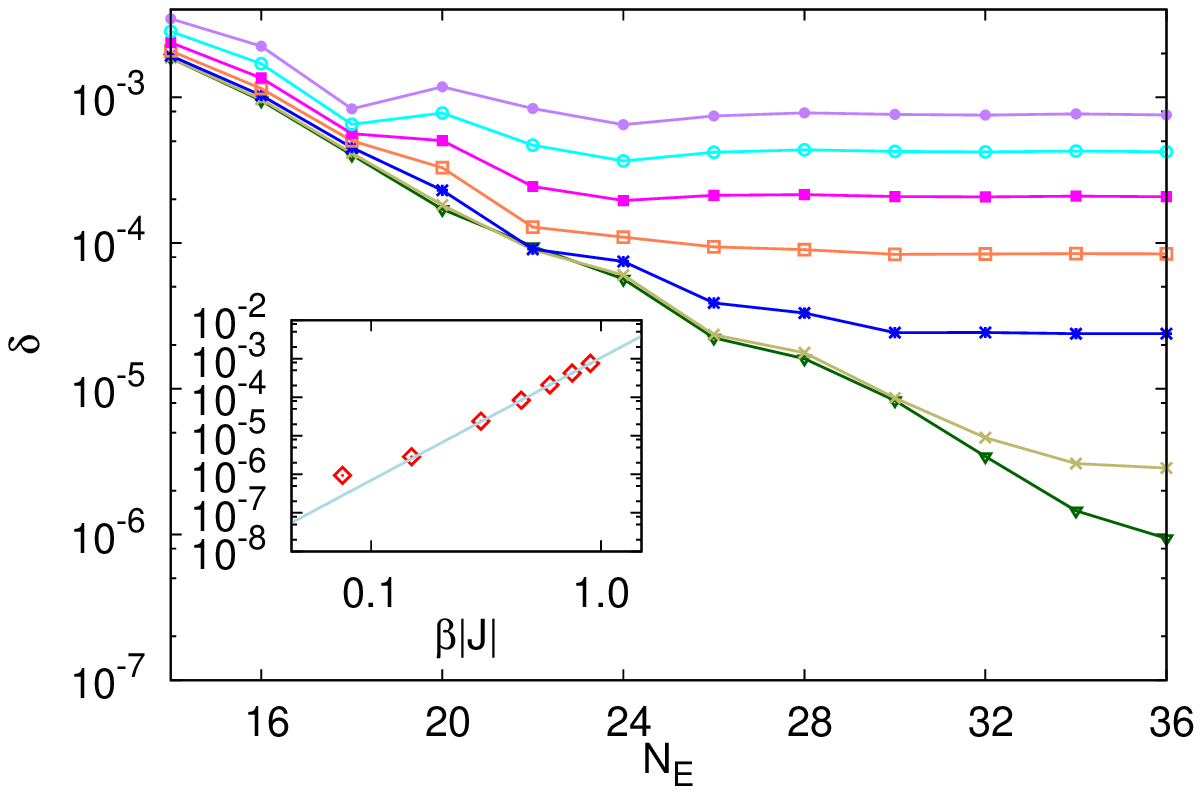}
\skip 0.1 true cm
\includegraphics[width=8cm]{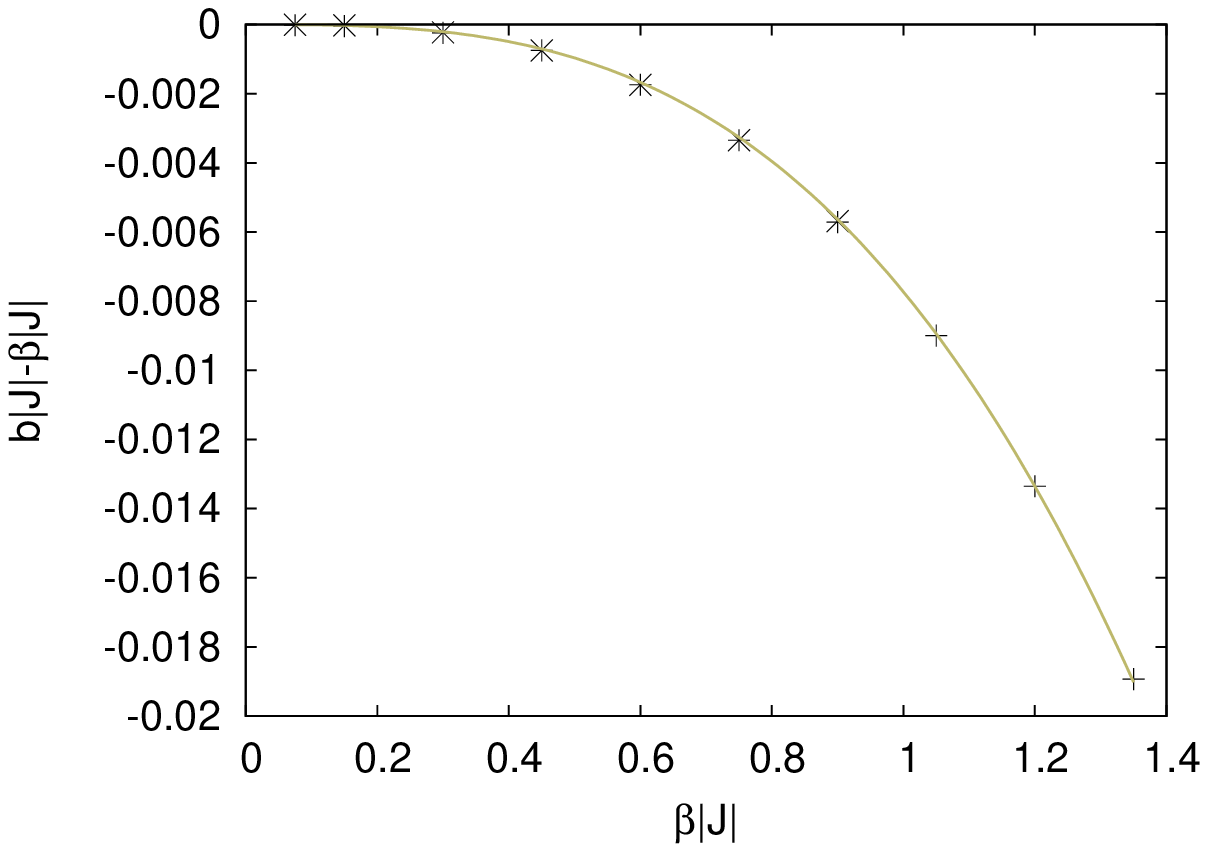}
\caption{
\label{figS4}
(Color online).  
Top: Simulation results for $\delta$ for a coupled ring entirety 
with $N_S=4$, $N_E=14$, 
$\ldots$, $36$ and $\lambda=1$ for different inverse temperatures $\beta$.
The initial states are canonical thermal states at different value of $\beta$, 
corresponding to curves from bottom to top with $\beta |J|=
0.075$, $0.15$, $0.30$, $0.45$, $0.60$, $0.75$, $0.90$.
Inset: $\delta$ as a function of $\beta |J|$ for $N_E=36$. 
The (light blue) solid line is a fitting curve
and gives $\delta \approx 0.00106(\beta |J|)^{3.18}$ for $\beta |J|\geq 0.15$. 
This figure corresponds to Fig.~\ref{SigmaNeBeta} in the main text. 
Bottom: Simulation results for the difference between the fitting 
temperature $b$ and the inverse temperature $\beta$ 
for entireties with $N_E=26$ (pluses) and 
$N_E=36$ (crosses). For $\beta |J|<1$, the data points
fit to $b|J|-\beta|J| \approx -0.00773\beta^3|J|^3 $ (solid curve).
}
\end{figure}

Figure~\ref{figS4} presents the simulation results for $\delta$ 
by varying the inverse temperature $\beta$ that is used in Eq.~(\ref{psi_ft}) 
to obtain the canonical thermal state from the state $X$.
Fig.~\ref{figS4} (top) corresponds to Fig.~\ref{SigmaNeBeta} in the main text.
We observe similar behavior for $\delta$ as we did for $\sigma$ in the main text, 
except there are larger fluctuations for the data points for $\delta$.
The convergent values of $\delta$ for $N_E=36$ is better fit to $(\beta |J|)^{3.18}$, 
which is slightly different from the fitting index for the convergent $\sigma$.
However, a definitive analysis of how robust the difference is would require 
high statistics calculations with averages over different times, different $H_E$, 
and different samples of the $X$ state.  
Figure~\ref{figS4} (bottom) shows the simulation results of 
the fitting temperature $b$ with $\beta$ subtracted.  
The data points for $\beta|J|<1$ fit well to $-(\beta|J|)^3$, just as did the the 
values in the main text for $\sigma$.  

\begin{figure}[t]
\includegraphics[width=8cm]{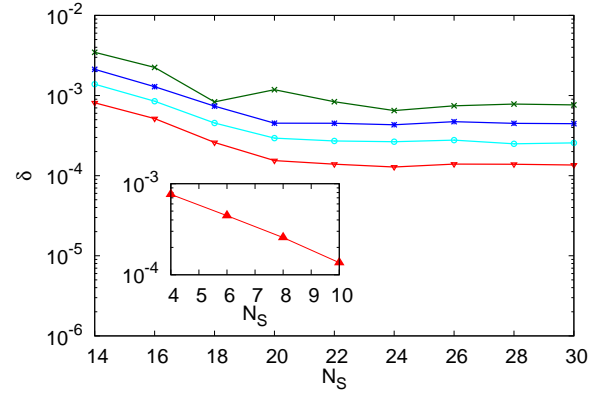}
\caption{
\label{figS5}
(Color online).  
Simulation results for $\delta$ for a coupled ring entirety with $N_S=4$, $6$, $8$, $10$ 
(symbols, top to bottom),
$N_E=14$, $\ldots$, $30$ and $\lambda=1$ for $\beta|J|=0.90$.
Inset: $\delta$ as a function of $N_S$ for $N_E=30$.
This figure corresponds to Fig.~\protect\ref{fig5}.
}
\end{figure}

Figure~\ref{figS5} presents the corresponding results for $\delta$ to compare with
results shown in Fig.~\ref{fig5} for $\sigma$.
We see similar convergent behavior for both $\sigma$ and $\delta$ 
when the environment size $N_E$ is larger than certain threshold value.
For $N_E$ is smaller than the threshold value, $\delta$ decreases approximately
exponentially with increasing $N_E$.
Unlike the data points of $\sigma$ which overlapped for this regime, 
the data points of $\delta$ do not overlap. 
This is because $\sigma$ is only related to the factor from the environment 
(see Eqs.~(\ref{Eq:Tinfinity}) and 
(\ref{sigma2}) in the main text), 
while $\delta$ is also related to the factor from the system itself 
(see Eqs.~(\ref{Eq:Tinfinity}) and (\ref{delta}) in the main text).

\begin{figure}[t]
\includegraphics[width=8cm]{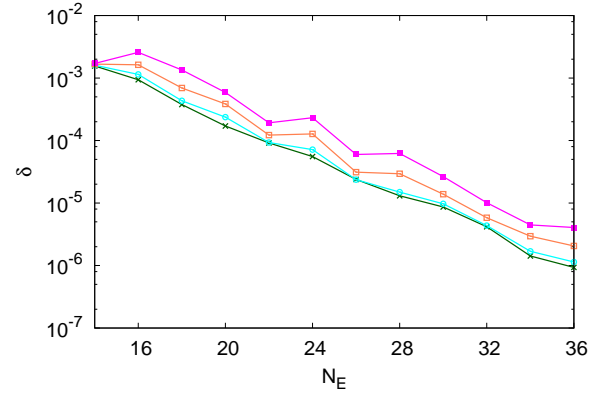}
\caption{
\label{figS6}
(Color online).  
Simulation results for $\delta$ for an uncoupled entirety ($\lambda=0$) 
with $N_S=4$ and $N_E=14$, $\ldots$, $36$
for different inverse temperatures. 
Curves from bottom to top correspond to $\beta|J|=0.075$,
$0.30$, $0.60$, and $0.90$. 
This figure corresponds to Fig.~\protect\ref{FigUncoupled} in the main text.
}
\end{figure}

Figure~\ref{figS6} presents the corresponding results for $\delta$ as shown 
in Fig.~\ref{FigUncoupled} for $\sigma$. 
It is clear that except for strong fluctuations $\delta$ for the uncoupled entirety 
($\lambda=0$) scales with the size of $N_E$.

\begin{figure}[t]
\includegraphics[width=8cm]{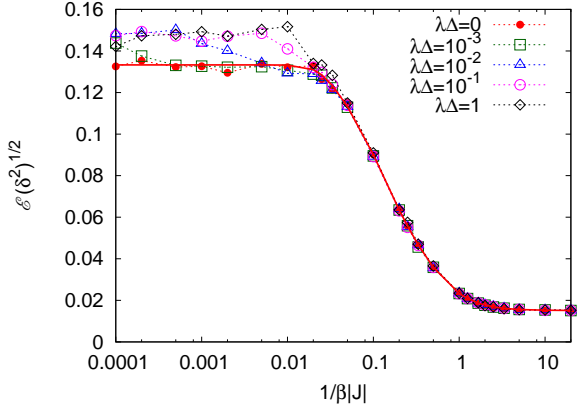}
\caption{
\label{figS7}
(Color online).  
Simulation results of $\sqrt{{\cal E} (\delta^2)}$ for ferromagnetic spin-$1/2$ 
chains with 
$N_S=4$ and $N_E=8$, $J=\Omega=1$, and various interaction strengths $\lambda\Delta$ as a function
of the temperature $T/J=1/(\beta J)$. The solid line (red) is obtained from 
Eq.~(\ref{delta}) by using numerical
values for the free energies $F_S(n\beta)$ and $F_E(n\beta)$.
The dotted lines are guides to the eye.
Note that the functional form of the $\lambda=0$ curve, as well as how data for finite 
$\lambda$ relate to this curve, are very similar to Fig.~\protect\ref{FigExactD} for $\sigma$.  
}
\end{figure}

\begin{figure}[t]
\includegraphics[width=8cm]{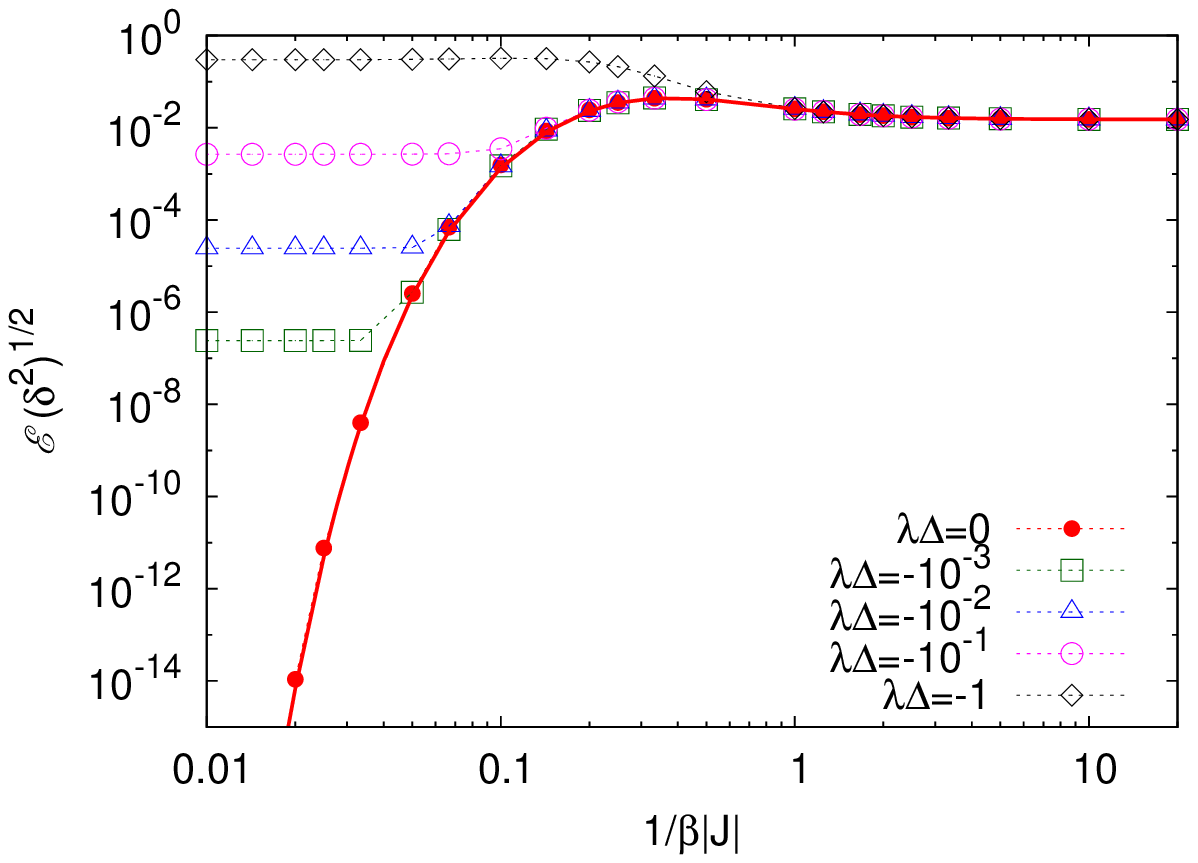}
\caption{
\label{figSIIIb}
(Color online). 
Simulation results for $\sqrt{{\cal E}\left(\delta^2\right)}$ for spin-$1/2$ chains 
with $N_S=4$, $N_E=8$, $J=-1$, $\Omega=1$ and various interaction strengths 
$\lambda \Delta$ as a function of temperature $T/|J|=1/\beta|J|$. 
The solid line (red) is obtained from Eq.~(\ref{delta}) 
by using numerical values for the
free energies $F_S(n\beta)$ and $F_E(n\beta)$.
The dotted lines are guides to the eyes. 
Note that this figure is for $g_{S}=1$, which looks very different compared to 
Fig.~\ref{figS7} for $g_{S}>1$. This figure for $\delta$ corresponds to 
Fig.~\protect\ref{figSIIIa} for $\sigma$.  
}
\end{figure}

Figures~\ref{figS7} and \ref{figSIIIb} present the simulation results for 
$\sqrt{{\cal E} (\delta^2)}$ obtained by exact diagonalization
for the entirety $S+E$ being a spin chain with $N_S=4$ and $N_E=8$.
These figures correspond to Figures~\ref{FigExactD} and \ref{figSIIIa} in the main text. 
The data points are averaged over $1000$ runs with different representations of the 
state $X$ at specific temperature $\beta$.
Therefore the simulation results shown in Figs.~\ref{figS7} and \ref{figSIIIb} have very good statistics.
We refer to the detailed discussion about these figures in the main text, as $\sigma$
and $\delta$ behave very similarly.  
We remind the reader that both Fig.~\ref{figS7} and Fig.~\ref{FigExactD} are for 
the case with the ground state degeneracy of the system being $g_{S}=5$.  
We remind the reader that both Fig.~\ref{figSIIIb} and Fig.~\ref{figSIIIa} are for 
the case with the ground state degeneracy of the system being $g_{S}=1$.  
Fig.~\ref{figSIIIb} for $g_{S}=1$ looks completely different from Fig.~\ref{figS7} 
for $g_{S}>1$.   Nevertheless, as the system-environment 
coupling strength $\lambda\Delta$ becomes small the data from 
the calculations fall nicely 
on the theoretical curve obtained from Eq.~(\ref{delta}) in the main text (red solid line).  
The theoretical curve for $\delta$ in the limit $T\rightarrow 0$, as seen in 
Eq.~(\ref{Eq:ggD}), is equal to zero.  
Note the extremely small values for $\sqrt{{\cal E}\left(\delta^2\right)}$ for low 
temperatures.  Calculating the theoretical curves (red solid lines) 
for these quantities at low temperatures required 
quadruple precision in the floating point numbers.   

\begin{figure}[t]
\includegraphics[width=8cm]{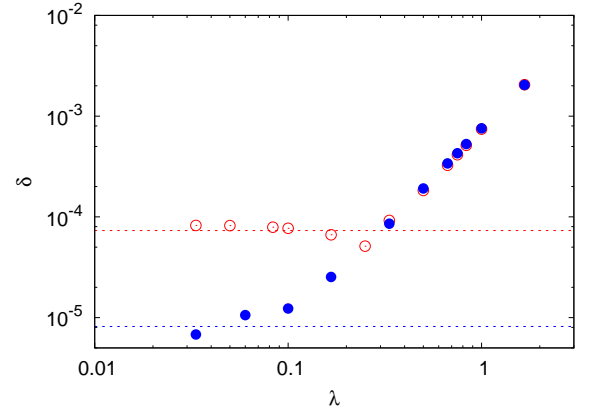}
\caption{
\label{figS8}
(Color online).  
Simulation results for $\delta$ for rings with $N_S=4$, $N_E=26$ (open circles) 
and $N_S=4$, $N_E=36$ (solid circles) as a function of the global interaction 
strength $\lambda$ for $\beta |J|=0.90$. The top (bottom) horizontal dashed 
line represents the value 
obtained by simulating
the non-interaction system, $\lambda=0$, with $30$ ($40$) spins.
This figure corresponds to Fig.~\protect\ref{fig8} in the main text.
}
\end{figure}

Figure~\ref{figS8} presents the corresponding simulation results for $\delta$ 
as shown in Fig.~\ref{fig8} for $\sigma$. 
Note that there is no fitting procedure for these data points.
The dashed lines, as in the main text, are for the uncoupled entirety, $\lambda=0$.  
The behavior for $\delta$ here is quite similar to 
the behavior of $\sigma$ in Fig.~\ref{fig8}.   

\section{Perturbation theory}

In this appendix the details of the perturbation theory calculations are presented.  
Additional definitions and important considerations are first given.  

\subsection{Hamiltonian}
The Hamiltonian has the form
\begin{equation}
\label{EqS:Ham}
H=H_S+H_E+\lambda H_{SE}=H_0+\lambda H_{I},
\end{equation}
where $\lambda$ is explicitly written as the perturbation parameter and the 
uncoupled Hamiltonian is 
$H_0=H_S+H_E$.
The dimension of the Hilbert space of the environment, the system and 
the entirety $S+E$ is $D_E$, $D_S$ and $D=D_S D_E$.  

\subsection{Random state}
Any state from the Hilbert space of $H$ can be written as the wave function 
\begin{equation}
\label{EqS:Psi0}
\left|\Psi_0\right\rangle = \sum_{k=1}^{D}  d_k \left| E_k\right\rangle,
\end{equation}
where $\{\left| E_k\right\rangle\}$ form the energy basis of $H$.  
Random states in the Hilbert space of the entirety Hamiltonian $H$ 
are obtained from Eq.~(\ref{EqS:Psi0})
if $\{d_k\}$ are random Gaussian coefficients, normalized to unity 
\begin{equation}
\sum_{k=1}^D d_k^* d_k \>=\> 1
\>. 
\end{equation}

In practice, in our computer program 
we generate the Gaussian random numbers $d_k=c_k+i b_k$ by using 
the Box-Muller method \cite{BoxM1958} to generate two Gaussian random numbers 
$c_k'$ and $b_k'$
\begin{eqnarray}
c_{k}'& = &\sqrt{-2 {\rm ln}\left(r_0\right)} \cos\left(2 \pi r_1\right) \nonumber \\
 &\> {\rm and} \> & \\
b_{k}'& = &\sqrt{-2 {\rm ln}\left(r_0\right)} \sin\left(2 \pi r_1\right) \>, \nonumber 
\end{eqnarray}
where $r_0$ and $r_1$ are two independent random numbers distributed 
uniformly on $[0,1)$, so that the Gaussian random number $d_k$ is given by simple normalization
\begin{equation}
\label{EqS:dk} 
d_k \>=\> c_k + i b_k \>=\> 
\frac{c_k' + i b_k'}
{\sqrt{\sum_{k'=1}^D \left[\left(c_{k'}'\right)^2+\left(b_{k'}'\right)^2\right]}}
\> = \> 
\sqrt{x_k} e^{i\phi_k}
\>.
\end{equation}
The ensemble of random states has been previously analyzed \cite{HAMS00} 
and has given predictions for measures of quantum decoherence and thermalization 
at infinite-temperature ($\beta$$=$$0$) \cite{JIN13a}.

\subsection{Canonical thermal state}
One forms a wave function at finite inverse temperature $\beta$ 
given by
\begin{equation}
\label{EqS:PsiBeta}
\left|\Psi_{\beta}\right\rangle = 
\frac{e^{-\frac{\beta{H}}{2}}\left|\Psi_0\right\rangle}
{\left\langle\Psi_0|e^{-\beta{  H}}|\Psi_0\right\rangle ^{1/2}} 
\> ,
\end{equation}
which defines the ensemble of canonical thermal states of Eq.~(\ref{psi_ft}).  
Here the inverse temperature is $\beta=1/k_{\rm B} T$ for temperature $T$, 
and we set Boltzmann's constant $k_{\rm B}$$=$$1$.  
Equation~(\ref{EqS:PsiBeta}) can be rewritten as 
\begin{eqnarray}
\left|\Psi_\beta\right\rangle 
 & \>=\> & 
\frac{\sum_{k=1}^D d_k e^{-\frac{\beta E_k}{2}}\left|E_k\right\rangle}
{\left[\sum_{k'=1}^D d_{k'}^* d_{k'} e^{-\beta E_{k'}}\right]^{\frac{1}{2}}}
\\
& \>=\> &
\frac{d_1 e^{-\frac{\beta E_1}{2}} \left|E_1\right\rangle +
\sum_{k=2}^D d_k e^{-\frac{\beta E_k}{2}}\left|E_k\right\rangle}
{\left[d_1^* d_1 e^{-\beta E_{1}}+
\sum_{k'=2}^D d_{k'}^* d_{k'} e^{-\beta E_{k'}}\right]^{\frac{1}{2}}} 
\\
\label{EqS:PsiBeta2}
& \>=\> & 
\frac{d_1 \left|E_1\right\rangle +
\sum_{k=2}^D d_k e^{-\frac{\beta \left(E_k-E_1\right)}{2}}\left|E_k\right\rangle}
{\left[d_1^* d_1 +
\sum_{k'=2}^D d_{k'}^* d_{k'} 
e^{-\beta \left(E_{k'}-E_1\right)}\right]^{\frac{1}{2}}} 
\>,
\end{eqnarray} 
so that it becomes obvious that in the 
infinite temperature ($\beta\rightarrow 0$) limit
\begin{equation}
\lim_{\beta\rightarrow 0} \> \left|\Psi_\beta\right\rangle 
\>\>=\>\>  
\left|\Psi_0\right\rangle 
\>.  
\end{equation}
A canonical thermal state is drawn from 
the distribution given by the canonical thermal state ensemble of Eq.~(\ref{EqS:PsiBeta}).  

The canonical thermal state can also be written as 
\begin{equation}
\left|\Psi_\beta\right\rangle 
\>=\> 
\sum_{k=1}^D 
\frac{d_k \> e^{-\beta E_k/2}\left|E_k\right\rangle}
{\sqrt{\sum_{k'=1}^D \left|d_{k'}\right|^2 e^{-\beta E_{k'}}}}
\>=\> 
\sum_{k=1}^D a_k \left|E_k\right\rangle
\end{equation}
with
\begin{eqnarray}
a_k 
& \>=\> & 
\frac{d_k \> e^{-\beta E_k/2}}{\sqrt{\sum_{k'=1}^D \left|d_{k'}\right|^2 e^{-\beta E_{k'}}}} \\
& \>=\> & 
\frac{d_k \> p_k^{1/2}}{\sqrt{\sum_{k'=1}^D \left|d_{k'}\right|^2 p_{k'}}} 
\end{eqnarray}
with the Boltzmann probability of being in state $k$ given by 
\begin{equation}
p_k \>=\> 
\frac{e^{-\beta E_k}}{\sum_{k'=1}^D e^{-\beta E_{k'}}} 
\>=\> 
\frac{e^{-\beta E_k}}{Z} 
\>.
\end{equation}
The partition function of the entirety $S+E$ is given by 
\begin{equation}
Z 
\> = \> 
{\rm Tr}_{S+E}\left(e^{-\beta H}\right)
\> = \> 
\sum_{k=1}^D e^{-\beta E_{k}}
\>.
\end{equation}

\subsection{Canonical thermal state for uncoupled entirety}

For the uncoupled case, $\lambda=0$, one has
\begin{widetext}
\begin{equation}
\left|\Psi_\beta\right\rangle 
\>=\>  
\frac{d_{1,1} \left|E_1^{(S)}\right\rangle\left|E_1^{(E)}\right\rangle +
\sum_{i=1}^{D_S} \sum_{p=1}^{D_E} d_{i,p} 
\left(1-\delta_{i,1}\delta_{p,1}\right)
e^{-\frac{\beta \left(E_i^{(S)}-E_1^{(S)}\right)}{2}}
e^{-\frac{\beta \left(E_p^{(E)}-E_1^{(E)}\right)}{2}}
\left|E_i^{(S)}\right\rangle\left|E_p^{(E)}\right\rangle}
{\left[d_{1,1}^* d_{1,1} +
\sum_{i'=1}^{D_S} \sum_{p'=1}^{D_E} d_{i',p'}^* d_{i',p'} 
\left(1-\delta_{i',1}\delta_{p',1}\right)
e^{-\beta \left(E_{i'}^{(S)}-E_1^{(S)}\right)}
e^{-\beta \left(E_{p'}^{(E)}-E_1^{(E)}\right)}
\right]^{\frac{1}{2}}} 
\>
\end{equation}
where $\left\{\left|E_i^{(S)}\right\rangle\right\}$ and 
$\left\{\left|E_p^{(E)}\right\rangle\right\}$ 
form the energy basis of $H_S$ and $H_E$, respectively.

The canonical thermal state for the uncoupled entirety can also be written as 
\begin{equation}
\label{EqS:Psi0a}
\left|\Psi_\beta\right\rangle 
\>\>=\>\> 
\sum_{i=1}^{D_S} \sum_{p=1}^{D_E} 
\frac{d_{i,p} \> e^{-\beta E_i^{(S)}/2} \> e^{-\beta E_p^{(E)}/2} \> 
\left|E_i^{(S)}\right\rangle\left|E_p^{(E)}\right\rangle}
{\sqrt{\sum_{i'=1}^{D_S} \sum_{p'=1}^{D_E} \left|d_{i',p'} \right|^2 
e^{-\beta E_{i'}^{(S)}}e^{-\beta E_{p'}^{(E)}}}}
\>\>=\>\> 
\sum_{i=1}^{D_S} \sum_{p=1}^{D_E} a_{i,p}
\left|E_i^{(S)}\right\rangle\left|E_i^{(E)}\right\rangle
\end{equation}
\end{widetext}
with
\begin{eqnarray}
a_{i,p}
& \>=\> & 
\frac{d_{i,p} \> e^{-\beta E_i^{(S)}/2} \> e^{-\beta E_p^{(E)}/2}}
{\sqrt{\sum_{i'=1}^{D_S} \sum_{p'=1}^{D_E} \left|d_{i',p'}\right|^2 
e^{-\beta E_{i'}^{(S)}}e^{-\beta E_{p'}^{(E)}}}} \\
& \>=\> & 
\frac{d_{i,p} \> \sqrt{p_i^{(S)}} \sqrt{p_p^{(E)}}}
{\sqrt{\sum_{i'=1}^{D_S} \sum_{p'=1}^{D_E} \left|d_{i',p'}\right|^2 p_{i'}^{(S)}p_{p'}^{(E)} }} 
\end{eqnarray}
where the Boltzmann probability of being in state $i$ of $H_S$ is given by 
\begin{equation}
p_i^{(S)} \>=\> 
\frac{e^{-\beta E_i^{(S)}}}{\sum_{i'=1}^{D_S} e^{-\beta E_{i'}^{(S)}}} 
\>=\> 
\frac{e^{-\beta E_i^{(S)}}}{Z_S} 
\end{equation}
and the Boltzmann probability of being in state $p$ of $H_E$ is given by 
\begin{equation}
p_p^{(E)} \>=\> 
\frac{e^{-\beta E_p^{(E)}}}{\sum_{p'=1}^{D_E} e^{-\beta E_{p'}^{(E)}}} 
\>=\> 
\frac{e^{-\beta E_p^{(E)}}}{Z_E} 
\>.
\end{equation}
The partition function of the system is given by 
\begin{equation}
Z_S(\beta) \>=\>
{\rm Tr}_{S}\left(e^{-\beta H_S}\right)
\> = \> 
\sum_{i=1}^{D_S} e^{-\beta E_i^{(S)}} 
\end{equation}
and the partition function of the environment is given by 
\begin{equation}
Z_E(\beta) \>=\>
{\rm Tr}_{E}\left(e^{-\beta H_E}\right)
\> = \> 
\sum_{p=1}^{D_E} e^{-\beta E_p^{(E)}} 
\>. 
\end{equation}
Important to note is that even though for the uncoupled case 
($\lambda =0$) the Hamiltonians $H_S$ and $H_E$ are uncoupled, the 
state of the entirety $S+E$ in Eq.~(\ref{EqS:Psi0a}) is entangled since 
$d_{i,p}\ne d_i d_p$ for the random Gaussian variables.  
As described in 
the main text, there are ways to achieve this condition physically, for example by using a 
much larger quantum bath that couples simultaneously to $S$ and $E$, 
and then slowly remove this large quantum bath.  

\subsection{Reduced density matrix}
The density matrix for the entirety $S+E$ is $\rho$.  
The reduced density matrix ${\widetilde\rho}$ for $S$, 
written in the basis $\left\{\left|E^{(S)}_i\right\rangle\right\}$ 
that diagonalizes $H_S$, is defined by a partial trace 
over the environment, and has matrix elements 
(for any $\lambda H_{SE}$) given by
\begin{widetext}
\begin{equation}
\left\langle E_i^{(S)}\right| {\widetilde\rho} \left| E_{i'}^{(S)}\right\rangle 
\>=\>
{\widetilde\rho}_{i,i'}
\>=\>
\left\langle E_i^{(S)}\right| {\rm Tr}_E\left(\rho\right) \left| E_{i'}^{(S)}\right\rangle 
\>=\>
\sum_{p=1}^{D_E} 
\left\langle E_i^{(S)}\right| \left(\left\langle p\right|
\rho 
\left|p\right\rangle\right) \left| E_{i'}^{(S)}\right\rangle 
\end{equation}
for any complete orthonormal basis 
$\left\{\left|p\right\rangle\right\}$ that spans the Hilbert space of the environment.  
The reduced density matrix elements ${\widetilde\rho}_{i,i'}$ in the energy basis 
that diagonalizes $H_S$ are thus 
\begin{eqnarray}
&&{\widetilde\rho}_{i,i'}
\>=\>   \cr 
&&\sum_{p=1}^{D_E} 
\frac{
\left(d_{1,1}^*\delta_{i,1}\delta_{p,1} + 
d_{i,p}^* \left(1-\delta_{i,1}\delta_{p,1}\right)
e^{-\frac{\beta \left(E_i^{(S)}-E_1^{(S)}\right)}{2}}
e^{-\frac{\beta \left(E_p^{(E)}-E_1^{(E)}\right)}{2}} \right)
\left(d_{1,1}\delta_{i',1}\delta_{p,1} + 
d_{i',p} \left(1-\delta_{i',1}\delta_{p,1}\right)
e^{-\frac{\beta \left(E_{i'}^{(S)}-E_1^{(S)}\right)}{2}}
e^{-\frac{\beta \left(E_p^{(E)}-E_1^{(E)}\right)}{2}} \right)
}
{d_{1,1}^* d_{1,1} +
\sum_{i''=1}^{D_S} \sum_{p''=1}^{D_E} d_{i'',p''}^* d_{i'',p''} 
\left(1-\delta_{i'',1}\delta_{p'',1}\right)
e^{-\beta \left(E_{i''}^{(S)}-E_1^{(S)}\right)}
e^{-\beta \left(E_{p''}^{(E)}-E_1^{(E)}\right)} } 
\>.\cr
&&
\label{EqS26}
\end{eqnarray}
\end{widetext}
Equation~(\ref{EqS26}) can be rewritten as
\begin{equation}
{\widetilde\rho}_{i,i'} \>=\> 
\sum_{p=1}^{D_E} \> 
\frac{d_{i,p}^* d_{i',p} e^{-\beta E_i^{(S)}/2} e^{-\beta E_{i'}^{(S)}/2} e^{-\beta E_p^{(E)}} }
{\sum_{i''=1}^{D_S} \sum_{p''=1}^{D_E} d_{i'',p''}^* d_{i'',p''} 
e^{-\beta E_{i''}^{(S)}} e^{-\beta E_{p''}^{(E)}} }
\>. 
\end{equation}

Care must be taken that for $d_{i,p}$, $d_{i',p}$ and $d_{i'',p''}$ 
the value of the random variable is the same wherever the indices are the same. For example 
the random number $d_{2,10}$ should be the same in both the numerator and denominator.  

\subsection{Expressions for the Random Gaussian Variables}
For the random Gaussian variables $d_k$, as defined in 
Eq.~(\ref{EqS:dk}), the $\phi_k$ for different $k$ are independent random variables 
distributed uniformly in $[0,2\pi)$.  
Furthermore, the probability density function (pdf) is given by 
\begin{equation}
\label{EqS:phiResults}
{\rm pdf}\left(\phi\right)
\>=\> 
\frac{1}{2\pi}
\end{equation}
so that the expectation values for the $\phi_k$ read
\begin{widetext}
\begin{equation}
\label{EqS:phiResults2}
\begin{array}{lclclcl}
{\cal E}\left(e^{i\phi}\right) & \> = \> & \int_0^{2\pi} e^{i\phi} \> {\rm pdf}(\phi) d\phi & 
       \>=\> & \frac{1}{2\pi} \int_0^{2\pi} \left[\cos(\phi)+i\sin(\phi)\right] \> 
       d\phi & \> = \> & 0 \\
{\cal E}\left(e^{i m \phi}\right) & \> = \> & \int_0^{2\pi} e^{i m \phi} \> {\rm pdf}(\phi) d\phi & 
       \>=\> & \frac{1}{2\pi} \int_0^{2\pi} \left[\cos(m\phi)+i\sin(m\phi)\right] \> 
       d\phi & \> = \> & 0 \\
{\cal E}\left(e^{i\phi_k} e^{+ i \phi_{k'}}\right) & \> = \> &&& 
       {\cal E}\left(e^{i\phi_k}\right) {\cal E}\left(e^{+ i \phi_{k'}}\right) 
       & = & 0 \quad {\rm for}\> k\ne k' \\
{\cal E}\left(e^{i\phi_k} e^{- i \phi_{k'}}\right) & \> = \> &&& 
       {\cal E}\left(e^{i\phi_k}\right) {\cal E}\left(e^{- i \phi_{k'}}\right) 
       & = & 0 \quad {\rm for}\> k\ne k' \\
{\cal E}\left(e^{i\phi_k} e^{- i \phi_{k'}}\right) & \> = \> &&& 
       {\cal E}\left(1\right) 
       & = & 1 \quad {\rm for}\> k= k' \\
\end{array}
\end{equation}
\end{widetext}
which greatly simplifies the perturbation calculations performed in this section.  
Note that all expectation values for $d_k$ are zero unless they are 
expectation values only for the absolute value $|d_k|^2= d_k^* d_k = x_k$ 
of the Gaussian random variables.  

For independent Gaussian random numbers (not our case, as we discuss 
below in this subsection), 
the distribution of the $\left|d\right|^2$ is given by a complete 
error function, defined by
\begin{equation}
{\rm erfc}(z) = 1 - {\rm erf}(z) =
\frac{2}{\sqrt{\pi}} \int_{z}^\infty e^{-t^2} dt 
\>.
\end{equation}
One can show this by using inverse transform sampling.  
In particular, the distribution for any $|d_1|^2$ is assumed to be, 
with the definition $x_1=|d_1|^2$, 
\begin{equation}
{\rm pdf}(x_1) = \frac{\pi D}{4} {\rm erfc}\left(\frac{D\sqrt{\pi}}{4} x_1 \right) 
\> .
\end{equation}
For independent $\{x_k\}$ the expectation values are
\begin{widetext}
\begin{equation}
\label{EqS:Ex}
\begin{array}{lclclcl}
{\cal E}\left(x\right) & \> = \> & \int_0^\infty x \> {\rm pdf}(x) dx & \>=\> & 
       \frac{\pi D}{4} \int_0^\infty x \> {\rm erfc}\left(\frac{D\sqrt{\pi}}{4} 
       dx \right) dx & \> = \> & \frac{1}{D} \\
{\cal E}\left(x^2\right) & \> = \> & \int_0^\infty x^2 \> {\rm pdf}(x) dx & \>=\> & 
       \frac{\pi D}{4} \int_0^\infty x^2 \> {\rm erfc}\left(\frac{D\sqrt{\pi}}{4} 
       dx \right) dx & \> = \> & \frac{16}{3 \pi D^2} \\
{\cal E}\left(x_i x_j\right) & \> = \> &&& {\cal E}\left(x_i\right) {\cal E}\left(x_j\right) 
       & = &\frac{1}{D^2} \\
{\cal E}\left(x^3\right) & \> = \> & \int_0^\infty x^3 \> {\rm pdf}(x) dx & \>=\> & 
       \frac{\pi D}{4} \int_0^\infty x^3 \> {\rm erfc}\left(\frac{D\sqrt{\pi}}{4} 
       dx \right) dx & \> = \> & \frac{12}{\pi D^3} \\
{\cal E}\left(x^4\right) & \> = \> & \int_0^\infty x^4 \> {\rm pdf}(x) dx & \>=\> & 
       \frac{\pi D}{4} \int_0^\infty x^4 \> {\rm erfc}\left(\frac{D\sqrt{\pi}}{4} 
       dx \right) dx & \> = \> & \frac{512}{5 \pi^2 D^4}\>. \\
\end{array}
\end{equation}
\end{widetext}

The expressions in Eq.~(\ref{EqS:Ex}) are only approximately correct for our case.
The reason is that the pdf for $D$ components of the random variables is
given by 
\begin{equation}
\label{EqS:pdfHamsHans}
\frac{1}{\rm Normalization} 
{\rm pdf}(x_1) {\rm pdf}(x_2) \cdots {\rm pdf}(x_D) 
\delta\left(x_1+x_2+\cdots + x_D -1\right) 
\end{equation}  
where the normalization is complicated.  
However,  Hams and De~Raedt~\cite{HAMS00} have calculated the 
correct expectation values for the pdf in Eq.~(\ref{EqS:pdfHamsHans}), namely 
\begin{equation}
\label{EqS:HamsEE}
\begin{array}{lcl}
{\cal E}(x) & \> = \> & \frac{1}{D} \> \\
{\cal E}(x^2) & = & \frac{2}{D(D+1)} \> \\
{\cal E}(x_i x_j) & = & \frac{1}{D(D+1)} \>. \\
\end{array}
\end{equation}
Therefore, we do not have to calculate these expectation values, but rather just use these results 
from \cite{HAMS00}.  

\begin{figure}[t]
\includegraphics[width=8cm]{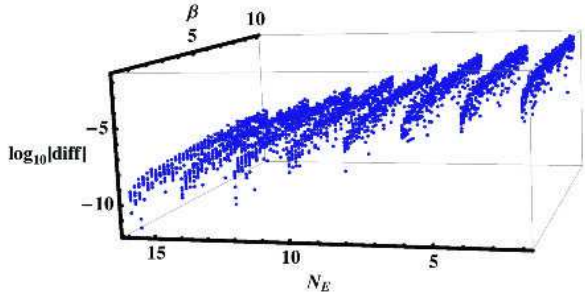}
\caption{\label{fig0}
Examples illustrating the approximation in Eq.~(\protect\ref{EqS:Dapprox}).  
The system is taken to have a Hilbert space of dimension $D_S=2^4$.  
The environment is taken to have a Hilbert space of dimension $D_S=2^{N_E}$, 
for $N_E=2, 4, 6, 8, 10, 12, 14 ,16$.  
The values of $\beta$ shown are from $\beta=0.25$ to $\beta=10$ in steps of $0.25$.  
Here ${\rm diff}= 
\left|\left(\sum_{i=1}^{D_S} \sum_{p=1}^{D_E} \left|d_{i,p}\right|^2 p_i^{(S)} p_p^{(E)}\right) 
-\frac{1}{D}\right|$.  
The eigenvalues for both $E$ and $S$ were taken to be random numbers uniformly 
distributed in $[-2,1]$.  
There are 10 points at each value of $N_E$ and $\beta$, each with different random 
eigenvalues for both $S$ and $E$ as well as different Gaussian random numbers $d_{i,p}$.  
}
\end{figure}

For sufficiently large $D$ we can use the approximation (see Fig.~\ref{fig0}) 
\begin{equation}
\label{EqS:Dapprox}
\sum_{k=1}^D \left|d_k\right|^2 p_k \approx \frac{1}{D}
\end{equation}
or by changing indices for the uncoupled case 
\begin{equation}
\sum_{i=1}^{D_S} \sum_{p=1}^{D_E} \left|d_{i,p}\right|^2 p_{i,p} 
\> = \> 
\sum_{i=1}^{D_S} \sum_{p=1}^{D_E} \left|d_{i,p}\right|^2 p_i^{(S)} p_p^{(E)} 
\approx \frac{1}{D}
\>.
\end{equation}
Note that Eq.~(\ref{EqS:Dapprox}) becomes exact in the infinite temperature limit 
($\beta\rightarrow 0$) where $p_k=1/D$ for all $k$ so 
\begin{equation}
\lim_{\beta\rightarrow 0} \sum_{k=1}^D \left|d_k\right|^2 p_k \>=\> 
\frac{1}{D} \sum_{k=1}^D \left|d_k\right|^2 \>=\> \frac{1}{D}
\>.
\end{equation}
In the zero temperature limit ($\beta\rightarrow +\infty$) 
Eq.~(\ref{EqS:Dapprox}) also becomes exact.  
Let $g_1$ be the ground state degeneracy of the entirety Hamiltonian $H$ associated 
with energy $E_1$.  
Then 
\begin{widetext}
\begin{equation}
\lim_{\beta\rightarrow\infty} p_k \>=\> 
\lim_{\beta\rightarrow\infty} \frac{e^{-\beta E_k}}{Z} \>=\> 
\lim_{\beta\rightarrow\infty} \frac{e^{-\beta E_k}}
{g_1 e^{-\beta E_1}+\sum_{k'=1+g_1}^D e^{-\beta E_{k'}}} \>=\> 
\left\{
\begin{array}{lcl}
\frac{1}{g_1} & \quad & k=1,2,\cdots,g_1 \\
0 & \quad & k=g_1+1, g_1+2, \cdots, D
\>. 
\end{array}
\right.
\end{equation}
Hence the expectation value is 
\begin{equation}
\lim_{\beta\rightarrow\infty} {\cal E}\left(\sum_{k=1}^D \left|d_k\right|^2 p_k \right) \>=\> 
\lim_{\beta\rightarrow\infty} \sum_{k=1}^D {\cal E}\left(\left|d_k\right|^2\right)  p_k \>=\> 
\frac{1}{g_1} \sum_{k=1}^{g_1} {\cal E}\left(\left|d_k\right|^2\right)  \>=\> 
\frac{1}{g_1} \> g_1\> \frac{1}{D} \>=\> \frac{1}{D}
\>.
\end{equation}
\end{widetext}

The approximation given by Eq.~(\ref{EqS:Dapprox}) is an uncontrolled approximation, and therefore 
we do not use it in our derivation of the perturbation theory for either $\sigma$ or $\delta$.  
We have included the results here because the approximation was discussed in the main paper 
as a way to motivate our perturbation results obtained without using the approximation.  

\subsection{General procedure for Taylor expansion: General function}

We need to calculate expectation values for the $x_i$ for a general function.  
We can do a Taylor 
expansion about $x_i=1/D$ and take the expectation value 
with respect to the probability distribution of the $x_i$ or $d_i$ denoted by ${\cal E}(\cdot)$
\begin{widetext}
\begin{equation}
\label{EqS:General1}
\begin{array}{lcl}
{\cal E}\left(f\left(\{x\}\right)\right) \!
& = & 
f\left(\frac{1}{D}, \frac{1}{D}, \cdots \frac{1}{D}\right)
\\
&  & 
+ \sum_{\ell=1}^D 
\left.\frac{\partial
f\left(x_1,x_2,\cdots,x_D\right)}{\partial x_\ell}\right|_{\{x\}=\frac{1}{D}}
{\cal E}\left(
\left(x_\ell-\frac{1}{D}\right)
\right)  
\\
&  & 
+ \frac{1}{2!} \sum_{\ell=1}^D 
\left.\frac{\partial^2
f\left(x_1,x_2,\cdots,x_D\right)}{\partial x_\ell^2}\right|_{\{x\}=\frac{1}{D}}
{\cal E}\left(
\left(x_\ell-\frac{1}{D}\right)^2
\right)
\\
&  & 
+ \frac{1}{2!} \sum_{\ell=1}^D \sum_{\ell'=1}^D 
\left(1-\delta_{\ell,\ell'}\right)
\left.\frac{\partial^2
f\left(x_1,x_2,\cdots,x_D\right)}{\partial x_\ell \partial x_{\ell'}}
\right|_{\{x\}=\frac{1}{D}}
{\cal E}\left(
\left(x_\ell-\frac{1}{D}\right) \left(x_{\ell'}-\frac{1}{D}\right)
\right)  
\\
&  & 
+ \frac{1}{3!} \sum_{\ell=1}^D 
\left.\frac{\partial^3
f\left(x_1,x_2,\cdots,x_D\right)}{\partial x_\ell^3}\right|_{\{x\}=\frac{1}{D}}
{\cal E}\left(
\left(x_\ell-\frac{1}{D}\right)^3
\right)
\\
&  & 
+ \frac{1}{3!} \sum_{\ell=1}^D \sum_{\ell'=1}^D \sum_{\ell''=1}^D 
\left(\delta_{\ell,\ell'}+\delta_{\ell,\ell''}+\delta_{\ell',\ell''}\right) 
\left(1-\delta_{\ell,\ell'}\delta_{\ell,\ell''}\delta_{\ell',\ell''}\right)
\>\times
\\
& & \quad \left.\frac{\partial^3
f\left(x_1,x_2,\cdots,x_D\right)}{\partial x_\ell \partial x_{\ell'} 
\partial x_{\ell''}}
\right|_{\{x\}=\frac{1}{D}}
{\cal E}\left(
\left(x_\ell-\frac{1}{D}\right)\left(x_\ell'-\frac{1}{D}\right)\left(x_\ell''-\frac{1}{D}\right)
\right)
\\
&  & 
+ \frac{1}{3!} \sum_{\ell=1}^D \sum_{\ell'=1}^D \sum_{\ell''=1}^D 
\left(1-\delta_{\ell,\ell'}\right)\left(1-\delta_{\ell,\ell''}\right)
\left(1-\delta_{\ell',\ell''}\right) 
\>\times 
\\
& & \quad 
\left.\frac{\partial^3
f\left(x_1,x_2,\cdots,x_D\right)}{\partial x_\ell \partial x_{\ell'}
\partial x_{\ell''}}
\right|_{\{x\}=\frac{1}{D}}
{\cal E}\left(
\left(x_\ell-\frac{1}{D}\right)\left(x_\ell'-\frac{1}{D}\right)\left(x_\ell''-\frac{1}{D}\right)
\right)
\\
&  & 
+ {\rm higher}\>{\rm order}\>{\rm terms}.
\\
\end{array}
\end{equation}
\end{widetext}
Note that since the expectation values for quantities such as ${\cal E}\left(x_{\ell}^2\right)$ and 
${\cal E}\left(x_{\ell}x_{\ell'}\right)$ are different, we had to write the second-order 
term as two terms: one for the same-$\ell$'s terms and one for the different-$\ell,\ell'$ terms.  
For the same reason,
the third-order term is written as three different terms, one with all-same $\ell$'s, one with 
all different $\ell$'s, and one with two and only two same-$\ell$'s.  
Then use the fact that the expectation values are known \cite{HAMS00} 
using Eq.~(\ref{EqS:HamsEE}), 
for example, up to second order,
\begin{widetext}
\begin{equation}
\begin{array}{lclcl}
{\cal E}\left( \left(x_\ell-\frac{1}{D}\right) \right) & \> = \> & 0  \\ 
{\cal E}\left( \left(x_\ell-\frac{1}{D}\right)^2 \right) & = & 
        {\cal E}\left(x_\ell^2\right) - \frac{2}{D}{\cal E}\left(x_\ell\right) + \frac{1}{D^2} 
        & \>=\> & \frac{D-1}{D^2(D+1)} \\ 
{\cal E}\left( \left(x_\ell-\frac{1}{D}\right) \left(x_{\ell'}-\frac{1}{D}\right) 
        \right) & = & 
        {\cal E}\left(x_\ell x_{\ell'}\right) - \frac{1}{D}{\cal E}\left(x_\ell\right) 
        - \frac{1}{D}{\cal E}\left(x_{\ell'}\right) + \frac{1}{D^2} 
        & \>=\> & -\>\frac{1}{D^2 (D+1)} \quad\quad \ell\ne\ell'\\ 
\end{array}
\end{equation}
\end{widetext}
and the derivatives of $f$ can be calculated, at least via Mathematica.  

\subsection{Derivation of ${\cal E}(\delta^2)$ for the uncoupled entirety}

We first derive the expectation value for ${\cal E}\left(\delta^2\right)$
since this is easier than the corresponding expectation value for $\sigma$.  
The ease is because only diagonal elements of $\widetilde\rho$ enter into the expression for 
$\delta$, since we have the definition
\begin{equation}
\delta^2 \> = \> 
\sum_{i=1}^{D_S} \left (
{\widetilde\rho}_{i,i} - 
\frac{e^{-b E_i^{(S)}}}{\sum_{i'=1}^{D_S} e^{-b E_{i'}^{(S)}}}
\right)^2
\end{equation}
with the fitting parameter $b$ given by 
\begin{equation}
b \>=\> 
\frac{\sum_{i<j, \> E_{i}^{(S)}\ne E_{j}^{(S)}} \> 
\frac{{\rm ln}\left({\widetilde\rho}_{i,i}\right) - {\rm ln}\left({\widetilde\rho}_{j,j}\right)}
{E_j^{(S)}-E_i^{(S)}}}
{\sum_{i'<j', \> E_{i'}^{(S)}\ne E_{j'}^{(S)}} \> 1}
\> .
\end{equation}
Therefore for $\delta^2$ there are no $\phi_k$ terms in the Gaussian random numbers in 
Eq.~(\ref{EqS:dk}).  This is because only the diagonal elements of the reduced density matrix 
given by 
\begin{equation}
{\widetilde\rho}_{i,i}(\beta,\{x_{i,p}\}) = 
\sum_{p=1}^{D_E} \frac{x_{i,p} p_{i,p}}
{\sum_{i''=1}^{D_S} \sum_{p''=1}^{D_E} x_{i'',p''} p_{i'',p''} }
\end{equation}
enter expressions for $\delta$ (while expressions for $\sigma$ involve the 
off-diagonal elements of ${\widetilde\rho}_{i,j}$).  
Remember, care must be taken that both for $x_{i,p}$ and $x_{i'',p''}$ wherever the 
indices are the same the value of the variable is the same. For example 
the random number $x_{1,1}$ is the same in both the numerator and denominator.  

Introduce $\Delta b = b-\beta$ with $b$ the fitting parameter, 
so $b=\beta + \Delta b$.  

The function we need to analyze is 
\begin{equation}
f_{\delta^2}(\beta,\Delta b,\{x_{i,p}\}) = 
\sum_{i=1}^{D_S}
\left[{\widetilde\rho}_{i,i}(\beta,\{x_{i,p}\})-p_i^S(\beta,\Delta b)\right]^2
\label{eqSf_delta}
\end{equation}
with the definition
\begin{equation}
p_i^{(S)}\left(\beta,\kappa\right)
\> = \>
\frac{e^{\left(\beta+\kappa\right)E_i^{(S)}}}
{\sum_{i'=1}^{D_S} e^{\left(\beta+\kappa\right)E_{i'}^{(S)}}}
\>.
\end{equation}

For the non-interacting case, $\lambda=0$, we need to analyze the function 
Eq.~(\ref{eqSf_delta}) with
\begin{equation}
{\widetilde\rho}_{i,i}(\beta,\{x_{i,p}\}) = 
\sum_{p=1}^{D_E} \frac{x_{i,p} p_{i}^{(S)}(\beta,0) p_{p}^{(E)}(\beta) }
{\sum_{i''=1}^{D_S} \sum_{p''=1}^{D_E} x_{i'',p''} 
p_{i''}^{(S)}(\beta,0) p_{p''}^{(E)}(\beta) }
\>.
\end{equation}

For the lowest-order (zeroth-order) term in the Taylor expansion we replace 
all $x_{i,p}$ by $1/D$.  This gives that 
\begin{widetext}
\begin{equation}
\begin{array}{lcl}
{\widetilde\rho}_{i,i}\left(\beta,\{x_{i,p}\}=\frac{1}{D}\right) 
& \> = \> & 
\sum_{p=1}^{D_E} \frac{ \frac{1}{D} p_{i}^{(S)}(\beta,0) p_{p}^{(E)}(\beta) }
{\sum_{i''=1}^{D_S} \sum_{p''=1}^{D_E} \frac{1}{D} 
p_{i''}^{(S)}(\beta,0) p_{p''}^{(E)}(\beta) }
 \> = \>  
\sum_{p=1}^{D_E} \frac{ p_{i}^{(S)}(\beta,0) p_{p}^{(E)}(\beta) }
{\sum_{i''=1}^{D_S} \sum_{p''=1}^{D_E} 
p_{i''}^{(S)}(\beta,0) p_{p''}^{(E)}(\beta) }
\\
& \> = \> & 
p_{i}^{(S)}(\beta,0) \sum_{p=1}^{D_E} p_{p}^{(E)}(\beta) 
 \> = \>  
p_{i}^{(S)}(\beta,0) 
\\
\end{array}
\end{equation}
since $\sum_{i=1}^{D_S} p_i^{(S)}(\beta,0) = 1$ 
and   $\sum_{p=1}^{D_E} p_p^{(E)}(\beta) = 1$.   
Thus one has 
\begin{equation}
f_{\delta^2}\left(\beta,\Delta b,\{x_{i,p}\}=\frac{1}{D}\right) = 
\sum_{i=1}^{D_S}
\left[p_i^{(S)}(\beta,0)-p_i^{(S)}(\beta,\Delta b)\right]^2
\end{equation}
which obviously has its minimum at 
$\Delta b=0$.  
Therefore, we perform a Taylor expansion also about $\Delta b=0$, as 
well as an expansion in the $\{x_{i,p}\}$ about $\frac{1}{D}$.   

For the first-order terms we make use of the chain rule.  This gives 
\begin{equation}
\frac{\partial f_{\delta^2}}{\partial\Delta b} 
\> = \> 
- 2 \sum_{i=1}^{D_S} 
\left( {\widetilde\rho_{i,i}}(\beta,\{x_{i,p}\}) - p_i^{(S)}(\beta,\Delta b) \right) 
\frac{\partial p_i^{(S)}(\beta,\Delta b)}{\partial\Delta b}
\end{equation}
and
\begin{equation}
\frac{\partial f_{\delta^2}}{\partial x_{j,q}} 
\> = \> 
2 \sum_{i=1}^{D_S} 
\left( {\widetilde\rho}_{i,i}(\beta,\{x_{i,p}\}) - p_i^{(S)}(\beta,\Delta b) \right) 
\frac{\partial {\widetilde\rho}_{i,i}(\beta,\{x_{i,p}\})}{\partial x_{j,q}}
\>.
\end{equation}
Note that 
\begin{equation}
\left.\frac{\partial f_{\delta^2}}{\partial\Delta b} 
\right|_{\Delta b=0,\> \{x_{i,p}\}=\frac{1}{D} }
\> = \> 
0
\end{equation}
and
\begin{equation}
\left. \frac{\partial f_{\delta^2}}{\partial x_{j,q}} 
\right|_{\Delta b=0,\> \{x_{i,p}\}=\frac{1}{D} }
\> = \> 
0
\>.
\end{equation}

Hence we need to go to the second order terms.  

For $\Delta b$, this is
\begin{equation}
\label{EqS:Deltab}
\frac{\partial^2 f_{\delta^2}}{\partial\left(\Delta b\right)^2} 
\> = \> 
2 \sum_{i=1}^{D_S} 
\left( \frac{\partial p_i^{(S)}(\beta,\Delta b)}{\partial\Delta b} \right)^2 
- 2 \sum_{i=1}^{D_S} 
\left( {\widetilde\rho}_{i,i}(\beta,\{x_{i,p}\}) - p_i^{(S)}(\beta,\Delta b) \right) 
\frac{\partial^2 p_i^{(S)}(\beta,\Delta b)}{\partial \left(\Delta b\right)^2}
\>.
\end{equation}
Evaluating at $\Delta b=0$ gives 
\begin{equation}
\left.\frac{\partial^2 f_{\delta^2}}{\partial\left(\Delta b\right)^2} 
\right|_{\Delta b=0,\> \{x_{i,p}\}=\frac{1}{D} }
\> = \> 
2 \sum_{i=1}^{D_S} 
\left.
\left( \frac{\partial p_i^{(S)}(\beta,\Delta b)}{\partial\Delta b} \right)^2 
\right|_{\Delta b=0,\> \{x_{i,p}\}=\frac{1}{D} }
\>.
\end{equation}
One has  
\begin{equation}
\begin{array}{lcl}
\sum_{i=1}^{D_S}
\left. \frac{\partial p_i^{(S)}(\beta,\Delta b)}{\partial\Delta b} 
\right|_{\Delta b=0}
& \> = \> & 
\left. 
\frac{\partial}{\partial\Delta b} 
\left(
\sum_{i=1}^{D_S} \frac{ e^{-\beta E_{i}^{(S)}} e^{-\Delta b E_{i}^{(S)}} }
{ \sum_{i'=1}^{D_S} e^{-\beta E_{i'}^{(S)}} e^{-\Delta b E_{i'}^{(S)}} }
\right)
\right|_{\Delta b=0}
\\
& \> = \> & 
\left. 
\frac{\partial}{\partial\Delta b} 
\left(
1\right)
\right|_{\Delta b=0}
\\
& = & 
0
\>.
\end{array}
\end{equation}
However, the term one needs to sum for the second order term of Eq.~(\ref{EqS:Deltab}) is   
\begin{eqnarray}
2 \sum_{i=1}^{D_S}
\left. \left[\frac{\partial p_i^{(S)}(\beta,\Delta b)}{\partial\Delta b} \right]^2 
\right|_{\Delta b=0}
& \> = \> & 
2 \left.
\sum_{i=1}^{D_S}
\left[  
\frac{\partial}{\partial\Delta b} 
\frac{ e^{-\beta E_{i}^{(S)}} e^{-\Delta b E_{i}^{(S)}} }
{ \sum_{i'=1}^{D_S} e^{-\beta E_{i'}^{(S)}} e^{-\Delta b E_{i'}^{(S)}} }
\right]^2 
\right|_{\Delta b=0}
\cr
& \> = \> & 
2 \sum_{i=1}^{D_S} 
\left.\left[
\frac{ e^{-\beta E_{i}^{(S)}} e^{-\Delta b E_{i}^{(S)}} 
\left(
\sum_{i''=1}^{D_S} E_{i''}^{(S)} e^{-\beta E_{i''}^{(S)}} e^{-\Delta b E_{i''}^{(S)}} 
\right)
} 
{\left(
\sum_{i'=1}^{D_S} 
e^{-\beta E_{i'}^{(S)}} e^{-\Delta b E_{i'}^{(S)}} 
\right)^2 }
-
\frac{ E_{i}^{(S)} e^{-\beta E_{i}^{(S)}} e^{-\Delta b E_{i}^{(S)}} }
{\left(\sum_{i'=1}^{D_S} e^{-\beta E_{i'}^{(S)}} e^{-\Delta b E_{i'}^{(S)}}\right) }
\right]^2 
\right|_{\Delta b=0}
\cr 
& \> = \> & 
2 \sum_{i=1}^{D_S} 
\left.
\left[
\frac{ e^{-\beta E_{i}^{(S)}} e^{-\Delta b E_{i}^{(S)}} 
\left(
\sum_{i''=1}^{D_S} E_{i''}^{(S)} e^{-\beta E_{i''}^{(S)}} e^{-\Delta b E_{i''}^{(S)}} 
\right)
} 
{\left(
\sum_{i'=1}^{D_S} 
e^{-\beta E_{i'}^{(S)}} e^{-\Delta b E_{i'}^{(S)}} 
\right)^2 }
\right]^2 
\right|_{\Delta b=0}
\cr 
& \>  \> & \quad 
- 4 \sum_{i=1}^{D_S} 
\left.
\left[
\frac{ e^{-\beta E_{i}^{(S)}} e^{-\Delta b E_{i}^{(S)}} 
\left(
\sum_{i''=1}^{D_S} E_{i''}^{(S)} e^{-\beta E_{i''}^{(S)}} e^{-\Delta b E_{i''}^{(S)}} 
\right)
} 
{\left(
\sum_{i'=1}^{D_S} 
e^{-\beta E_{i'}^{(S)}} e^{-\Delta b E_{i'}^{(S)}} 
\right)^2 }
\> 
\frac{ E_{i}^{(S)} e^{-\beta E_{i}^{(S)}} e^{-\Delta b E_{i}^{(S)}} }
{ \left(\sum_{i'=1}^{D_S} e^{-\beta E_{i'}^{(S)}} e^{-\Delta b E_{i'}^{(S)}} \right) }
\right]
\right|_{\Delta b=0}
\cr 
& \>  \> & \quad 
+2 \sum_{i=1}^{D_S} 
\left.
\left[
\frac{ E_{i}^{(S)} e^{-\beta E_{i}^{(S)}} e^{-\Delta b E_{i}^{(S)}} }
{ \sum_{i'=1}^{D_S} e^{-\beta E_{i'}^{(S)}} e^{-\Delta b E_{i'}^{(S)}} }
\right]^2 
\right|_{\Delta b=0}
\cr 
& \> = \> & 
2 \sum_{i=1}^{D_S} 
\left[
\frac{ e^{-\beta E_{i}^{(S)}}  
\left(
\sum_{i''=1}^{D_S} E_{i''}^{(S)} e^{-\beta E_{i''}^{(S)}}  
\right)
} 
{\left(
\sum_{i'=1}^{D_S} 
e^{-\beta E_{i'}^{(S)}}  
\right)^2 }
\right]^2 
\cr 
& \>  \> & \quad 
- 4 \sum_{i=1}^{D_S} 
\left[
\frac{ e^{-\beta E_{i}^{(S)}}  
\left(
\sum_{i''=1}^{D_S} E_{i''}^{(S)} e^{-\beta E_{i''}^{(S)}}  
\right)
} 
{\left(
\sum_{i'=1}^{D_S} 
e^{-\beta E_{i'}^{(S)}} 
\right)^2 }
\>
\frac{ E_{i}^{(S)} e^{-\beta E_{i}^{(S)}}  }
{ \left(\sum_{i'=1}^{D_S} e^{-\beta E_{i'}^{(S)}} \right) }
\right]
+2 \sum_{i=1}^{D_S} 
\left[
\frac{ E_{i}^{(S)} e^{-\beta E_{i}^{(S)}} }
{ \sum_{i'=1}^{D_S} e^{-\beta E_{i'}^{(S)}} }
\right]^2 
\cr 
& \> = \> & 
2 \frac{1}{ Z_S^4(\beta) } \>
\sum_{i=1}^{D_S} 
\left[
e^{-\beta E_{i}^{(S)}}  
\left(
\sum_{i''=1}^{D_S} E_{i''}^{(S)} e^{-\beta E_{i''}^{(S)}}  
\right)
\right]^2 
\cr
& \>  \> & \quad 
- 4 \frac{1}{ Z_S^3(\beta) }
\sum_{i=1}^{D_S} 
\left[
e^{-\beta E_{i}^{(S)}}  
\left(
\sum_{i''=1}^{D_S} E_{i''}^{(S)} e^{-\beta E_{i''}^{(S)}}  
\right)
\>
E_{i}^{(S)} e^{-\beta E_{i}^{(S)}}  
\right]
\cr
& \>  \> & \quad 
+2 \frac{1}{ Z_S^2(\beta) } \> 
\sum_{i=1}^{D_S} 
\left[
E_{i}^{(S)} e^{-\beta E_{i}^{(S)}} 
\right]^2 
\cr
& \> = \> & 
2 \frac{\left(\left\langle E(\beta)\right\rangle_S\right)^2\>Z_S(2\beta)}
{ Z_S^2(\beta) } 
- 4 
\frac{ \left\langle E(\beta)\right\rangle_S \> 
\left\langle E(2\beta)\right\rangle_S \> 
Z_S(2\beta)}{ Z_S^2(\beta) }
+2 \frac{ \left\langle E^2(2\beta) \right\rangle_S \> Z_S(2\beta)}
{ Z_S^2(\beta) } 
\>.
\end{eqnarray}
Therefore, the result for the first non-zero term for $\Delta b$ is  
\begin{equation}
\frac{1}{2!} \> 
\left.
\frac{\partial^2 f_{\delta^2}}{\partial\left(\Delta b\right)^2} 
\right|_{\Delta b=0,\> \{x_{i,p}\}=\frac{1}{D} }
\> \left(\Delta b\right)^2 
\> = \> 
\frac{ Z_S(2\beta) }{ Z_S^2(\beta) } 
\>
\left[
\left(\left\langle E(\beta)\right\rangle_S\right)^2\>
- 2 
\left\langle E(\beta)\right\rangle_S \> \left\langle E(2\beta)\right\rangle_S  
+ 
\left\langle E^2(2\beta) \right\rangle_S 
\right]
\left(\Delta b\right)^2 
\> + \> 
{\rm higher}\>{\rm order}\>{\rm terms} 
\>.
\end{equation}
Initially one would anticipate that one needs to calculate terms such as
\begin{equation}
\frac{\partial^2 f_{\delta^2}}{\partial\left(\Delta b\right) \partial x_{j,q} } 
\end{equation}
and evaluate them at $\Delta b=0,\> \{x_{i,p}\}=\frac{1}{D}$.  However, 
all such terms will be multiplied by $\left(x_{j,q}-\frac{1}{D}\right)$, which has 
an expectation value which vanishes.  
Therefore one has
\begin{eqnarray}
{\cal E}\left(\delta^2\right) \> &=& \>
\frac{ Z_S(2\beta) }{ Z_S^2(\beta) } 
\>
\left[
\left(\left\langle E(\beta)\right\rangle_S\right)^2\>
- 2 
\left\langle E(\beta)\right\rangle_S \> \left\langle E(2\beta)\right\rangle_S  
+ 
\left\langle E^2(2\beta) \right\rangle_S 
\right]
\left(\Delta b\right)^2 
\> \cr
&&
+ {\cal O}\left( (\Delta b)^3 \right) 
+ {\cal O}\left( (\Delta b) \left\{x_{j,q}\right\}^2 \right) 
+ {\cal O}\left( \left\{x_{j,q}\right\} \left\{x_{j',q'}\right\}  
\left(1-\delta_{j,j'}\delta_{q,q'}\right) \right) 
+ {\cal O}\left( \left\{x_{j,q}\right\}^2 \right) 
\>. 
\end{eqnarray} 
One can also use that the specific heat (at constant volume) is 
$C_v(\beta) = k_{\rm B} \beta^2 
\left\langle \left(\Delta E(\beta)\right)^2\right\rangle$,
so 
\begin{equation}
\left\langle E^2(2\beta)\right\rangle 
\> = \>
\frac{C_v(2\beta)}{4 k_{\rm B} \beta^2} 
+ \left( \left\langle E(2\beta)\right\rangle \right)^2
\>.
\end{equation}
The final result is consequently 
\begin{eqnarray}
{\cal E}\left(\delta^2\right) \> &=& \>
\frac{ Z_S(2\beta) }{ Z_S^2(\beta) } 
\>
\left[
\frac{1}{4 k_{\rm B} \beta^2} C_v^{(S)}(2\beta) 
+ \> \left(
\left\langle E(2\beta) \right\rangle_S 
\>-\>\left\langle E(\beta)\right\rangle_S 
\right)^2
\right]
\left(\Delta b\right)^2 
\> \cr
&&
+ {\cal O}\left( (\Delta b)^3 \right) 
+ {\cal O}\left( (\Delta b) \left\{x_{j,q}\right\}^2 \right) 
+ {\cal O}\left( \left\{x_{j,q}\right\} \left\{x_{j',q'}\right\}  
\left(1-\delta_{j,j'}\delta_{q,q'}\right) \right) 
+ {\cal O}\left( \left\{x_{j,q}\right\}^2 \right) 
\>. 
\end{eqnarray} 
Thus equilibrating the system, in particular fitting for $\Delta b$, 
is difficult to do near a phase transition where $C_v$ diverges.    

For the second order terms for the $\{x_{i,p}\}$ one has 
\begin{eqnarray}
\frac{\partial^2 f_{\delta^2}}{\partial x_{j,q} \partial x_{j',q'} } 
& \> = \> & 
2 \sum_{i=1}^{D_S} 
\left( {\widetilde\rho}_{i,i}(\beta,\{x_{i,p}\}) - p_i^{(S)}(\beta,\Delta b) \right) 
\frac{\partial^2 {\widetilde\rho}_{i,i}(\beta,\{x_{i,p}\})}{\partial x_{j,q}\partial x_{j',q'}}
\cr
& & \quad 
+ \> 2 \sum_{i=1}^{D_S} 
\frac{\partial {\widetilde\rho}_{i,i}(\beta,\{x_{i,p}\})}{\partial x_{j,q}}
\frac{\partial {\widetilde\rho}_{i,i}(\beta,\{x_{i,p}\})}{\partial x_{j',q'}}
\>.
\end{eqnarray}
The derivative of ${\widetilde\rho}_{i,i}$ with respect to $\{x_{j,q}\}$ is given by
\begin{eqnarray} 
\frac{\partial {\widetilde\rho}_{i,i}}{\partial x_{j,q} } 
& \> = \> & 
\frac{\partial}{\partial x_{j,q} } 
\left(
\sum_{p=1}^{D_E} \frac{x_{i,p} \> p_{i}^{(S)}(\beta,0) \> p_{p}^{(E)}(\beta) }
{\sum_{i''=1}^{D_S} \sum_{p''=1}^{D_E} x_{i'',p''} 
p_{i''}^{(S)}(\beta,0) p_{p''}^{(E)}(\beta) }
\right)
\cr
& \> = \> & 
\delta_{i,j} \frac{ p_{i}^{(S)}(\beta,0) \> p_{q}^{(E)}(\beta) }
{\sum_{i''=1}^{D_S} \sum_{p''=1}^{D_E} x_{i'',p''} \> 
p_{i''}^{(S)}(\beta,0) p_{p''}^{(E)}(\beta) }
-
\sum_{p=1}^{D_E} 
\frac{x_{i,p} \> p_{i}^{(S)}(\beta,0) \> p_{p}^{(E)}(\beta) \> 
p_{j}^{(S)}(\beta,0) \> p_{q}^{(E)}(\beta) }
{\left(\sum_{i''=1}^{D_S} \sum_{p''=1}^{D_E} x_{i'',p''} \> 
p_{i''}^{(S)}(\beta,0) p_{p''}^{(E)}(\beta) \right)^2}
\>.
\end{eqnarray}
Evaluating at $\{x_{i,p}\}=\frac{1}{D}$ gives
\begin{eqnarray}
\left. \frac{\partial {\widetilde\rho}_{i,i}}{\partial x_{j,q} } 
\right|_{\left\{x\right\}=\frac{1}{D}} 
&  =  & 
\delta_{i,j} \frac{ p_{i}^{(S)}(\beta,0) \> p_{q}^{(E)}(\beta) }
{\sum_{i''=1}^{D_S} \sum_{p''=1}^{D_E} \frac{1}{D} \> 
p_{i''}^{(S)}(\beta,0) p_{p''}^{(E)}(\beta) }
-
\sum_{p=1}^{D_E} 
\frac{\frac{1}{D} \> p_{i}^{(S)}(\beta,0) \> p_{p}^{(E)}(\beta) \> 
p_{j}^{(S)}(\beta,0) \> p_{q}^{(E)}(\beta) }
{\left(\sum_{i''=1}^{D_S} \sum_{p''=1}^{D_E} \frac{1}{D} \> 
p_{i''}^{(S)}(\beta,0) p_{p''}^{(E)}(\beta) \right)^2}
\cr
& = & 
D \delta_{i,j} p_{i}^{(S)}(\beta,0) \> p_{q}^{(E)}(\beta) 
- D p_{i}^{(S)}(\beta,0) \> p_{j}^{(S)}(\beta,0) \> p_{q}^{(E)}(\beta) 
\sum_{p=1}^{D_E} p_p^{(E)}(\beta) 
\cr
& = & 
D \delta_{i,j} p_{i}^{(S)}(\beta,0) \> p_{q}^{(E)}(\beta) 
- D p_{i}^{(S)}(\beta,0) \> p_{j}^{(S)}(\beta,0) \> p_{q}^{(E)}(\beta) 
\cr
& = & 
D p_{i}^{(S)}(\beta,0) \> p_{q}^{(E)}(\beta)
\left( 
\delta_{i,j} \>  - \> p_{j}^{(S)}(\beta,0) 
\right)
\end{eqnarray}
since
$\sum_{i''=1}^{D_S} \sum_{p''=1}^{D_E} \> 
p_{i''}^{(S)}(\beta,0) p_{p''}^{(E)}(\beta) = 1$
and
$\sum_{p=1}^{D_E} p_p^{(E)}(\beta)=1$.

The second order term for the same $x_{j,q}$ is 
\begin{eqnarray} 
\frac{\partial^2 {\widetilde\rho}_{i,i}}{\partial x_{j,q}^2 } 
& \> = \> & 
- \delta_{i,j} \frac{ \left(p_{i}^{(S)}(\beta,0)\right)^2 \> 
\left(p_{q}^{(E)}(\beta)\right)^2 }
{\left(\sum_{i''=1}^{D_S} \sum_{p''=1}^{D_E} x_{i'',p''} \> 
p_{i''}^{(S)}(\beta,0) p_{p''}^{(E)}(\beta) \right)^2}
-
\delta_{i,j}  
\frac{ \left(p_{i}^{(S)}(\beta,0)\right)^2 \> \left(p_{q}^{(E)}(\beta)\right)^2  
}
{\left(\sum_{i''=1}^{D_S} \sum_{p''=1}^{D_E} x_{i'',p''} \> 
p_{i''}^{(S)}(\beta,0) p_{p''}^{(E)}(\beta) \right)^2}
\cr
& & \quad
+ 2 
\sum_{p=1}^{D_E} 
\frac{x_{i,p} \> p_{i}^{(S)}(\beta,0) \> p_{p}^{(E)}(\beta) \> 
\left(p_{j}^{(S)}(\beta,0)\right)^2 \> \left(p_{q}^{(E)}(\beta)\right)^2 }
{\left(\sum_{i''=1}^{D_S} \sum_{p''=1}^{D_E} x_{i'',p''} \> 
p_{i''}^{(S)}(\beta,0) p_{p''}^{(E)}(\beta) \right)^3}
\>.
\end{eqnarray}
However, one does not need to calculate this term, since it only multiplies a 
terms which is zero when $\Delta b=0$ and $\{x_{i,p}\}=\frac{1}{D}$.  

For the second order term twice for the $\{x_{i,p}\}$ one has 
\begin{equation}
\frac{\partial^2 f_{\delta^2}}{\partial \left(x_{j,q}\right)^2 } 
 =  
2 \sum_{i=1}^{D_S} 
\left( {\widetilde\rho}_{i,i}(\beta,\{x_{i,p}\}) - p_i^{(S)}(\beta,\Delta b) \right) 
\frac{\partial^2 {\widetilde\rho}_{i,i}(\beta,\{x_{i,p}\})}{\partial \left(x_{j,q}\right)^2 }
+ \> 2 \sum_{i=1}^{D_S} 
\left(\frac{\partial {\widetilde\rho}_{i,i}(\beta,\{x_{i,p}\})}{\partial x_{j,q}} \right)^2
\>.
\end{equation}
Hence 
\begin{eqnarray}
\left. \frac{\partial^2 f_{\delta^2}}{\partial \left(x_{j,q}\right)^2 } 
\right|_{\Delta b=0,\>\{x\}=\frac{1}{D}}
& \> = \> & 
\left. 2 \sum_{i=1}^{D_S} 
\left( {\widetilde\rho}_{i,i}(\beta,\{x_{i,p}\}) - p_i^{(S)}(\beta,\Delta b) \right) 
\frac{\partial^2 {\widetilde\rho}_{i,i}(\beta,\{x_{i,p}\})}{\partial \left(x_{j,q}\right)^2 }
\right|_{\Delta b=0,\>\{x\}=\frac{1}{D}}
\cr
& & \quad 
+ \> \left. 2 \sum_{i=1}^{D_S} 
\left(\frac{\partial {\widetilde\rho}_{i,i}(\beta,\{x_{i,p}\})}{\partial x_{j,q}} \right)^2
\right|_{\Delta b=0,\>\{x\}=\frac{1}{D}}
\cr
& = &  
\left. 2 \sum_{i=1}^{D_S} 
\left(\frac{\partial {\widetilde\rho}_{i,i}(\beta,\{x_{i,p}\})}{\partial x_{j,q}} \right)^2
\right|_{\Delta b=0,\>\{x\}=\frac{1}{D}}
\cr
& = &  
2 \sum_{i=1}^{D_S} 
\left(
D \> p_{i}^{(S)}(\beta,0) \> p_{q}^{(E)}(\beta)
\left( 
\delta_{i,j} \>  - \> p_{j}^{(S)}(\beta,0) 
\right)
\right)^2
\cr
& = &  
2 D^2 \> \left(p_{q}^{(E)}(\beta)\right)^2 \> 
\sum_{i=1}^{D_S} 
\left( p_{i}^{(S)}(\beta,0) \right)^2 \> 
\left( 
\delta_{i,j} \>  - \> p_{j}^{(S)}(\beta,0) 
\right)^2
\cr
& = &  
2 D^2 \> \left(p_{q}^{(E)}(\beta)\right)^2 \> 
\sum_{i=1}^{D_S} 
\left( p_{i}^{(S)}(\beta,0) \right)^2 \> 
\left(
\delta_{i,j} \>  - \> 2 \delta_{i,j} \> p_{j}^{(S)}(\beta,0) 
\>+\> \left( p_{j}^{(S)}(\beta,0) \right)^2 
\right)
\cr
& = &  
2 D^2 \> \left(p_{q}^{(E)}(\beta)\right)^2 \> 
\left(
\sum_{i=1}^{D_S} \delta_{i,j} \left( p_{i}^{(S)}(\beta,0) \right)^2 \> 
\right.  
- \> 2 
\sum_{i=1}^{D_S} \delta_{i,j} \> \left( p_{i}^{(S)}(\beta,0) \right)^2 \> 
p_{j}^{(S)}(\beta,0) 
\cr
& & \quad
+\> 
\left.
\left( \left( p_{j}^{(S)}(\beta,0) \right)^2 
\sum_{i=1}^{D_S} \left( p_{i}^{(S)}(\beta,0) \right)^2 \> \right) 
\right)
\cr
& = &  
2 D^2 \> \left(p_{q}^{(E)}(\beta)\right)^2 \> 
\left[
\left( p_{j}^{(S)}(\beta,0) \right)^2 \> \>  - 
\> 2 \> \left( p_{j}^{(S)}(\beta,0) \right)^3  
+ 
\left(p_{j}^{(S)}(\beta,0)\right)^2 \> \left( 
\sum_{i=1}^{D_S} \left( p_{i}^{(S)}(\beta,0) \right)^2 \> \right) 
\right].
\end{eqnarray}
We have to sum over all the same-second-partial terms to get the term that 
multiplies 
\begin{equation}
{\cal E}\left( \left(x_{i,p}-\frac{1}{D}\right)^2  \right)
\> = \>
{\cal E}\left( \left(x-\frac{1}{D}\right)^2  \right)
\> = \>
\frac{D-1}{D^2 (D+1)}
\end{equation}
since these expectation values are the same for all $x_{i,p}$.  
One has
\begin{eqnarray}
\sum_{j=1}^{D_S} \sum_{q=1}^{D_E} 
\left. \frac{\partial^2 f_{\delta^2}}{\partial \left(x_{j,q}\right)^2 } 
\right|_{\Delta b=0,\>\{x\}=\frac{1}{D}}
& = & 
2 D^2 \> \sum_{j=1}^{D_S} \sum_{q=1}^{D_E} 
\> \left(p_{q}^{(E)}(\beta)\right)^2 \> 
\left[
\left( p_{j}^{(S)}(\beta,0) \right)^2 \>  - 
\> 2 \> \left( p_{j}^{(S)}(\beta,0) \right)^3  
\right.
\cr
& & \quad 
\>+\> 
\left. 
\left(p_{j}^{(S)}(\beta,0)\right)^2 \> \left( 
\sum_{i=1}^{D_S} \left( p_{i}^{(S)}(\beta,0) \right)^2 \> \right) 
\right]
\cr
& \> = \> & 
2 D^2 \> \left(\sum_{q=1}^{D_E} 
\> \left(p_{q}^{(E)}(\beta)\right)^2 \right) \> \times 
\cr
& & \quad 
\sum_{j=1}^{D_S} \left[
\left( p_{j}^{(S)}(\beta,0) \right)^2 \> \>  - 
\> 2 \> \left( p_{j}^{(S)}(\beta,0) \right)^3  
\>+\> 
\left(p_{j}^{(S)}(\beta,0)\right)^2 \> \left( 
\sum_{i=1}^{D_S} \left( p_{i}^{(S)}(\beta,0) \right)^2 \> \right) 
\right]
\cr
& \> = \> & 
2 D^2 \> 
\left(\frac{Z_E(2\beta)}{Z_E^2(\beta)}\right) \>
\sum_{j=1}^{D_S} \left[
\left( p_{j}^{(S)}(\beta,0) \right)^2 \> \>  - 
\> 2 \> \left( p_{j}^{(S)}(\beta,0) \right)^3  
\right.
\cr
& & \quad 
\left.
\>+\> 
\left(p_{j}^{(S)}(\beta,0)\right)^2 \> \left( 
\sum_{i=1}^{D_S} \left( p_{i}^{(S)}(\beta,0) \right)^2 \> \right) 
\right]
\cr
& \> = \> & 
2 D^2 \> 
\left(\frac{Z_E(2\beta)}{Z_E^2(\beta)}\right) \>
\left[
\sum_{j=1}^{D_S} \left( p_{j}^{(S)}(\beta,0) \right)^2 \> \>  - 
\> 2 \> \sum_{j=1}^{D_S} \left( p_{j}^{(S)}(\beta,0) \right)^3  
\right.
\cr
& & \quad 
\left.
\>+\> 
\left(\sum_{j=1}^{D_S} \left(p_{j}^{(S)}(\beta,0)\right)^2 \right) 
\> \left( \sum_{i=1}^{D_S} \left( p_{i}^{(S)}(\beta,0) \right)^2 \> \right) 
\right]
\cr
& \> = \> & 
2 D^2 \> 
\left(\frac{Z_E(2\beta)}{Z_E^2(\beta)}\right) \>
\left[
\frac{Z_S(2\beta)}{Z_S^2(\beta)} \> \>  - 
\> 2 \> \frac{Z_S(3\beta)}{Z_S^3(\beta)} 
\right.
\cr
& & \quad 
\left.
\>+\> 
\left(\sum_{j=1}^{D_S} \left(p_{j}^{(S)}(\beta,0)\right)^2 \right) 
\> \left( \frac{Z_S(2\beta)}{Z_S^2(\beta)} \right) 
\right]
\cr
& \> = \> & 
2 \> D^2 \> \left(\frac{Z_E(2\beta)}{Z_E^2(\beta)}\right) \>
\left[
\frac{Z_S(2\beta)}{Z_S^2(\beta)} 
-2 \frac{Z_S(3\beta)}{Z_S^3(\beta)} 
+\frac{Z_S^2(2\beta)}{Z_S^4(\beta)}
\right].
\end{eqnarray}
Therefore, for these second-order terms the final result is that 
\begin{equation}
{\cal E}\left( \left(x_{i,p}-\frac{1}{D}\right)^2  \right)
\sum_{j=1}^{D_S} \sum_{q=1}^{D_E} 
\left. \frac{\partial^2 f_{\delta^2}}{\partial \left(x_{j,q}\right)^2 } 
\right|_{\Delta b=0,\>\{x\}=\frac{1}{D}}
\> = \>
2 \> \frac{D-1}{D+1} \> \left(\frac{Z_E(2\beta)}{Z_E^2(\beta)}\right) \>
\left[
\frac{Z_S(2\beta)}{Z_S^2(\beta)} 
-2 \frac{Z_S(3\beta)}{Z_S^3(\beta)} 
+\frac{Z_S^2(2\beta)}{Z_S^4(\beta)}
\right]
\>.
\end{equation}

For the second order terms with two different $\{x_{i,p}\}$ one has 
\begin{eqnarray}
\frac{\partial^2 f_{\delta^2}}{\partial x_{j,q} \partial x_{j',q'} } 
& \> = \> & 
2 \sum_{i=1}^{D_S} 
\left( {\widetilde\rho}_{i,i}(\beta,\{x_{i,p}\}) - p_i^{(S)}(\beta,\Delta b) \right) 
\frac{\partial^2 {\widetilde\rho}_{i,i}(\beta,\{x_{i,p}\})}{\partial x_{j,q}\partial x_{j',q'}}
\cr
& & \quad 
+ \> 2 \sum_{i=1}^{D_S} 
\frac{\partial {\widetilde\rho}_{i,i}(\beta,\{x_{i,p}\})}{\partial x_{j,q}}
\frac{\partial {\widetilde\rho}_{i,i}(\beta,\{x_{i,p}\})}{\partial x_{j',q'}}
\>.
\end{eqnarray}
Evaluating at $\Delta b=0$ and $\{x\}=\frac{1}{D}$ gives
\begin{eqnarray}
\left. 
\frac{\partial^2 f_{\delta^2}}{\partial x_{j,q} \partial x_{j',q'} } 
\right|_{\Delta b=0,\> \{x\}=\frac{1}{D}}
& \> = \> & 
2 \sum_{i=1}^{D_S} 
\left. \left( {\widetilde\rho}_{i,i}(\beta,\{x_{i,p}\}) - p_i^{(S)}(\beta,\Delta b) \right) 
\frac{\partial^2 {\widetilde\rho}_{i,i}(\beta,\{x_{i,p}\})}{\partial x_{j,q}\partial x_{j',q'}}
\right|_{\Delta b=0,\> \{x\}=\frac{1}{D}}
\cr
& & \quad 
+ \> 2 \sum_{i=1}^{D_S} 
\left. \frac{\partial {\widetilde\rho}_{i,i}(\beta,\{x_{i,p}\})}{\partial x_{j,q}}
\frac{\partial {\widetilde\rho}_{i,i}(\beta,\{x_{i,p}\})}{\partial x_{j',q'}}
\right|_{\Delta b=0,\> \{x\}=\frac{1}{D}}
\cr
& = &  
2 \sum_{i=1}^{D_S} 
\left. \frac{\partial {\widetilde\rho}_{i,i}(\beta,\{x_{i,p}\})}{\partial x_{j,q}}
\frac{\partial {\widetilde\rho}_{i,i}(\beta,\{x_{i,p}\})}{\partial x_{j',q'}}
\right|_{\Delta b=0,\> \{x\}=\frac{1}{D}}
\cr
& = &  
2 \sum_{i=1}^{D_S} 
\left(
D p_{i}^{(S)}(\beta,0) \> p_{q}^{(E)}(\beta)
\left( 
\delta_{i,j} \>  - \> p_{j}^{(S)}(\beta,0) 
\right)
\right)
\left(
D p_{i}^{(S)}(\beta,0) \> p_{q'}^{(E)}(\beta)
\left( 
\delta_{i,j'} \>  - \> p_{j'}^{(S)}(\beta,0) 
\right)
\right)
\cr
& = &  
2 \> D^2 \> \sum_{i=1}^{D_S} 
\left( p_{i}^{(S)}(\beta,0) \right)^2
p_{q}^{(E)}(\beta)
p_{q'}^{(E)}(\beta)
\left( 
\delta_{i,j} \>  - \> p_{j}^{(S)}(\beta,0) 
\right)
\left(
\delta_{i,j'} \>  - \> p_{j'}^{(S)}(\beta,0) 
\right)
\>.
\end{eqnarray}
We have to sum over all the different-$x_{i,p}$-second-partial terms to get the term that 
multiplies 
\begin{equation}
{\cal E}\left( \left(x_{i,p}-\frac{1}{D}\right)\left(x_{i',p'}-\frac{1}{D}\right)  \right)
\> = \>
{\cal E}\left( \left(x-\frac{1}{D}\right)  \left(x'-\frac{1}{D}\right)  \right)
\> = \>
-\> \frac{1}{D^2 (D+1)}
\end{equation}
since these expectation values are the same for all pairs $x_{i,p}$ and $x_{i',p'}$.  
One has
\begin{eqnarray}
&&\sum_{j,j'=1}^{D_S} \sum_{q,q'=1}^{D_E} 
\left(1-\delta_{j,j'} \delta_{q,q'}\right)
\left. 
\frac{\partial^2 f_{\delta^2}}{\partial x_{j,q} \partial x_{j',q'} } 
\right|_{\Delta b=0,\>\{x\}=\frac{1}{D}}
\cr
&&
\begin{array}{lll}
 =  & 
2 \> D^2 \> \sum_{j,j'=1}^{D_S} \sum_{q,q'=1}^{D_E} 
\left(1-\delta_{j,j'} \delta_{q,q'}\right)
\>
\sum_{i=1}^{D_S} 
\left[ \left( p_{i}^{(S)}(\beta,0) \right)^2
p_{q}^{(E)}(\beta)
p_{q'}^{(E)}(\beta)
\> \times 
\right.
&\\
 & 
\quad 
\left.
\left( 
\delta_{i,j} \>  - \> p_{j}^{(S)}(\beta,0) 
\right)
\left(
\delta_{i,j'} \>  - \> p_{j'}^{(S)}(\beta,0) 
\right)
\right]
&\\
 =  & 
2 \> D^2 \> \sum_{j,j'=1}^{D_S} \sum_{q,q'=1}^{D_E} 
\>
\sum_{i=1}^{D_S} 
\left[ \left( p_{i}^{(S)}(\beta,0) \right)^2
p_{q}^{(E)}(\beta)
p_{q'}^{(E)}(\beta)
\>
\left( 
\delta_{i,j} \>  - \> p_{j}^{(S)}(\beta,0) 
\right)
\left(
\delta_{i,j'} \>  - \> p_{j'}^{(S)}(\beta,0) 
\right)
\right]
&\\
 & 
\quad - \> 
2 \> D^2 \> \sum_{j=1}^{D_S} \sum_{q=1}^{D_E} 
\>
\sum_{i=1}^{D_S} 
\left[ \left( p_{i}^{(S)}(\beta,0) \right)^2
\left( p_{q}^{(E)}(\beta) \right)^2
\>
\left( 
\delta_{i,j} \>  - \> p_{j}^{(S)}(\beta,0) 
\right)^2
\right]
&\\
\end{array}
\cr
&&
\begin{array}{lll}
 = & 
2 \> D^2 \> \sum_{j,j'=1}^{D_S} \sum_{q,q'=1}^{D_E} 
\sum_{i=1}^{D_S} 
\left( p_{i}^{(S)}(\beta,0) \right)^2
p_{q}^{(E)}(\beta)
p_{q'}^{(E)}(\beta)
\>
\delta_{i,j} 
\delta_{i,j'} 
&\\
 & \quad
- 2 \> D^2 \> \sum_{j,j'=1}^{D_S} \sum_{q,q'=1}^{D_E} 
\sum_{i=1}^{D_S} 
\left( p_{i}^{(S)}(\beta,0) \right)^2
p_{q}^{(E)}(\beta)
p_{q'}^{(E)}(\beta)
\>
\delta_{i,j}  p_{j'}^{(S)}(\beta,0) 
&\\
& \quad
- 2 \> D^2 \> \sum_{j,j'=1}^{D_S} \sum_{q,q'=1}^{D_E} 
\sum_{i=1}^{D_S} 
\left( p_{i}^{(S)}(\beta,0) \right)^2
p_{q}^{(E)}(\beta)
p_{q'}^{(E)}(\beta)
\>
\delta_{i,j'} p_{j}^{(S)}(\beta,0) 
&\\
 & \quad 
+ 2 \> D^2 \> \sum_{j,j'=1}^{D_S} \sum_{q,q'=1}^{D_E} 
\sum_{i=1}^{D_S} 
\left( p_{i}^{(S)}(\beta,0) \right)^2
p_{q}^{(E)}(\beta)
p_{q'}^{(E)}(\beta)
\>
\> p_{j}^{(S)}(\beta,0) 
\> p_{j'}^{(S)}(\beta,0) 
&\\
 & 
\quad - \> 
2 \> D^2 \> \sum_{j=1}^{D_S} \sum_{q=1}^{D_E} 
\sum_{i=1}^{D_S} 
\left( p_{i}^{(S)}(\beta,0) \right)^2
\left( p_{q}^{(E)}(\beta) \right)^2
\delta_{i,j} 
&\\
 & \quad + \> 
4 \> D^2 \> \sum_{j=1}^{D_S} \sum_{q=1}^{D_E} 
\sum_{i=1}^{D_S} 
\left( p_{i}^{(S)}(\beta,0) \right)^2
\left( p_{q}^{(E)}(\beta) \right)^2
\delta_{i,j} p_{j}^{(S)}(\beta,0)
&\\
 & 
\quad - \> 
2 \> D^2 \> \sum_{j=1}^{D_S} \sum_{q=1}^{D_E} 
\sum_{i=1}^{D_S} 
\left( p_{i}^{(S)}(\beta,0) \right)^2
\left( p_{q}^{(E)}(\beta) \right)^2
\left( p_{j}^{(S)}(\beta,0) \right)^2
&\\
\end{array}
\cr
&&
\begin{array}{lll}
 =  & 
2 \> D^2 \> \sum_{j,j'=1}^{D_S}  
\sum_{i=1}^{D_S} 
\left( p_{i}^{(S)}(\beta,0) \right)^2
\>
\delta_{i,j} 
\delta_{i,j'} 
&\\
 & \quad
- 2 \> D^2 \> \sum_{j,j'=1}^{D_S}  
\sum_{i=1}^{D_S} 
\left( p_{i}^{(S)}(\beta,0) \right)^2
\>
\delta_{i,j}  p_{j'}^{(S)}(\beta,0) 
&\\
 & \quad
- 2 \> D^2 \> \sum_{j,j'=1}^{D_S}  
\sum_{i=1}^{D_S} 
\left( p_{i}^{(S)}(\beta,0) \right)^2
\>
\delta_{i,j'} p_{j}^{(S)}(\beta,0) 
&\\
 & \quad 
+ 2 \> D^2 \> \sum_{j,j'=1}^{D_S}  
\sum_{i=1}^{D_S} 
\left( p_{i}^{(S)}(\beta,0) \right)^2
\>
\> p_{j}^{(S)}(\beta,0) 
\> p_{j'}^{(S)}(\beta,0) 
&\\
 & 
\quad - \> 
2 \> D^2 \> \sum_{j=1}^{D_S}  
\frac{Z_E(2\beta)}{Z_E^2(\beta)}
\sum_{i=1}^{D_S} 
\left( p_{i}^{(S)}(\beta,0) \right)^2
\delta_{i,j} 
&\\
 & 
\quad + \> 
4 \> D^2 \> \sum_{j=1}^{D_S}  
\frac{Z_E(2\beta)}{Z_E^2(\beta)}
\sum_{i=1}^{D_S} 
\left( p_{i}^{(S)}(\beta,0) \right)^2
\delta_{i,j} p_{j}^{(S)}(\beta,0)
&\\
 & 
\quad - \> 
2 \> D^2 \> \sum_{j=1}^{D_S} 
\frac{Z_E(2\beta)}{Z_E^2(\beta)}
\sum_{i=1}^{D_S} 
\left( p_{i}^{(S)}(\beta,0) \right)^2
\left( p_{j}^{(S)}(\beta,0) \right)^2
&\\
\end{array}
\cr
&&
\begin{array}{lll}
 =  & 
2 \> D^2 \>   
\sum_{i=1}^{D_S} 
\left( p_{i}^{(S)}(\beta,0) \right)^2
- 2 \> D^2 \>   
\sum_{i=1}^{D_S} 
\left( p_{i}^{(S)}(\beta,0) \right)^2
- 2 \> D^2 \> 
\sum_{i=1}^{D_S} 
\left( p_{i}^{(S)}(\beta,0) \right)^2
&\\
 & \quad
+ 2 \> D^2 \> 
\sum_{i=1}^{D_S} 
\left( p_{i}^{(S)}(\beta,0) \right)^2
 - \> 
2 \> D^2 \> 
\frac{Z_E(2\beta)}{Z_E^2(\beta)}
\sum_{i=1}^{D_S} 
\left( p_{i}^{(S)}(\beta,0) \right)^2
 + \> 
4 \> D^2 \>  
\frac{Z_E(2\beta)}{Z_E^2(\beta)}
\sum_{i=1}^{D_S} 
\left( p_{i}^{(S)}(\beta,0) \right)^3
&\\ 
&\quad - \> 
2 \> D^2 \> \sum_{j=1}^{D_S} 
\frac{Z_E(2\beta)}{Z_E^2(\beta)}
\sum_{i=1}^{D_S} 
\left( p_{i}^{(S)}(\beta,0) \right)^2
\left( p_{j}^{(S)}(\beta,0) \right)^2
&\\
\end{array}
\cr
&&
\begin{array}{lll}
 = & 
 - \> 
2 \> D^2 \> 
\frac{Z_E(2\beta)}{Z_E^2(\beta)}\>
\left[
\frac{Z_S(2\beta)}{Z_S^2(\beta)} 
-2 \frac{Z_S(3\beta)}{Z_S^3(\beta)} 
+\frac{Z_S^2(2\beta)}{Z_S^4(\beta)} 
\right]
&\\
\end{array}
\end{eqnarray}
since 
$\sum_{q=1}^{D_E} p_q^{(E)} = 1$  
and 
$\sum_{j=1}^{D_S} p_j^{(S)} = 1$.  

Therefore, for these second-order terms the final result is that 
\begin{eqnarray}
&&{\cal E}\left( \left(x-\frac{1}{D}\right)  \left(x'-\frac{1}{D}\right)  \right)
\sum_{j,j'=1}^{D_S} \sum_{q,q'=1}^{D_E} 
\left(1-\delta_{j,j'} \delta_{q,q'}\right)
\left. 
\frac{\partial^2 f_{\delta^2}}{\partial x_{j,q} \partial x_{j',q'} } 
\right|_{\Delta b=0,\>\{x\}=\frac{1}{D}}
\cr
 &  = & 
\left(\> -\> \frac{1}{D^2 (D+1)} \right)
\left[ 
-\> 2 \> D^2 \> 
\frac{Z_E(2\beta)}{Z_E^2(\beta)} 
\>
\left( 
\frac{Z_S(2\beta)}{Z_S^2(\beta)} 
\>  - \> 
2 
\frac{Z_S(3\beta)}{Z_S^3(\beta)} 
\> + \> 
\frac{Z_S^2(2\beta)}{Z_S^4(\beta)} 
\right)
\right]
\cr
 &  =  & 
\frac{2}{D+1} 
\> 
\frac{Z_E(2\beta)}{Z_E^2(\beta)} 
\>
\left( 
\frac{Z_S(2\beta)}{Z_S^2(\beta)} 
\>  - \> 
2 
\frac{Z_S(3\beta)}{Z_S^3(\beta)} 
\> + \> 
\frac{Z_S^2(2\beta)}{Z_S^4(\beta)} 
\right)
\end{eqnarray}

Thus the complete answer for ${\cal E}(\delta^2)$, to second order 
in $\Delta b$ and all the $\{x\}$, is
\begin{eqnarray}
{\cal E}\left(\delta^2\right) 
&  =  & 
\frac{1}{2!}\> \left[
2 \> \frac{D-1}{D+1} \> \left(\frac{Z_E(2\beta)}{Z_E^2(\beta)}\right) \>
\left(
\frac{Z_S(2\beta)}{Z_S^2(\beta)} 
-2 \frac{Z_S(3\beta)}{Z_S^3(\beta)} 
+\frac{Z_S^2(2\beta)}{Z_S^4(\beta)}
\right)
\right]
\cr
& & \quad
\> + \> 
\frac{1}{2!}\> \left[
\frac{2}{D+1} 
\> 
\frac{Z_E(2\beta)}{Z_E^2(\beta)} 
\>
\left( 
\frac{Z_S(2\beta)}{Z_S^2(\beta)} 
\>  - \> 
2 
\frac{Z_S(3\beta)}{Z_S^3(\beta)} 
\> + \> 
\frac{Z_S^2(2\beta)}{Z_S^4(\beta)} 
\right)
\right]
\cr
&  =  & 
\left(\frac{Z_E(2\beta)}{Z_E^2(\beta)}\right) \>
\left(
\frac{Z_S(2\beta)}{Z_S^2(\beta)} 
-2 \frac{Z_S(3\beta)}{Z_S^3(\beta)} 
+\frac{Z_S^2(2\beta)}{Z_S^4(\beta)}
\right)
\left[
\frac{D-1}{D+1} + \frac{1}{D+1}
\right]
\cr
&  =  & 
\frac{D}{D+1}
\> 
\left(\frac{Z_E(2\beta)}{Z_E^2(\beta)}\right) \>
\left(
\frac{Z_S(2\beta)}{Z_S^2(\beta)} 
-2 \frac{Z_S(3\beta)}{Z_S^3(\beta)} 
+\frac{Z_S^2(2\beta)}{Z_S^4(\beta)}
\right)
\>.
\end{eqnarray}

In the infinite temperature limit ($\beta$=0),
one has that 
$Z_E(\beta\rightarrow 0) = D_E$ and 
$Z_S(\beta\rightarrow 0) = D_S$.  
Our expression then gives that 
\begin{equation}
\begin{array}{lcl}
\lim_{\beta\rightarrow 0} {\cal E}\left(\delta^2\right) 
& \> = \> & 
\frac{D}{D+1} \> \frac{D_E}{ D_E^2} \> 
\left(\frac{1}{D_S}-\frac{2}{D_S^2}+\frac{1}{D_S^2}\right)
\\
& \> = \> & 
\frac{D}{D+1} \> \frac{1}{ D_E} \> 
\left( \frac{D_S-1}{D_S^2}\right)
\\
& \> = \> & 
\frac{D}{D+1} \> \frac{1}{D_E} \> \frac{1}{D_S} \> 
\left( \frac{D_S-1}{D_S}\right)
\\
& \> = \> & 
\frac{1}{D+1} \>  
\frac{D_S-1}{D_S}
\,
\\
\end{array}
\end{equation}
which is the same expression as we published in our 2013 paper~\cite{JIN13a}, 
Eq.~(C3).  

One can also calculate how the low temperature (high $\beta$) limit 
of ${\cal E}(\delta^2)$ is approached.  
However, one has to be cautious about the low-temperature ($\beta\rightarrow +\infty$) 
limit, since the analysis requires that $\beta \langle {  H}_{SE}\rangle$ be small.  
Let $g_{S}$ and $g_{E}$ be the ground state degeneracies of the Hamiltonians 
$H_S$ and $H_E$ associated with ground state energies $E_1^{(S)}$ and $E_2^{(E)}$, 
respectively.  Use that
\begin{equation}
\begin{array}{lcl}
\lim_{\beta\rightarrow\infty} \frac{Z_E(2\beta)}{Z_E^2(\beta)} 
& \> = \> &
\lim_{\beta\rightarrow\infty} 
\frac{g_{E} e^{-2 \beta E_1^{(E)}} + \sum_{p=1+g_{E}}^{D_E} e^{-2\beta E_p^{(E)}} }
{\left(g_{E} e^{-\beta E_1^{(E)}} + \sum_{p'=1+g_{E}}^{D_E} e^{-\beta E_{p'}^{(E)}}\right)^2}
\\
& \> = \> &
\lim_{\beta\rightarrow\infty} 
\frac{g_{E} + \sum_{p=1+g_{E}}^{D_E} e^{-2\beta \left(E_p^{(E)}-E_1^{(E)} \right)}}
{\left(g_{E} + \sum_{p'=1+g_{E}}^{D_E} e^{-\beta \left(E_{p'}^{(E)}-E_1^{(E)}\right)}\right)^2}
\\
& \> = \> &
\frac{g_{E}}{g_{E}^2} 
\\
& \> = \> & 
\frac{1}{g_{E}} 
\>.
\end{array}
\end{equation}
\end{widetext}
Similarly one has the limits
\begin{equation}
\begin{array}{lcl}
\lim_{\beta\rightarrow\infty} \frac{Z_S(2\beta)}{Z_S^2(\beta)} & \> = \> &  \frac{1}{g_{S}} \\
\lim_{\beta\rightarrow\infty} \frac{Z_S(3\beta)}{Z_S^3(\beta)} & \> = \> &  \frac{1}{g_{S}^2} \\
\lim_{\beta\rightarrow\infty} \frac{Z_S^2(2\beta)}{Z_S^4(\beta)} & \> = \> &  \frac{1}{g_{S}^2} 
\>. 
\end{array}
\end{equation}
Hence one has the low-temperature limit
\begin{equation}
\begin{array}{lcl}
\lim_{\beta\rightarrow\infty} {\cal E}\left(\delta^2\right) 
& \> = \> & 
\frac{1}{g_{E}} \> \frac{D}{D+1} \> 
\left(\frac{1}{g_{S}}-\frac{2}{g_{S}^2}+\frac{1}{g_{S}^2}\right)
\\
& \> = \> & 
\frac{1}{g_{S} \> g_{E}} \> \frac{D}{D+1} \> \left(1-\frac{1}{g_{S}}\right) \\
& \> = \> & 
\frac{g_{S}-1}{g_{S}^2 \> g_{E}} \> \frac{D}{D+1} \\
& \> = \> & 
\frac{g_{S}-1}{g_{S}^2 \> g_{E}} \> \frac{1}{1+\frac{1}{D}} 
\>.
\end{array}
\end{equation}
In the limit of large $D$ this becomes 
\begin{equation}
\label{EqS:EdeltaLowT}
\lim_{\beta\rightarrow\infty} {\cal E}\left(\delta^2\right) 
\> \approx \> 
\frac{g_{S}-1}{g_{S}^2 \> g_{E}} 
\>. 
\end{equation}
Therefore in the low temperature limit the expectation value goes to zero for 
$g_{S}=1$ and goes to a finite value for a degenerate ground state ($g_{S}>1$).  
In principle, one could use any system with $g_{S}>1$ and 
for a large bath $D\rightarrow +\infty$ at very low temperature measure 
${\cal E}(\delta^2)$ in the system and from that deduce the 
degeneracy $g_{E}$ of the ground state of the bath.  

We also have  
${\cal O}\left( (\Delta b)^2 \left\{x_{j,q}\right\} \right) = 0 $. 
Putting everything together with the $(\Delta b)^2$ term gives our final perturbation expression, 
\begin{widetext}
\begin{equation}
\begin{array}{lcl}
{\cal E}\left(\delta^2\right) 
& \> = \> & 
\frac{D}{D+1}
\> 
\left(\frac{Z_E(2\beta)}{Z_E^2(\beta)}\right) \>
\left(
\frac{Z_S(2\beta)}{Z_S^2(\beta)} 
-2 \frac{Z_S(3\beta)}{Z_S^3(\beta)} 
+\frac{Z_S^2(2\beta)}{Z_S^4(\beta)}
\right)
\\
& & \quad
+\> \frac{ Z_S(2\beta) }{ Z_S^2(\beta) } 
\>
\left[
\frac{1}{4 k_{\rm B} \beta^2} C_v^{(S)}(2\beta) 
+ \> \left(
\left\langle E(2\beta) \right\rangle_S 
\>-\>\left\langle E(\beta)\right\rangle_S 
\right)^2
\right]
\left(\Delta b\right)^2 
\\
& & \quad\quad
+ {\cal O}\left( (\Delta b)^3 \right) 
+ {\cal O}\left( (\Delta b) \left\{x_{j,q}\right\} \left\{x_{j',q'}\right\} \right) 
+ {\cal O}\left( \left\{x_{j,q}\right\} \left\{x_{j',q'}\right\} 
\left\{x_{j'',q''}\right\} \right) .
\\
\end{array}
\label{EqS_delta}
\end{equation}
\end{widetext}
Equation~(\ref{EqS_delta}) is written as Eq.~(\ref{delta}) in the main text, but is 
written in terms of free energies rather than partition functions.  

\subsection{Derivation of ${\cal E}(2\sigma^2)$ for the uncoupled entirety}

In this subsection we derive the result for ${\cal E}\left(2\sigma^2\right)$, 
starting from the general expression of Eq.~(\ref{EqS:General1}) and the definition
\begin{equation}
\sigma \> = \> 
\sqrt{\sum_{i=1}^{D_S-1} \sum_{j=i+1}^{D_S} \left|{\widetilde\rho}_{i,j}\right|^2}
\end{equation}
which can be rewritten as 
\begin{equation}
\sigma^2 \> = \> 
\frac{1}{2} \> \sum_{i=1}^{D_S} \sum_{j=1}^{D_S} \left(1-\delta_{i,j}\right) 
\> \left|{\widetilde\rho}_{i,j}\right|^2
\>. 
\end{equation}
To second order one has the expression for $2\sigma^2$, 
\begin{widetext}
\begin{equation}
\begin{array}{lcl}
{\cal E}\left(f_{2\sigma^2}\right) 
& \> = \> &
{\cal E}\left(
\left.f_{2\sigma^2}\right|_{\{x\}=\frac{1}{D}} 
\right) 
\\
& \>  \> & \quad
+ \> 
\frac{1}{2!} \>
{\cal E}\left( \left(x-\frac{1}{D}\right)^2\right) 
\>
\sum_{k=1}^{D_S} \sum_{q=1}^{D_E} 
\left.
\frac{\partial^2 f_{2\sigma^2} }
{\partial x_{k,q}^2 }
\right|_{\{x\}=\frac{1}{D}} 
\\
& \>  \> & \quad
+ \> 
\frac{1}{2!} \>
{\cal E}\left(\left(x-\frac{1}{D}\right) \left(x'-\frac{1}{D}\right) \right) 
\>
\sum_{k=1}^{D_S} \sum_{k'=1}^{D_S} \sum_{q=1}^{D_E} \sum_{q'=1}^{D_E} 
\left(1-\delta_{k,k'}\delta_{q,q'}\right)
\left.
\frac{\partial^2 f_{2\sigma^2} }
{\partial x_{k,q} x_{k',q'} }
\right|_{\{x\}=\frac{1}{D}} 
\\
\end{array}
\end{equation}
so there are three terms to calculate.  
The expectation value involves a sum over all $\phi_{j,p}$ and hence ample use will be made of the 
properties of Eq.~(\ref{EqS:phiResults2}).  

We want to calculate without any approximations
\begin{equation}
{\cal E}\left(2\sigma^2\right) 
\> = \> 
{\cal E}\left(
\sum_{j=1}^{D_S} 
\sum_{j'=1}^{D_S} 
\left(1-\delta_{j,j'}\right)
{\widetilde\rho}_{j,j'}^* 
{\widetilde\rho}_{j,j'}
\right )
\>.
\end{equation}
Let
\begin{equation}
d_{j,p} \> = > \sqrt{x_{j,p}} \> e^{i \phi_{j,p} }
\quad {\rm and\/} \quad
d_{j,p}^* \> = > \sqrt{x_{j,p}} \> e^{-i \phi_{j,p} }
\>.
\end{equation}

For the case with $\lambda=0$, one has the reduced density matrix is
\begin{equation}
\begin{array}{clc}
{\widetilde\rho}_{j,j'}\left(\beta,\{x\},\{\phi\}\right)
& \> = \>  &
\sum_{p=1}^{D_E}
\left\langle E_{j}^{(S)}\right| 
\left\langle E_{p}^{(E)}\right. 
\left|\Psi_{\beta}\right\rangle
\left\langle\Psi_{\beta}\right|
\left. E_{p}^{(E)}\right\rangle 
\left| E_{j'}^{(S)}\right\rangle 
\\
& = & 
\sum_{p=1}^{D_E}
\frac{
\sqrt{x_{j,p}} \sqrt{x_{j',p}} 
e^{i \phi_{j,p} } e^{-i \phi_{j',p} }
\sqrt{p_{j}^{(S)}} \sqrt{p_{j'}^{(S)}} p_{p}^{(E)}
}
{
\sum_{j''=1}^{D_S} \sum_{p''=1}^{D_E} x_{j'',p''} 
p_{j''}^{(S)} p_{p''}^{(E)} }
\>.
\end{array}
\end{equation}
The complex conjugate (not the adjoint) is
\begin{equation}
\begin{array}{clc}
{\widetilde\rho}_{j,j'}^*\left(\beta,\{x\},\{\phi\}\right)
& \> = \>  &
\sum_{p'=1}^{D_E}
\frac{
\sqrt{x_{j,p'}} \sqrt{x_{j',p'}} 
e^{-i \phi_{j,p'} } e^{i \phi_{j',p'} }
\sqrt{p_{j}^{(S)}} \sqrt{p_{j'}^{(S)}} p_{p'}^{(E)}
}
{
\sum_{j'''=1}^{D_S} \sum_{p'''=1}^{D_E} x_{j''',p'''} 
p_{j'''}^{(S)} p_{p'''}^{(E)} }
\>.
\end{array}
\end{equation}
Extreme care must be taken that both for $x_{j,p}$, $x_{j',p}$ and $x_{j'',p''}$ 
as well as for $\phi_{j,p}$ and $\phi_{j',p}$  
wherever the 
indices are the same the value of the variable is the same. For example 
the value of $x_{3,13}$ is the same in both the numerator and denominator.

\subsubsection{\textbf{Zero-th order term of ${\cal E}(2\sigma^2)$}}

We expand about all $x_{j,p}=\frac{1}{D}$, but will perform the 
exact average over all $\phi_{j,p}$.  

The reduced density matrix evaluated at the expansion point $\{x\}=\frac{1}{D}$ is
\begin{equation}
\begin{array}{clc}
\left.
{\widetilde\rho}_{j,j'}\left(\beta,\{x\},\{\phi\}\right)
\right|_{\{x\}=\frac{1}{D}}
& \> = \> & 
\sqrt{p_{j}^{(S)}} \sqrt{p_{j'}^{(S)}} 
\> 
\sum_{p=1}^{D_E}
e^{i \phi_{j,p} } e^{-i \phi_{j',p} } \>
p_{p}^{(E)}
\>.
\end{array}
\end{equation}
Similarly, the zero-th order term also uses the complex conjugate, which is 
\begin{equation}
\begin{array}{clc}
\left.
{\widetilde\rho}_{j,j'}\left(\beta,\{x\},\{-\phi\}\right)
\right|_{\{x\}=\frac{1}{D}}
& \> = \> & 
\sqrt{p_{j}^{(S)}} \sqrt{p_{j'}^{(S)}} 
\> 
\sum_{p=1}^{D_E}
e^{-i \phi_{j,p} } e^{i \phi_{j',p} } \>
p_{p}^{(E)}
\>.
\end{array}
\end{equation}

The zero-th order equation is given by
\begin{equation}
\begin{array}{l}
\begin{array}{lcl}
&\left. f_{2\sigma^2}\left(\{x\},\>\{\phi\}\right) 
\right|_{\{x\}=\frac{1}{D}} 
\\
\quad &\> = \> &
\left.
\left[
\sum_{j=1}^{D_S} \sum_{j'=1}^{D_S}
\left(1-\delta_{j,j'}\right)
{\widetilde\rho}_{j,j'}^*\left(\beta,\{x\},\{\phi\}\right)
{\widetilde\rho}_{j,j'}\left(\beta,\{x\},\{\phi\}\right)
\right]
\right|_{\{x\}=\frac{1}{D}}
\\
\end{array}
\\
\begin{array}{lll}
\quad & \> = \> & 
\left[
\sum_{j=1}^{D_S} \sum_{j'=1}^{D_S}
\left(1-\delta_{j,j'}\right)
\frac{
\left(
\sum_{p=1}^{D_E}
\frac{1}{D}
e^{-i \phi_{j,p} } e^{i \phi_{j',p} } 
\sqrt{p_{j}^{(S)}} \sqrt{p_{j'}^{(S)}} 
p_{p}^{(E)}
\right)
\left(
\sum_{p'=1}^{D_E}
\frac{1}{D}
e^{-i \phi_{j,p'} } e^{i \phi_{j',p'} } 
\sqrt{p_{j}^{(S)}} \sqrt{p_{j'}^{(S)}} p_{p'}^{(E)}
\right)
}
{
\left(\sum_{j''=1}^{D_S} \sum_{p''=1}^{D_E} \frac{1}{D} 
p_{j''}^{(S)} p_{p''}^{(E)} \right)^2
}
\right]
\\
\quad & \> = \> & 
\sum_{p=1}^{D_E} \sum_{p'=1}^{D_E} 
\sum_{j=1}^{D_S} \sum_{j'=1}^{D_S}
\left(1-\delta_{j,j'}\right)
p_{j}^{(S)} p_{j'}^{(S)} 
p_{p}^{(E)} p_{p'}^{(E)} 
\> 
e^{i \phi_{j,p}} \>
e^{-i \phi_{j',p}} \>
e^{-i \phi_{j,p'}} \>
e^{i \phi_{j',p'}} 
\\
\quad & \> = \> & 
\sum_{p=1}^{D_E} \sum_{p'=1}^{D_E} 
\sum_{j=1}^{D_S} \sum_{j'=1}^{D_S}
\left(1-\delta_{j,j'}\right)
p_{j}^{(S)} p_{j'}^{(S)} 
p_{p}^{(E)} p_{p'}^{(E)} 
\> \delta_{p,p'} 
\> \delta_{p,p'} 
\\
\quad & \> = \> & 
\left(\sum_{j=1}^{D_S} \sum_{j'=1}^{D_S}
\left(1-\delta_{j,j'}\right)
p_{j}^{(S)} p_{j'}^{(S)} 
\right)
\> \left(
\sum_{p=1}^{D_E}  
\left(p_{p}^{(E)}\right)^2 
\right)
\\
\quad & \> = \> & 
\left(
1-\frac{Z_S(2\beta)}{Z_S^2(\beta)}
\right)
\> \left(
\frac{Z_E(2\beta)}{Z_E^2(\beta)}
\right)
\>
\\
\end{array}
\\
\end{array}
\end{equation}
since
$\sum_{p=1}^{D_E} p_p^{(E)} = 1$ and
$\sum_{j=1}^{D_S} p_j^{(S)} = 1$.  
Use has been made of Eq.~(\ref{EqS:phiResults}) with 
\begin{equation}
\frac{1}{2\pi} \int_{-\pi}^\pi e^{i\left(\phi_{j,p}-\phi_{j,p'}\right)} d\phi 
\> = \> 
\delta_{p,p'}
\label{ave_phi}
\end{equation}
since
\begin{equation}
\frac{1}{2\pi} \int_{-\pi}^\pi e^{i \phi} d\phi 
\> = \> 
\frac{1}{2\pi\>i} \left.e^{i\phi}\right|_{\phi=-\pi}^{\pi} 
\> = \> 
\frac{1}{2\pi\>i} \left(e^{i\pi} - e^{-i\pi}\right) 
\> = \> 
0.  
\end{equation}
In the limits one has
\begin{equation}
\begin{array}{lllllll}
\left. f_{2\sigma^2}\left(\{x\},\>\{\phi\}\right) 
\right|_{\{x\}=\frac{1}{D}}
& \rightarrow & 
\> \frac{1}{D_E} \> \frac{D_S-1}{D_S}
& \quad
& \beta & \rightarrow & 0 \\
\left. f_{2\sigma^2}\left(\{x\},\>\{\phi\}\right) 
\right|_{\{x\}=\frac{1}{D}}
& \rightarrow & 
\frac{g_{S}-1}{g_{S}g_{E}}
& \quad
& \beta & \rightarrow & +\infty 
\> 
\\
\end{array}
\end{equation}
where $g_{S}$ and $g_{E}$ are degeneracy of the ground state of $H_S$ and $H_E$, respectively.

\subsubsection{\textbf{First order term of ${\cal E}(2\sigma^2)$}}

The first partial derivative of ${\widetilde\rho}$ with respect to $x_{k,q}$ is
\begin{equation}
\begin{array}{clc}
\left(1-\delta_{j,j'}\right)
\frac{\partial\>{\widetilde\rho}_{j,j'}\left(\beta,\{x\},\{\phi\}\right)}
{\partial x_{k,q} }
& = & 
\left(1-\delta_{j,j'}\right)
\left[
\frac{
\frac{1}{2} \frac{1}{\sqrt{x_{j,q}}} \sqrt{x_{j',q}} 
e^{i \phi_{j,q} } e^{-i \phi_{j',q} }
\sqrt{p_{j}^{(S)}} \sqrt{p_{j'}^{(S)}} p_{q}^{(E)}
}
{
\sum_{j''=1}^{D_S} \sum_{p''=1}^{D_E} x_{j'',p''} 
p_{j''}^{(S)} p_{p''}^{(E)} }
\>\delta_{k,j}
\right.
\\
&  & \quad
+ \> 
\frac{
\frac{1}{2} \sqrt{x_{j,q}} \frac{1}{\sqrt{x_{j',q}}} 
e^{i \phi_{j,q} } e^{-i \phi_{j',q} }
\sqrt{p_{j}^{(S)}} \sqrt{p_{j'}^{(S)}} p_{q}^{(E)}
}
{
\sum_{j''=1}^{D_S} \sum_{p''=1}^{D_E} x_{j'',p''} 
p_{j''}^{(S)} p_{p''}^{(E)} }
\>\delta_{k,j'}
\\
&  & \quad
-\>
\left.
\frac{
p_{k}^{(S)} p_{q}^{(E)} \> 
\left(\sum_{p=1}^{D_E}
\sqrt{x_{j,p}} \sqrt{x_{j',p}} 
e^{i \phi_{j,p} } e^{-i \phi_{j',p} }
\sqrt{p_{j}^{(S)}} \sqrt{p_{j'}^{(S)}} p_{p}^{(E)}
\right)
}
{
\left(
\sum_{j''=1}^{D_S} \sum_{p''=1}^{D_E} x_{j'',p''} 
p_{j''}^{(S)} p_{p''}^{(E)} \right)^2 }
\right]
\\
\end{array}
\end{equation}
and evaluating about the expansion point $\{x\}=\frac{1}{D}$ gives
\begin{equation}
\begin{array}{clc}
\left.
\left(1-\delta_{j,j'}\right)
\frac{\partial\>{\widetilde\rho}_{j,j'}\left(\beta,\{x\},\{\phi\}\right)}
{\partial x_{k,q} }
\right|_{\{x\}=\frac{1}{D}}
& = & 
\left(1-\delta_{j,j'}\right) 
\>
\left[
\frac{D}{2} \>
\sqrt{p_{j}^{(S)}} \sqrt{p_{j'}^{(S)}} p_{q}^{(E)}
\> \delta_{k,j}
\> e^{i \phi_{j,q}} \> e^{-i \phi_{j',q}}  
\right.
\\
&  & \quad
+ \> 
\frac{D}{2} \>
\sqrt{p_{j}^{(S)}} \sqrt{p_{j'}^{(S)}} p_{q}^{(E)}
\> \delta_{k,j'}
\> e^{i \phi_{j,q}} \> e^{-i \phi_{j',q}}  
\\
&  & \quad
-\>
D \>
\left.
p_{k}^{(S)} 
\sqrt{p_{j}^{(S)}} \sqrt{p_{j'}^{(S)}} 
\> 
p_{q}^{(E)}  
\>
\left(
\sum_{p=1}^{D_E} 
\> e^{i \phi_{j,p}} \> e^{-i \phi_{j',p}}  
\>
p_p^{(E)}
\right)
\right]
\\

& = & 
\left(1-\delta_{j,j'}\right) 
\>
\left[
\frac{D}{2} \>
\sqrt{p_{j}^{(S)}} \sqrt{p_{j'}^{(S)}} p_{q}^{(E)}
\> e^{i \phi_{j,q}} \> e^{-i \phi_{j',q}}  
\>
\left(\delta_{k,j}+\delta_{k,j'}\right)
\right.
\\
&  & \quad
-\>
D \>
\left.
p_{k}^{(S)} 
\sqrt{p_{j}^{(S)}} \sqrt{p_{j'}^{(S)}} 
\> 
p_{q}^{(E)}  
\>
\left(
\sum_{p=1}^{D_E} 
\> e^{i \phi_{j,p}} \> e^{-i \phi_{j',p}}  
\>
p_p^{(E)}
\right)
\right].
\\
\end{array}
\end{equation}

\subsubsection{\textbf{Second order ({\bf same\/}) term of ${\cal E}(2\sigma^2)$}}

The second partial derivative with respect to the same $x_{k,q}$, 
evaluated about $\{x\}=\frac{1}{D}$ is 
\begin{equation}
\begin{array}{clc}
\left.
\left(1-\delta_{j,j'}\right)
\frac{\partial^2\>{\widetilde\rho}_{j,j'}\left(\beta,\{x\},\{\phi\}\right)}
{\partial x_{k,q}^2 }
\right|_{\{x\}=\frac{1}{D}}
& = & 
\left(1-\delta_{j,j'}\right)
\left[
-\> \frac{D^2}{4} \> 
e^{i \phi_{j,q} } e^{-i \phi_{j',q} }
\sqrt{p_{j}^{(S)}} \sqrt{p_{j'}^{(S)}} p_{q}^{(E)}
\>
\delta_{k,j} 
\right.
\\
&  & \quad
- \> \frac{D^2}{2} \> 
e^{i \phi_{j,q} } e^{-i \phi_{j',q} }
\left(p_{j}^{(S)}\right)^{\frac{3}{2}} \sqrt{p_{j'}^{(S)}} \left(p_{q}^{(E)}\right)^2
\>
\delta_{k,j} 
\\
&  & \quad
- \> \frac{D^2}{4} \> 
e^{i \phi_{j,q} } e^{-i \phi_{j',q} }
\sqrt{p_{j}^{(S)}} \sqrt{p_{j'}^{(S)}} p_{q}^{(E)}
\>
\delta_{k,j'} 
\\
&  & \quad
- \> \frac{D^2}{2} \> 
\sqrt{p_{j}^{(S)}} \left(p_{j'}^{(S)}\right)^{\frac{3}{2}} \left(p_{q}^{(E)}\right)^2
\>
e^{i \phi_{j,q} } e^{-i \phi_{j',q} } 
\>
\delta_{k,j'} 
\\
&  & \quad
- \> \frac{D^2}{2} \> 
p_{k}^{(S)} 
\> 
\sqrt{p_{j}^{(S)}} \sqrt{p_{j'}^{(S)}} 
\> 
\left( p_{q}^{(E)}\right)^2
\> 
e^{i \phi_{j,q} } e^{-i \phi_{j',q} }
\left(\delta_{k,j} + \delta_{k,j'} \right)
\\
&  & \quad
+\> 2 \> D^2
\left.
\left(p_{k}^{(S)}\right)^2 \>
\sqrt{p_{j}^{(S)}} \sqrt{p_{j'}^{(S)}} \> 
\left(p_{q}^{(E)}\right)^2 \> 
\left(\sum_{p=1}^{D_E}
e^{i \phi_{j,p} } e^{-i \phi_{j',p} } \>
p_{p}^{(E)}
\right)
\right]
\>.
\\
\end{array}
\end{equation}
One needs to sum over all possible derivatives.  
Putting together this for the same-$x_{k,q}$ second derivatives gives
\begin{equation}
\begin{array}{lcl}
\left.
\frac{1}{2!} \>
\sum_{k=1}^{D_S} 
\sum_{q=1}^{D_E} 
\frac{\partial^2 f_{2\sigma^2}}{\partial x_{k,q}^2} 
\right|_{\{x\}=\frac{1}{D}}
& \> = \> & 
\frac{1}{2!} \>
\sum_{k=1}^{D_S} 
\sum_{q=1}^{D_E} 
\sum_{j,j'}^{D_S} \left(1-\delta_{j,j'}\right) \times
\\
& & \quad
\left[
\left.
\frac{\partial^2 \widetilde\rho\left(\{x\},\> \phi_1, \phi_2, \cdots \phi_D\right)}
{\partial x_{k,q}^2}
\widetilde\rho\left(\{x\},\> -\phi_1, -\phi_2, \cdots -\phi_D\right) 
\right|_{\{x\}=\frac{1}{D}}
\right.
\\
& & \quad\quad
+ 2
\left.
\frac{\partial \widetilde\rho\left(\{x\},\> \phi_1, \phi_2, \cdots \phi_D\right)}
{\partial x_{k,q}}
\frac{\partial \widetilde\rho\left(\{x\},\> -\phi_1, -\phi_2, \cdots -\phi_D\right)} 
{\partial x_{k,q}}
\right|_{\{x\}=\frac{1}{D}}
\\
& & \quad\quad
\left.
+
\left.
\widetilde\rho\left(\{x\},\> \phi_1, \phi_2, \cdots \phi_D\right) 
\frac{\partial^2 \widetilde\rho\left(\{x\},\> -\phi_1, -\phi_2, \cdots -\phi_D\right) }
{\partial x_{k,q}^2}
\right|_{\{x\}=\frac{1}{D}}
\right]
\>.
\\
\end{array}
\end{equation}

The first term to calculate for the same-$x_{k,q}$ is
\begin{eqnarray}
 & & \frac{1}{2!} \>
\sum_{k=1}^{D_S} 
\sum_{q=1}^{D_E} 
\sum_{j,j'}^{D_S} \left(1-\delta_{j,j'}\right) \>
\left.
\frac{\partial^2 \widetilde\rho\left(\{x\},\> \phi_1, \phi_2, \cdots \phi_D\right)}
{\partial x_{k,q}^2}
\right|_{\{x\}=\frac{1}{D}}
\left.
\widetilde\rho\left(\{x\},\> -\phi_1, -\phi_2, \cdots -\phi_D\right) 
\right|_{\{x\}=\frac{1}{D}}
\cr
&& \begin{array}{lll}
=& 
\frac{1}{2!} \>
\sum_{k=1}^{D_S} 
\sum_{q=1}^{D_E} 
\sum_{j,j'}^{D_S} \left(1-\delta_{j,j'}\right) 
\> 
\left[
\sqrt{p_{j}^{(S)}} \sqrt{p_{j'}^{(S)}} 
\> 
\sum_{p=1}^{D_E}
e^{-i \phi_{j,p} } e^{i \phi_{j',p} } \>
p_{p}^{(E)}
\right] 
\> \times 
\\
& \quad
\left[
-\> \frac{D^2}{4} \> 
e^{i \phi_{j,q} } e^{-i \phi_{j',q} }
\sqrt{p_{j}^{(S)}} \sqrt{p_{j'}^{(S)}} p_{q}^{(E)}
\>
\delta_{k,j} 
- \> \frac{D^2}{2} \> 
e^{i \phi_{j,q} } e^{-i \phi_{j',q} }
\left(p_{j}^{(S)}\right)^{\frac{3}{2}} \sqrt{p_{j'}^{(S)}} \left(p_{q}^{(E)}\right)^2
\>
\delta_{k,j} 
\right . 
& \\
& \quad
- \> \frac{D^2}{4} \> 
e^{i \phi_{j,q} } e^{-i \phi_{j',q} }
\sqrt{p_{j}^{(S)}} \sqrt{p_{j'}^{(S)}} p_{q}^{(E)}
\>
\delta_{k,j'} 
- \> \frac{D^2}{2} \> 
\sqrt{p_{j}^{(S)}} \left(p_{j'}^{(S)}\right)^{\frac{3}{2}} \left(p_{q}^{(E)}\right)^2
\>
e^{i \phi_{j,q} } e^{-i \phi_{j',q} } 
\>
\delta_{k,j'} 
& \\
 & \quad
- \> \frac{D^2}{2} \> 
p_{k}^{(S)} 
\> 
\sqrt{p_{j}^{(S)}} \sqrt{p_{j'}^{(S)}} 
\> 
\left( p_{q}^{(E)}\right)^2
\> 
e^{i \phi_{j,q} } e^{-i \phi_{j',q} }
\left(\delta_{k,j} + \delta_{k,j'} \right)
\left.
+\> 2 \> D^2
\left(p_{k}^{(S)}\right)^2 \>
\sqrt{p_{j}^{(S)}} \sqrt{p_{j'}^{(S)}} \> 
\left(p_{q}^{(E)}\right)^2 \> 
\left(\sum_{p=1}^{D_E}
e^{i \phi_{j,p} } e^{-i \phi_{j',p} } \>
p_{p}^{(E)}
\right)
\right]
& \\
\end{array}
\cr
&&
\begin{array}{lll}
=& 
-\> \frac{D^2}{8} \> 
\sum_{k,j,j'}^{D_S} \sum_{q=1}^{D_E} \left(1-\delta_{j,j'}\right)
p_{j}^{(S)} p_{j'}^{(S)} \> 
\left(p_{q}^{(E)}\right)^2
\>
\delta_{k,j} 
- \> \frac{D^2}{4} \> 
\sum_{k,j,j'}^{D_S} \sum_{q=1}^{D_E} \left(1-\delta_{j,j'}\right)
\left(p_{j}^{(S)}\right)^{2} p_{j'}^{(S)} \left(p_{q}^{(E)}\right)^3
\>
\delta_{k,j} 
& \\
& 
- \> \frac{D^2}{8} \> 
\sum_{k,j,j'}^{D_S} \sum_{q=1}^{D_E} \left(1-\delta_{j,j'}\right)
p_{j}^{(S)} p_{j'}^{(S)} \left(p_{q}^{(E)}\right)^2
\>
\delta_{k,j'} 
- \> \frac{D^2}{4} \> 
\sum_{k,j,j'}^{D_S} \sum_{q=1}^{D_E} \left(1-\delta_{j,j'}\right)
p_{j}^{(S)} \left(p_{j'}^{(S)}\right)^{2} \left(p_{q}^{(E)}\right)^3
\>
\delta_{k,j'} 
& \\
& 
- \> \frac{D^2}{4} \> 
\sum_{k,j,j'}^{D_S} \sum_{q=1}^{D_E} \left(1-\delta_{j,j'}\right)
p_{k}^{(S)} 
\> 
p_{j}^{(S)} p_{j'}^{(S)} 
\> 
\left( p_{q}^{(E)}\right)^3
\> 
\left(\delta_{k,j} + \delta_{k,j'} \right)
&\\
&+\> D^2
\sum_{k,j,j'}^{D_S} \sum_{q=1}^{D_E} \left(1-\delta_{j,j'}\right)
\left(p_{k}^{(S)}\right)^2 \>
p_{j}^{(S)}  p_{j'}^{(S)} \> 
\left(p_{q}^{(E)}\right)^2 \> 
\left(\sum_{p'=1}^{D_E}
\left(p_{p'}^{(E)}\right)^2
\right)
& \\
\end{array} 
\cr
&&
\begin{array}{lll}
=& 
-\> \frac{D^2}{8} \> 
\frac{Z_E(2\beta)}{Z_E^2(\beta)} \>
\left( 1 - \frac{Z_S(2\beta)}{Z_S^2(\beta)} \right) 
- \> \frac{D^2}{4} \> 
\frac{Z_E(3\beta)}{Z_E^3(\beta)} \>
\left( \frac{Z_S(2\beta)}{Z_S^2(\beta)} - \frac{Z_S(3\beta)}{Z_S^3(\beta)} \right) 
- \> \frac{D^2}{8} \> 
\frac{Z_E(2\beta)}{Z_E^2(\beta)} \>
\left( 1 - \frac{Z_S(2\beta)}{Z_S^2(\beta)} \right) 
&\\
&  
- \> \frac{D^2}{4} \> 
\frac{Z_E(3\beta)}{Z_E^3(\beta)} \>
\left( \frac{Z_S(2\beta)}{Z_S^2(\beta)} - \frac{Z_S(3\beta)}{Z_S^3(\beta)} \right) 
- \> \frac{D^2}{2} \> 
\frac{Z_E(3\beta)}{Z_E^3(\beta)} \>
\left( \frac{Z_S(2\beta)}{Z_S^2(\beta)} - \frac{Z_S(3\beta)}{Z_S^3(\beta)} \right) 
+\> D^2 \>
\frac{Z_E^2(2\beta)}{Z_E^4(\beta)} \>
\left( \frac{Z_S(2\beta)}{Z_S^2(\beta)} - \frac{Z_S^2(2\beta)}{Z_S^4(\beta)} \right) 
&\\ 
\end{array}
\cr
&&\begin{array}{lll}
= & 
-\> \frac{D^2}{4} \> 
\frac{Z_E(2\beta)}{Z_E^2(\beta)} \>
\left( 1 - \frac{Z_S(2\beta)}{Z_S^2(\beta)} \right) 
- \> D^2 \>
\frac{Z_E(3\beta)}{Z_E^3(\beta)} \>
\left( \frac{Z_S(2\beta)}{Z_S^2(\beta)} - \frac{Z_S(3\beta)}{Z_S^3(\beta)} \right) 
+\> D^2 \>
\frac{Z_E^2(2\beta)}{Z_E^4(\beta)} \>
\left( \frac{Z_S(2\beta)}{Z_S^2(\beta)} - \frac{Z_S^2(2\beta)}{Z_S^4(\beta)} \right) 
& \\
\end{array}
\end{eqnarray}
and the middle term to calculate is
\begin{eqnarray}
&&
\frac{1}{2!} \>
\sum_{k=1}^{D_S} 
\sum_{q=1}^{D_E} 
\sum_{j,j'}^{D_S} \left(1-\delta_{j,j'}\right) \times
\> 2 \> 
\left.
\frac{\partial \widetilde\rho\left(\{x\},\> \phi_1, \phi_2, \cdots \phi_D\right)}
{\partial x_{k,q}}
\right|_{\{x\}=\frac{1}{D}}
\left.
\frac{\partial \widetilde\rho\left(\{x\},\> -\phi_1, -\phi_2, \cdots -\phi_D\right)}
{\partial x_{k,q}}
\right|_{\{x\}=\frac{1}{D}}
\cr
&&\begin{array}{lll}
= & 
\sum_{k=1}^{D_S} 
\sum_{q=1}^{D_E} 
\sum_{j,j'} \left(1-\delta_{j,j'}\right) \times
&\\
&
\left[
\frac{D}{2} \>
\sqrt{p_{j}^{(S)}} \sqrt{p_{j'}^{(S)}} p_{q}^{(E)}
\> e^{i \phi_{j,q}} \> e^{-i \phi_{j',q}}  
\>
\left(\delta_{k,j}+\delta_{k,j'}\right)

-\>
D 
p_{k}^{(S)} 
\sqrt{p_{j}^{(S)}} \sqrt{p_{j'}^{(S)}} 
\> 
p_{q}^{(E)}  
\>
\left(
\sum_{p=1}^{D_E} 
\> e^{i \phi_{j,p}} \> e^{-i \phi_{j',p}}  
\>
p_p^{(E)}
\right)
\right]
\> \times
& \\
& 
\left[
\frac{D}{2} \>
\sqrt{p_{j}^{(S)}} \sqrt{p_{j'}^{(S)}} p_{q}^{(E)}
\> e^{-i \phi_{j,q}} \> e^{i \phi_{j',q}}  
\>
\left(\delta_{k,j}+\delta_{k,j'}\right)
-\>
D 
p_{k}^{(S)} 
\sqrt{p_{j}^{(S)}} \sqrt{p_{j'}^{(S)}} 
\> 
p_{q}^{(E)}  
\>
\left(
\sum_{p'=1}^{D_E} 
\> e^{-i \phi_{j,p'}} \> e^{i \phi_{j',p'}}  
\>
p_{p'}^{(E)}
\right)
\right]
& \\
= & 
\frac{D^2}{4} \>
\sum_{k,j,j'}^{D_S} \left(1-\delta_{j,j'}\right) \> 
\sum_{q=1}^{D_E}
\>
p_{j}^{(S)} p_{j'}^{(S)} 
\left(p_{q}^{(E)}\right)^2
\>
\left(\delta_{k,j}+\delta_{k,j'}\right)
& \\
&  
- \>
\> D^2 \>
\sum_{k,j,j'}^{D_S} \left(1-\delta_{j,j'}\right) \> 
\sum_{q=1}^{D_E}
\>
p_k^{(S)} \>
p_{j}^{(S)} p_{j'}^{(S)} 
\left(p_{q}^{(E)}\right)^3
\>
\left(\delta_{k,j}+\delta_{k,j'}\right)
&\\
&  
+ \> D^2 \>
\sum_{k,j,j'}^{D_S} \left(1-\delta_{j,j'}\right) \> 
\sum_{q=1}^{D_E}
\>
\left(p_k^{(S)}\right)^2 \>
p_{j}^{(S)} p_{j'}^{(S)} \> 
\left(p_{q}^{(E)}\right)^2
\>
\left(
\sum_{p=1}^{D_E} \left(p_p^{(E)}\right)^2
\right)
&\\
\end{array}
\cr
&&\begin{array}{lll}
 = & 
\frac{D^2}{2} \> 
\frac{Z_E(2\beta)}{Z_E^2(\beta)} \>
\left(1 - \frac{Z_S(2\beta)}{Z_S^2(\beta)} \right) 
- \> 2 \> D^2 \>
\frac{Z_E(3\beta)}{Z_E^3(\beta)} \>
\left( \frac{Z_S(2\beta)}{Z_S^2(\beta)}  - \frac{Z_S(3\beta)}{Z_S^3(\beta)} \right)
+ \> D^2 \>
\frac{Z_E^2(2\beta)}{Z_E^4(\beta)} \>
\left( \frac{Z_S(2\beta)}{Z_S^2(\beta)}  
- \frac{Z_S^2(2\beta)}{Z_S^4(\beta)} \right).
& \\
\end{array}
\end{eqnarray}
Putting this all together for the same-$x_{k,q}$ gives
\begin{eqnarray}
&&\left.
\frac{1}{2!} \>
\sum_{k=1}^{D_S} 
\sum_{q=1}^{D_E} 
\frac{\partial^2 f_{2\sigma^2}}{\partial x_{k,q}^2} 
\right|_{\{x\}=\frac{1}{D}}
\cr
&&\begin{array}{lll}
=& 
-\> \frac{D^2}{2} \> 
\frac{Z_E(2\beta)}{Z_E^2(\beta)} \>
\left( 1 - \frac{Z_S(2\beta)}{Z_S^2(\beta)} \right) 
- \> 2 \> D^2 \>
\frac{Z_E(3\beta)}{Z_E^3(\beta)} \>
\left( \frac{Z_S(2\beta)}{Z_S^2(\beta)} - \frac{Z_S(3\beta)}{Z_S^3(\beta)} \right) 
&\\
&  
+\> 2 \> D^2 \>
\frac{Z_E^2(2\beta)}{Z_E^4(\beta)} \>
\left( \frac{Z_S(2\beta)}{Z_S^2(\beta)} - \frac{Z_S^2(2\beta)}{Z_S^4(\beta)} \right) 
+ \frac{D^2}{2} \> 
\frac{Z_E(2\beta)}{Z_E^2(\beta)} \>
\left(1 - \frac{Z_S(2\beta)}{Z_S^2(\beta)} \right)
&\\
&   
- \> 2 \> D^2 \>
\frac{Z_E(3\beta)}{Z_E^3(\beta)} \>
\left( \frac{Z_S(2\beta)}{Z_S^2(\beta)}  - \frac{Z_S(3\beta)}{Z_S^3(\beta)} \right)
+ \> D^2 \>
\frac{Z_E^2(2\beta)}{Z_E^4(\beta)} \>
\left( \frac{Z_S(2\beta)}{Z_S^2(\beta)}  
- \frac{Z_S^2(2\beta)}{Z_S^4(\beta)} \right)
&\\
\end{array}
\cr
&&\begin{array}{lll}
= & 
- \> 4 \> D^2 \>
\frac{Z_E(3\beta)}{Z_E^3(\beta)} \>
\left( \frac{Z_S(2\beta)}{Z_S^2(\beta)}  - \frac{Z_S(3\beta)}{Z_S^3(\beta)} \right) 
+ \> 3 \> D^2 \>
\frac{Z_E^2(2\beta)}{Z_E^4(\beta)} \>
\left( \frac{Z_S(2\beta)}{Z_S^2(\beta)}  
- \frac{Z_S^2(2\beta)}{Z_S^4(\beta)} \right)
\>.
\end{array}
\end{eqnarray}

\subsubsection{\textbf{Second order ({\bf different\/}) term of ${\cal E}(2\sigma^2)$}}

The different-$x_{k,q}$ second partial derivatives, evaluated 
about $\{x\}=\frac{1}{D}$ is
\begin{eqnarray}
&&
\left.
\left(1-\delta_{j,j'}\right)
\left(1-\delta_{k,k'} \delta_{q,q'}\right)
\frac{\partial^2\>{\widetilde\rho}_{j,j'}\left(\beta,\{x\},\{\phi\}\right)}
{\partial x_{k,q} \> \partial x_{k',q'} }
\right|_{\{x\}=\frac{1}{D}}
\cr
&&\begin{array}{lll}
= & 
\left(1-\delta_{j,j'}\right)
\left(1-\delta_{k,k'} \delta_{q,q'}\right)
\> \times
&\\
& \quad
\left[
\frac{D^2}{4} \>
\sqrt{p_{j}^{(S)}} \sqrt{p_{j'}^{(S)}} 
\>
p_{q}^{(E)}
\>
e^{i \phi_{j,q} } e^{-i \phi_{j',q} }
\> \delta_{k,j} \delta_{k',j'} \delta_{q,q'}
\right.
&\\
& \quad 
- \>  \frac{D^2}{2} \> 
p_{k'}^{(S)} 
\> 
\sqrt{p_{j}^{(S)}} \sqrt{p_{j'}^{(S)}} 
\> 
p_{q}^{(E)} p_{q'}^{(E)} 
\>
e^{i \phi_{j,q} } e^{-i \phi_{j',q} }
\> \delta_{k,j}
&\\
& \quad
+ \> \frac{D^2}{4} \>  
\sqrt{p_{j}^{(S)}} \sqrt{p_{j'}^{(S)}} 
\>
p_{q}^{(E)}
\>
e^{i \phi_{j,q} } e^{-i \phi_{j',q} }
\> \delta_{k,j'} \delta_{k',j} \delta_{q,q'} 
&\\
& \quad
- \> \frac{D^2}{2} \>
p_{k'}^{(S)} 
\> 
\sqrt{p_{j}^{(S)}} \sqrt{p_{j'}^{(S)}}  
\>
p_{q}^{(E)} p_{q'}^{(E)} 
\>
e^{i \phi_{j,q} } e^{-i \phi_{j',q} }
\> \delta_{k,j'}
&\\
& \quad
-\> \frac{D^2}{2} \> 
p_{k}^{(S)} 
\>
\sqrt{p_{j}^{(S)}} \sqrt{p_{j'}^{(S)}} 
\>
p_{q}^{(E)} p_{q'}^{(E)}  
\> 
e^{i \phi_{j,q'} } e^{-i \phi_{j',q'} }
\> \delta_{k',j}
&\\
& \quad
-\> \frac{D^2}{2} \> 
p_{k}^{(S)} 
\> 
\sqrt{p_{j}^{(S)}} \sqrt{p_{j'}^{(S)}} 
\>
p_{q}^{(E)} p_{q'}^{(E)}  
\> 
e^{i \phi_{j,q'} } e^{-i \phi_{j',q'} }
\> \delta_{k',j'}
&\\
& \quad
+\>
2 \> D^2 \>
\left.
p_{k}^{(S)} p_{k'}^{(S)}  
\>
\sqrt{p_{j}^{(S)}} \sqrt{p_{j'}^{(S)}} 
\>
p_{q}^{(E)} p_{q'}^{(E)}  
\> 
\left(\sum_{p=1}^{D_E}
\>
e^{i \phi_{j,p} } e^{-i \phi_{j',p} }
\>
p_{p}^{(E)}
\right)
\right]
&\\
\end{array}
\cr
&&\begin{array}{lll}
= & 
\left(1-\delta_{j,j'}\right)
\left(1-\delta_{k,k'} \delta_{q,q'}\right)
\> \times
&\\
& \quad
\left[
\frac{D^2}{4} \>
\sqrt{p_{j}^{(S)}} \sqrt{p_{j'}^{(S)}} 
\>
p_{q}^{(E)}
\>
e^{i \phi_{j,q} } e^{-i \phi_{j',q} }
\> \left(\delta_{k,j} \delta_{k',j'}+\delta_{k,j'} \delta_{k',j}\right) \delta_{q,q'}
\right.
&\\
& \quad
- \> \frac{D^2}{2} \>
p_{k'}^{(S)} 
\> 
\sqrt{p_{j}^{(S)}} \sqrt{p_{j'}^{(S)}}  
\>
p_{q}^{(E)} p_{q'}^{(E)} 
\>
e^{i \phi_{j,q} } e^{-i \phi_{j',q} }
\> \left( \delta_{k,j}+\delta_{k,j'}\right)
&\\
& \quad
-\> \frac{D^2}{2} \> 
p_{k}^{(S)} 
\>
\sqrt{p_{j}^{(S)}} \sqrt{p_{j'}^{(S)}} 
\>
p_{q}^{(E)} p_{q'}^{(E)}  
\> 
e^{i \phi_{j,q} } e^{-i \phi_{j',q} }
\> \left(\delta_{k',j}+\delta_{k',j'}\right)
&\\
& \quad
+\>
2 \> D^2 \>
\left.
p_{k}^{(S)} p_{k'}^{(S)}  
\>
\sqrt{p_{j}^{(S)}} \sqrt{p_{j'}^{(S)}} 
\>
p_{q}^{(E)} p_{q'}^{(E)}  
\> 
\left(\sum_{p=1}^{D_E}
\>
e^{i \phi_{j,p} } e^{-i \phi_{j',p} }
\>
p_{p}^{(E)}
\right)
\right]
&\\
\end{array}
\end{eqnarray}
where the terms have been combined.  

One needs to sum over all possible derivatives.  
Putting together this for the different-$x_{k,q}$ second derivatives gives
\begin{eqnarray}
&&
\frac{1}{2!} \>
\sum_{k=1}^{D_S} \sum_{k'=1}^{D_S} 
\sum_{q=1}^{D_E} \sum_{q'=1}^{D_E} 
\left(1-\delta_{k,k'} \delta_{q,q'}\right)
\left.
\frac{\partial^2 f_{2\sigma^2}}{\partial x_{k,q} \> \partial x_{k',q'}} 
\right|_{\{x\}=\frac{1}{D}}
\cr
&&\begin{array}{lll}
= &  
\frac{1}{2!} \>
\sum_{k=1}^{D_S} \sum_{k'=1}^{D_S} 
\sum_{q=1}^{D_E} \sum_{q'=1}^{D_E} 
\left(1-\delta_{k,k'} \delta_{q,q'}\right)
\sum_{j,j'}^{D_S} \left(1-\delta_{j,j'}\right)
\left.
\frac{\partial^2 \>
\left[\widetilde\rho\left(\{x\},\> \phi_1, \phi_2, \cdots \phi_D\right) 
\widetilde\rho\left(\{x\},\> -\phi_1, -\phi_2, \cdots -\phi_D\right) 
\right] }
{\partial x_{k,q} \> \partial x_{k',q'}}
\right|_{\{x\}=\frac{1}{D}}
&\\
= & 
\frac{1}{2!} \>
\sum_{k=1}^{D_S} \sum_{k'=1}^{D_S} 
\sum_{q=1}^{D_E} \sum_{q'=1}^{D_E} 
\left(1-\delta_{k,k'} \delta_{q,q'}\right) \>
\sum_{j,j'}^{D_S} \left(1-\delta_{j,j'}\right)
\> \times
&\\
& \quad
\left[
\left.
\frac{\partial^2 \widetilde\rho\left(\{x\},\> \phi_1, \phi_2, \cdots \phi_D\right)}
{\partial x_{k,q} \> \partial x_{k',q'}}
\widetilde\rho\left(\{x\},\> -\phi_1, -\phi_2, \cdots -\phi_D\right) 
\right|_{\{x\}=\frac{1}{D}}
\right.
&\\
 & \quad\quad
+
\left.
\frac{\partial \widetilde\rho\left(\{x\},\> \phi_1, \phi_2, \cdots \phi_D\right)}
{\partial x_{k,q}}
\frac{\partial \widetilde\rho\left(\{x\},\> -\phi_1, -\phi_2, \cdots -\phi_D\right) }
{\partial x_{k',q'}}
\right|_{\{x\}=\frac{1}{D}}
&\\
 & \quad\quad
+
\left.
\frac{\partial \widetilde\rho\left(\{x\},\> \phi_1, \phi_2, \cdots \phi_D\right) }
{\partial x_{k',q'}}
\frac{\partial \widetilde\rho\left(\{x\},\> -\phi_1, -\phi_2, \cdots -\phi_D\right) }
{\partial x_{k,q}}
\right|_{\{x\}=\frac{1}{D}}
&\\
 & \quad\quad
\left.
+
\left.
\widetilde\rho\left(\{x\},\> \phi_1, \phi_2, \cdots \phi_D\right) 
\frac{\partial^2 \widetilde\rho\left(\{x\},\> -\phi_1, -\phi_2, \cdots -\phi_D\right) }
{\partial x_{k,q} \> \partial x_{k',q'}}
\right|_{\{x\}=\frac{1}{D}}
\right]
\>.
&\\
\end{array}
\end{eqnarray}

We need to sum over all possible derivatives.  
The first term to analyze for different-$x_{k,q}$ is
\begin{eqnarray}
&&
\frac{1}{2!} \>
\sum_{k=1}^{D_S} \sum_{k'=1}^{D_S} 
\sum_{q=1}^{D_E} \sum_{q'=1}^{D_E} 
\left(1-\delta_{k,k'} \delta_{q,q'}\right)
\sum_{j,j'} \left(1-\delta_{j,j'}\right)
\left.
\frac{\partial^2 \widetilde\rho\left(\{x\},\> \phi_1, \phi_2, \cdots \phi_D\right)}
{\partial x_{k,q} \> \partial x_{k',q'}}
\right|_{\{x\}=\frac{1}{D}}
\left.
\widetilde\rho\left(\{x\},\> -\phi_1, -\phi_2, \cdots -\phi_D\right) 
\right|_{\{x\}=\frac{1}{D}}
\cr
&&\begin{array}{lll}
= & 
\frac{1}{2!} \>
\sum_{k=1}^{D_S} \sum_{k'=1}^{D_S} 
\sum_{q=1}^{D_E} \sum_{q'=1}^{D_E} 
\left(1-\delta_{k,k'} \delta_{q,q'}\right)
\sum_{j,j'}^{D_S} \left(1-\delta_{j,j'}\right)
\> \times
&\\
 & \quad 
\left[
\sqrt{p_j^{(S)}} \sqrt{p_{j'}^{(S)}}  
\left(
\sum_{p'=1}^{D_E} e^{-i\phi_{j,p'}} e^{i\phi_{j',p'}} p_{p'}^{(E)} 
\right)
\right] \times
&\\
 & \quad 
\left[
\frac{D^2}{4} \>
\sqrt{p_{j}^{(S)}} \sqrt{p_{j'}^{(S)}} 
\>
p_{q}^{(E)}
\>
e^{i \phi_{j,q} } e^{-i \phi_{j',q} }
\> \left(\delta_{k,j} \delta_{k',j'}+\delta_{k,j'} \delta_{k',j}\right) \delta_{q,q'}
\right.
&\\
  & \quad
- \> \frac{D^2}{2} \>
p_{k'}^{(S)} 
\> 
\sqrt{p_{j}^{(S)}} \sqrt{p_{j'}^{(S)}}  
\>
p_{q}^{(E)} p_{q'}^{(E)} 
\>
e^{i \phi_{j,q} } e^{-i \phi_{j',q} }
\> \left( \delta_{k,j}+\delta_{k,j'}\right)
&\\
 & \quad
-\> \frac{D^2}{2} \> 
p_{k}^{(S)} 
\>
\sqrt{p_{j}^{(S)}} \sqrt{p_{j'}^{(S)}} 
\>
p_{q}^{(E)} p_{q'}^{(E)}  
\> 
e^{i \phi_{j,q} } e^{-i \phi_{j',q} }
\> \left(\delta_{k',j}+\delta_{k',j'}\right)
&\\
 & \quad
+\>
2 \> D^2 \>
\left.
p_{k}^{(S)} p_{k'}^{(S)}  
\>
\sqrt{p_{j}^{(S)}} \sqrt{p_{j'}^{(S)}} 
\>
p_{q}^{(E)} p_{q'}^{(E)}  
\> 
\left(\sum_{p=1}^{D_E}
\>
e^{i \phi_{j,p} } e^{-i \phi_{j',p} }
\>
p_{p}^{(E)}
\right)
\right]
&\\
\end{array}
\cr
&&
\begin{array}{lll}
= & 
\frac{1}{2!} \>
\sum_{k=1}^{D_S} \sum_{k'=1}^{D_S} 
\sum_{q=1}^{D_E} \sum_{q'=1}^{D_E} 
\left(1-\delta_{k,k'} \delta_{q,q'}\right)
\sum_{j,j'}^{D_S} \left(1-\delta_{j,j'}\right)
\> p_j^{(S)} p_{j'}^{(S)}  
\times
&\\
 & \quad 
\left[
\sum_{p'=1}^{D_E} e^{-i\phi_{j,p'}} e^{i\phi_{j',p'}} p_{p'}^{(E)} 
\right] \times
&\\
 & \quad 
\left[
\frac{D^2}{4} \>
p_{q}^{(E)}
\>
e^{i \phi_{j,q} } e^{-i \phi_{j',q} }
\> \left(\delta_{k,j} \delta_{k',j'}+\delta_{k,j'} \delta_{k',j}\right) \delta_{q,q'}
\right.
&\\
 & \quad
- \> \frac{D^2}{2} \>
p_{k'}^{(S)} 
\> 
p_{q}^{(E)} p_{q'}^{(E)} 
\>
e^{i \phi_{j,q} } e^{-i \phi_{j',q} }
\> \left( \delta_{k,j}+\delta_{k,j'}\right)
&\\
 & \quad
-\> \frac{D^2}{2} \> 
p_{k}^{(S)} 
\>
p_{q}^{(E)} p_{q'}^{(E)}  
\> 
e^{i \phi_{j,q} } e^{-i \phi_{j',q} }
\> \left(\delta_{k',j}+\delta_{k',j'}\right)
&\\
  & \quad
+\>
2 \> D^2 \>
\left.
p_{k}^{(S)} p_{k'}^{(S)}  
\>
p_{q}^{(E)} p_{q'}^{(E)}  
\> 
\left(\sum_{p=1}^{D_E}
\>
e^{i \phi_{j,p} } e^{-i \phi_{j',p} }
\>
p_{p}^{(E)}
\right)
\right]
&\\
\end{array}
\cr
&&
\begin{array}{lll}
= & 
\frac{1}{2!} \>
\sum_{k=1}^{D_S} \sum_{k'=1}^{D_S} 
\sum_{q=1}^{D_E} \sum_{q'=1}^{D_E} 
\left(1-\delta_{k,k'} \delta_{q,q'}\right)
\sum_{j,j'}^{D_S} \left(1-\delta_{j,j'}\right)
\> p_j^{(S)} p_{j'}^{(S)}  
\times
&\\
 & \quad 
\left[
\frac{D^2}{4} \>
\left(p_{q}^{(E)}\right)^2
\> \left(\delta_{k,j} \delta_{k',j'}+\delta_{k,j'} \delta_{k',j}\right) \delta_{q,q'}
\right.
&\\
  & \quad
- \> \frac{D^2}{2} \>
p_{k'}^{(S)} 
\> 
\left(p_{q}^{(E)}\right)^2 p_{q'}^{(E)} 
\> \left( \delta_{k,j}+\delta_{k,j'}\right)
&\\
  & \quad
-\> \frac{D^2}{2} \> 
p_{k}^{(S)} 
\>
\left(p_{q}^{(E)}\right)^2 p_{q'}^{(E)}  
\> \left(\delta_{k',j}+\delta_{k',j'}\right)
&\\
  & \quad
+\>
2 \> D^2 \>
\left.
p_{k}^{(S)} p_{k'}^{(S)}  
\>
p_{q}^{(E)} p_{q'}^{(E)}  
\> 
\left(\sum_{p=1}^{D_E}
\>
\left(p_{p}^{(E)}\right)^2
\right)
\right]
&\\
\end{array}
\cr
&&
\begin{array}{lll}
= & 
\frac{1}{2!} \>
\sum_{k=1}^{D_S} \sum_{k'=1}^{D_S} 
\sum_{j=1}^{D_S} \sum_{j'=1}^{D_S} \left(1-\delta_{j,j'}\right)
\> p_j^{(S)} p_{j'}^{(S)}  
\times
&\\
 & \quad 
\left[
\frac{D^2}{4} \>
\> \left(\delta_{k,j} \delta_{k',j'}+\delta_{k,j'} \delta_{k',j}\right) 
\> \left(1-\delta_{k,k'}\right)
\> \frac{Z_E(2\beta)}{Z_E^2(\beta)} 
\right.
&\\
  & \quad
- \> \frac{D^2}{2} \>
p_{k'}^{(S)} 
\> \left( \delta_{k,j}+\delta_{k,j'}\right)
\> \left(\frac{Z_E(2\beta)}{Z_E^2(\beta)}-
\delta_{k,k'}\frac{Z_E(3\beta)}{Z_E^3(\beta)}\right)
&\\
  & \quad
-\> \frac{D^2}{2} \> 
p_{k}^{(S)} 
\>
\> \left(\delta_{k',j}+\delta_{k',j'}\right)
\> \left(\frac{Z_E(2\beta)}{Z_E^2(\beta)}-
\delta_{k,k'}\frac{Z_E(3\beta)}{Z_E^3(\beta)}\right)
&\\
  & \quad
+\>
2 \> D^2 \>
\left.
p_{k}^{(S)} p_{k'}^{(S)}  
\>
\frac{Z_E(2\beta)}{Z_E^2(\beta)}
\> \left(1-\delta_{k,k'}\frac{Z_E(2\beta)}{Z_E^2(\beta)}\right)
\right]
&\\
\end{array}
\end{eqnarray}
which multiplying out gives
\begin{eqnarray}
&&
\frac{1}{2!} \>
\sum_{k=1}^{D_S} \sum_{k'=1}^{D_S} 
\sum_{q=1}^{D_E} \sum_{q'=1}^{D_E} 
\left(1-\delta_{k,k'} \delta_{q,q'}\right)
\sum_{j,j'} \left(1-\delta_{j,j'}\right)
\left.
\frac{\partial^2 \widetilde\rho\left(\{x\},\> \phi_1, \phi_2, \cdots \phi_D\right)}
{\partial x_{k,q} \> \partial x_{k',q'}}
\right|_{\{x\}=\frac{1}{D}}
\left.
\widetilde\rho\left(\{x\},\> -\phi_1, -\phi_2, \cdots -\phi_D\right) 
\right|_{\{x\}=\frac{1}{D}}
\cr
&&\begin{array}{lll}
 = & 
\frac{1}{2!} \>
\sum_{k=1}^{D_S} \sum_{k'=1}^{D_S} 
\sum_{j=1}^{D_S} \sum_{j'=1}^{D_S} 
\> p_j^{(S)} p_{j'}^{(S)}  
\times
&\\
 & \quad 
\left[
\frac{D^2}{4} \>
\>\left(
\delta_{k,j} \delta_{k',j'}+\delta_{k,j'} \delta_{k',j}
-\delta_{j,j'}\delta_{k,j} \delta_{k',j'}-\delta_{j,j'}\delta_{k,j'} \delta_{k',j}
\right.
\right.
&\\
 & \quad
\left.
-\delta_{k,k'}\delta_{k,j} \delta_{k',j'}-\delta_{k,k'}\delta_{k,j'} \delta_{k',j}
+\delta_{k,k'}\delta_{j,j'}\delta_{k,j} \delta_{k',j'}
+\delta_{k,k'}\delta_{j,j'}\delta_{k,j'} \delta_{k',j}
\right)
\> \frac{Z_E(2\beta)}{Z_E^2(\beta)} 
&\\
  & \quad
- \> \frac{D^2}{2} \>
p_{k'}^{(S)} 
\>
\left(
\delta_{k,j}\frac{Z_E(2\beta)}{Z_E^2(\beta)}
-\delta_{k,j}\delta_{k,k'}\frac{Z_E(3\beta)}{Z_E^3(\beta)}
+\delta_{k,j'}\frac{Z_E(2\beta)}{Z_E^2(\beta)}
-\delta_{k,j'}\delta_{k,k'}\frac{Z_E(3\beta)}{Z_E^3(\beta)}
\right.
&\\
  & \quad
\left.
-\delta_{j,j'}\delta_{k,j}\frac{Z_E(2\beta)}{Z_E^2(\beta)}
+\delta_{j,j'}\delta_{k,j}\delta_{k,k'}\frac{Z_E(3\beta)}{Z_E^3(\beta)}
-\delta_{j,j'}\delta_{k,j'}\frac{Z_E(2\beta)}{Z_E^2(\beta)}
+\delta_{j,j'}\delta_{k,j'}\delta_{k,k'}\frac{Z_E(3\beta)}{Z_E^3(\beta)}
\right)
&\\
  & \quad
-\> \frac{D^2}{2} \> 
p_{k}^{(S)} 
\>
\left(
\delta_{k',j}\frac{Z_E(2\beta)}{Z_E^2(\beta)}
-\delta_{k',j}\delta_{k,k'}\frac{Z_E(3\beta)}{Z_E^3(\beta)}
+\delta_{k',j'}\frac{Z_E(2\beta)}{Z_E^2(\beta)}
-\delta_{k',j'}\delta_{k,k'}\frac{Z_E(3\beta)}{Z_E^3(\beta)}
\right.
&\\
  & \quad\quad
\left.
-\delta_{j,j'}\delta_{k',j}\frac{Z_E(2\beta)}{Z_E^2(\beta)}
+\delta_{j,j'}\delta_{k',j}\delta_{k,k'}\frac{Z_E(3\beta)}{Z_E^3(\beta)}
-\delta_{j,j'}\delta_{k',j'}\frac{Z_E(2\beta)}{Z_E^2(\beta)}
+\delta_{j,j'}\delta_{k',j'}\delta_{k,k'}\frac{Z_E(3\beta)}{Z_E^3(\beta)}
\right)
&\\
  & \quad
+\>
2 \> D^2 \>
\left.
p_{k}^{(S)} p_{k'}^{(S)}  
\>
\frac{Z_E(2\beta)}{Z_E^2(\beta)}
\> \left(
1-\delta_{j,j'}
-\delta_{k,k'}\frac{Z_E(2\beta)}{Z_E^2(\beta)}
+\delta_{j,j'}\delta_{k,k'}\frac{Z_E(2\beta)}{Z_E^2(\beta)}
\right)
\right]
&\\
\end{array}
\cr
&&
\begin{array}{lll}
=  & 
\left[
\frac{D^2}{8} \>
\>\left(
1+1
-\frac{Z_S(2\beta)}{Z_S^2(\beta)}-\frac{Z_S(2\beta)}{Z_S^2(\beta)}
\right.
\right.
&\\
 & \quad
\left.
-\frac{Z_S(2\beta)}{Z_S^2(\beta)}-\frac{Z_S(2\beta)}{Z_S^2(\beta)}
+\frac{Z_S(2\beta)}{Z_S^2(\beta)}
+\frac{Z_S(2\beta)}{Z_S^2(\beta)}
\right)
\> \frac{Z_E(2\beta)}{Z_E^2(\beta)} 
&\\
  & \quad
- \> \frac{D^2}{4} \>
\left(
\frac{Z_E(2\beta)}{Z_E^2(\beta)}
-\frac{Z_S(2\beta)}{Z_S^2(\beta)}\frac{Z_E(3\beta)}{Z_E^3(\beta)}
+\frac{Z_E(2\beta)}{Z_E^2(\beta)}
-\frac{Z_S(2\beta)}{Z_S^2(\beta)}\frac{Z_E(3\beta)}{Z_E^3(\beta)}
\right.
&\\
  & \quad\quad
\left.
-\frac{Z_S(2\beta)}{Z_S^2(\beta)}\frac{Z_E(2\beta)}{Z_E^2(\beta)}
+\frac{Z_S(3\beta)}{Z_S^3(\beta)}\frac{Z_E(3\beta)}{Z_E^3(\beta)}
-\frac{Z_S(2\beta)}{Z_S^2(\beta)}\frac{Z_E(2\beta)}{Z_E^2(\beta)}
+\frac{Z_S(3\beta)}{Z_S^3(\beta)}\frac{Z_E(3\beta)}{Z_E^3(\beta)}
\right)
&\\
  & \quad
-\> \frac{D^2}{4} \> 
\left(
\frac{Z_E(2\beta)}{Z_E^2(\beta)}
-\frac{Z_S(2\beta)}{Z_S^2(\beta)}\frac{Z_E(3\beta)}{Z_E^3(\beta)}
+\frac{Z_E(2\beta)}{Z_E^2(\beta)}
-\frac{Z_S(2\beta)}{Z_S^2(\beta)}\frac{Z_E(3\beta)}{Z_E^3(\beta)}
\right.
&\\
  & \quad\quad
\left.
-\frac{Z_S(2\beta)}{Z_S^2(\beta)}\frac{Z_E(2\beta)}{Z_E^2(\beta)}
+\frac{Z_S(3\beta)}{Z_S^3(\beta)}\frac{Z_E(3\beta)}{Z_E^3(\beta)}
-\frac{Z_S(2\beta)}{Z_S^2(\beta)}\frac{Z_E(2\beta)}{Z_E^2(\beta)}
+\frac{Z_S(3\beta)}{Z_S^3(\beta)}\frac{Z_E(3\beta)}{Z_E^3(\beta)}
\right)
&\\
  & \quad
+\>
D^2 \>
\left.
\frac{Z_E(2\beta)}{Z_E^2(\beta)}
\> \left(
1-\frac{Z_S(2\beta)}{Z_S^2(\beta)}
-\frac{Z_S(2\beta)}{Z_S^2(\beta)}\frac{Z_E(2\beta)}{Z_E^2(\beta)}
+\frac{Z_S^2(2\beta)}{Z_S^4(\beta)}\frac{Z_E(2\beta)}{Z_E^2(\beta)}
\right)
\right]
&\\
\end{array}
\cr
&&
\begin{array}{lll}
 =  & 
D^2 \> 
\left[
\frac{1}{4} \>
\> \left( 1 -\frac{Z_S(2\beta)}{Z_S^2(\beta)} \right)
\> \frac{Z_E(2\beta)}{Z_E^2(\beta)} 
\right.
&\\
  & \quad
- \>
\left(
\frac{Z_E(2\beta)}{Z_E^2(\beta)}
-\frac{Z_S(2\beta)}{Z_S^2(\beta)}\frac{Z_E(3\beta)}{Z_E^3(\beta)}
-\frac{Z_S(2\beta)}{Z_S^2(\beta)}\frac{Z_E(2\beta)}{Z_E^2(\beta)}
+\frac{Z_S(3\beta)}{Z_S^3(\beta)}\frac{Z_E(3\beta)}{Z_E^3(\beta)}
\right)
&\\
  & \quad
+ \>
\left.
\frac{Z_E(2\beta)}{Z_E^2(\beta)}
\> \left(
1-\frac{Z_S(2\beta)}{Z_S^2(\beta)}
-\frac{Z_S(2\beta)}{Z_S^2(\beta)}\frac{Z_E(2\beta)}{Z_E^2(\beta)}
+\frac{Z_S^2(2\beta)}{Z_S^4(\beta)}\frac{Z_E(2\beta)}{Z_E^2(\beta)}
\right)
\right]
&\\
\end{array}
\cr
&&
\begin{array}{lll}
 =  & 
\frac{D^2}{4}\> \frac{Z_E(2\beta)}{Z_E^2(\beta)} \>
\left( 1-\frac{Z_S(2\beta)}{Z_S^2(\beta)} \right)
\> + \> 
D^2 \> 
\frac{Z_E(3\beta)}{Z_E^3(\beta)} \> 
\left(\frac{Z_S(2\beta)}{Z_S^2(\beta)}-\frac{Z_S(3\beta)}{Z_S^3(\beta)}\right)
\> - \> 
D^2 \>
\frac{Z_E^2(2\beta)}{Z_E^4(\beta)} \>
\frac{Z_S(2\beta)}{Z_S^2(\beta)} \>
\left( 1 - \frac{Z_S(2\beta)}{Z_S^2(\beta)} \right) 
&\\
\end{array}
\end{eqnarray}
which is not too pretty of an expression.  

The second term (first middle term) to calculate is
\begin{eqnarray}
&&
\frac{1}{2!} \>
\sum_{k=1}^{D_S} \sum_{k'=1}^{D_S} 
\sum_{q=1}^{D_E} \sum_{q'=1}^{D_E} 
\left(1-\delta_{k,k'} \delta_{q,q'}\right)
\sum_{j,j'}^{D_S} \left(1-\delta_{j,j'}\right) 
\>
\left.
\frac{\partial \widetilde\rho\left(\{x\},\> \phi_1, \phi_2, \cdots \phi_D\right)}
{\partial x_{k,q}}
\right|_{\{x\}=\frac{1}{D}}
\left.
\frac{\partial \widetilde\rho\left(\{x\},\> -\phi_1, -\phi_2, \cdots -\phi_D\right)}
{\partial x_{k',q'}}
\right|_{\{x\}=\frac{1}{D}}
\cr
&&\begin{array}{lll}
 =  & 
\frac{1}{2!} \>
\sum_{k=1}^{D_S} \sum_{k'=1}^{D_S} 
\sum_{q=1}^{D_E} \sum_{q'=1}^{D_E} 
\sum_{j,j'}^{D_S} 
\left(1-\delta_{k,k'} \delta_{q,q'}\right)
\left(1-\delta_{j,j'}\right) \times
&\\
 & \quad
\left[
\frac{D}{2} \>
\sqrt{p_{j}^{(S)}} \sqrt{p_{j'}^{(S)}} p_{q}^{(E)}
\> e^{i \phi_{j,q}} \> e^{-i \phi_{j',q}}  
\>
\left(\delta_{k,j}+\delta_{k,j'}\right)
\right.
&\\
  & \quad
-\>
D \>
\left.
p_{k}^{(S)} 
\sqrt{p_{j}^{(S)}} \sqrt{p_{j'}^{(S)}} 
\> 
p_{q}^{(E)}  
\>
\left(
\sum_{p=1}^{D_E} 
\> e^{i \phi_{j,p}} \> e^{-i \phi_{j',p}}  
\>
p_p^{(E)}
\right)
\right]
\> \times
&\\
 & \quad
\left[
\frac{D}{2} \>
\sqrt{p_{j}^{(S)}} \sqrt{p_{j'}^{(S)}} p_{q'}^{(E)}
\> e^{-i \phi_{j,q'}} \> e^{i \phi_{j',q'}}  
\>
\left(\delta_{k',j}+\delta_{k',j'}\right)
\right.
&\\
  & \quad
-\>
D \>
\left.
p_{k'}^{(S)} 
\sqrt{p_{j}^{(S)}} \sqrt{p_{j'}^{(S)}} 
\> 
p_{q'}^{(E)}  
\>
\left(
\sum_{p'=1}^{D_E} 
\> e^{-i \phi_{j,p'}} \> e^{i \phi_{j',p'}}  
\>
p_{p'}^{(E)}
\right)
\right]
&\\
\end{array}
\cr
&&
\begin{array}{lll}
 =  & 
\frac{1}{2!} \>
\sum_{k=1}^{D_S} \sum_{k'=1}^{D_S} 
\sum_{q=1}^{D_E} \sum_{q'=1}^{D_E} 
\sum_{j=1}^{D_S} \sum_{j'=1}^{D_S} 
\left(1-\delta_{k,k'} \delta_{q,q'}\right)
\left(1-\delta_{j,j'}\right) \times
&\\
 & \quad
\left[
\frac{D^2}{4} \>
p_{j}^{(S)} p_{j'}^{(S)} \> 
p_{q}^{(E)} p_{q'}^{(E)} \> 
\delta_{q,q'} \>
\left(\delta_{k,j}+\delta_{k,j'}\right)
\left(\delta_{k',j}+\delta_{k',j'}\right)
\right.
&\\
 & \quad
\> - \> 
\frac{D^2}{2} \>
p_{k'}^{(S)}  
p_{j}^{(S)} p_{j'}^{(S)} 
\> 
\left(p_{q}^{(E)}\right)^2 p_{q'}^{(E)}
\>
\left(\delta_{k,j}+\delta_{k,j'}\right)
&\\
 & \quad
\> - \> 
\frac{D^2}{2} \>
p_{k}^{(S)}  
p_{j}^{(S)} p_{j'}^{(S)} 
\> 
p_{q}^{(E)} \left(p_{q'}^{(E)}\right)^2
\>
\left(\delta_{k',j}+\delta_{k',j'}\right)
&\\
 & \quad
\> + \> 
\left.
D^2 \>
p_{k}^{(S)} p_{k'}^{(S)}  
p_{j}^{(S)} p_{j'}^{(S)} 
\> 
p_{q}^{(E)} p_{q'}^{(E)}
\>
\left(\sum_{p=1}^{D_E} \left(p_p^{(E)}\right)^2 \right)
\right]
&\\
\end{array}
\cr
&&\begin{array}{lll}
 =  & 
\frac{1}{2!} \>
\sum_{k=1}^{D_S} \sum_{k'=1}^{D_S} 
\sum_{j=1}^{D_S} \sum_{j'=1}^{D_S} 
\left(1-\delta_{j,j'}\right) \>
p_{j}^{(S)} p_{j'}^{(S)} \> 
\times
&\\
 & \quad
\left[
\frac{D^2}{4} \>
\frac{Z_E(2\beta)}{Z_E^2(\beta)} \> 
\left(\delta_{k,j}+\delta_{k,j'}\right)
\left(\delta_{k',j}+\delta_{k',j'}\right)
\left(1 - \delta_{k,k'} \right) \>
\right.
&\\
 & \quad
\> - \> 
\frac{D^2}{2} \>
p_{k'}^{(S)}  
\> 
\left(\frac{Z_E(2\beta)}{Z_E^2(\beta)} - 
\delta_{k,k'} \frac{Z_E(3\beta)}{Z_E^3(\beta)} \right)
\>
\left(\delta_{k,j}+\delta_{k,j'}\right)
&\\
 & \quad
\> - \> 
\frac{D^2}{2} \>
p_{k}^{(S)}  
\> 
\left(\frac{Z_E(2\beta)}{Z_E^2(\beta)} - 
\delta_{k,k'} \frac{Z_E(3\beta)}{Z_E^3(\beta)} \right)
\>
\left(\delta_{k',j}+\delta_{k',j'}\right)
&\\
 & \quad
\> + \> 
\left.
D^2 \>
p_{k}^{(S)} p_{k'}^{(S)}  
\> 
\left(1 - \delta_{k,k'} \frac{Z_E(2\beta)}{Z_E^2(\beta)} \right)
\>
\left(\frac{Z_E(2\beta)}{Z_E^2(\beta)}\right) 
\right]
&\\
\end{array}
\cr
&&
\begin{array}{lll}
 =  & \quad 
\frac{D^2}{2!} \>
\sum_{k=1}^{D_S} \sum_{k'=1}^{D_S} 
\sum_{j=1}^{D_S} \sum_{j'=1}^{D_S} 
\> p_j^{(S)} p_{j'}^{(S)}\> \times
&\\
 & \quad
\left[
\frac{1}{4} \>
\frac{Z_E(2\beta)}{Z_E^2(\beta)} \> 
\left(
\delta_{k,j}\delta_{k',j}+\delta_{k,j'}\delta_{k',j} + 
\delta_{k,j}\delta_{k',j'}+\delta_{k,j'}\delta_{k',j'}
\right)
\>
\left(
1 - \delta_{k,k'} - \delta_{j,j'} + \delta_{k,k'}\delta_{j,j'} 
\right)
\right.
&\\
 & \quad
- \frac{1}{2} \> p_{k'}^{(S)} \>
\left(
\delta_{k,j}\frac{Z_E(2\beta)}{Z_E^2(\beta)} 
-\delta_{k,j}\delta_{k,k'}\frac{Z_E(3\beta)}{Z_E^3(\beta)} 
+\delta_{k,j'}\frac{Z_E(2\beta)}{Z_E^2(\beta)} 
-\delta_{k,j'}\delta_{k,k'}\frac{Z_E(3\beta)}{Z_E^3(\beta)} 
\right.
&\\
 & \quad
\left.
-\delta_{j,j'}\delta_{k,j}\frac{Z_E(2\beta)}{Z_E^2(\beta)} 
+\delta_{j,j'}\delta_{k,j}\delta_{k,k'}\frac{Z_E(3\beta)}{Z_E^3(\beta)} 
-\delta_{j,j'}\delta_{k,j'}\frac{Z_E(2\beta)}{Z_E^2(\beta)} 
+\delta_{j,j'}\delta_{k,j'}\delta_{k,k'}\frac{Z_E(3\beta)}{Z_E^3(\beta)} 
\right) \>
&\\
 & \quad
- \frac{1}{2} \> p_{k}^{(S)} \>
\left(
\delta_{k',j}\frac{Z_E(2\beta)}{Z_E^2(\beta)} 
-\delta_{k',j}\delta_{k,k'}\frac{Z_E(3\beta)}{Z_E^3(\beta)} 
+\delta_{k',j'}\frac{Z_E(2\beta)}{Z_E^2(\beta)} 
-\delta_{k',j'}\delta_{k,k'}\frac{Z_E(3\beta)}{Z_E^3(\beta)} 
\right.
&\\
& \quad
\left.
-\delta_{j,j'}\delta_{k',j}\frac{Z_E(2\beta)}{Z_E^2(\beta)} 
+\delta_{j,j'}\delta_{k',j}\delta_{k,k'}\frac{Z_E(3\beta)}{Z_E^3(\beta)} 
-\delta_{j,j'}\delta_{k',j'}\frac{Z_E(2\beta)}{Z_E^2(\beta)} 
+\delta_{j,j'}\delta_{k',j'}\delta_{k,k'}\frac{Z_E(3\beta)}{Z_E^3(\beta)} 
\right) \>
&\\
&\quad
\left.
+ \> p_{k}^{(S)} p_{k'}^{(S)} \>
\left(
1 - \delta_{k,k'} \frac{Z_E(2\beta)}{Z_E^2(\beta)}
-\delta_{j,j'} + \delta_{j,j'}\delta_{k,k'} \frac{Z_E(2\beta)}{Z_E^2(\beta)}
\right)
\> \left(\frac{Z_E(2\beta)}{Z_E^2(\beta)}\right) 
\right]
&\\
\end{array}
\end{eqnarray}
which is simplified to
\begin{eqnarray}
&&
\frac{1}{2!} \>
\sum_{k=1}^{D_S} \sum_{k'=1}^{D_S} 
\sum_{q=1}^{D_E} \sum_{q'=1}^{D_E} 
\left(1-\delta_{k,k'} \delta_{q,q'}\right)
\sum_{j,j'}^{D_S} \left(1-\delta_{j,j'}\right) 
\>
\left.
\frac{\partial \widetilde\rho\left(\{x\},\> \phi_1, \phi_2, \cdots \phi_D\right)}
{\partial x_{k,q}}
\right|_{\{x\}=\frac{1}{D}}
\left.
\frac{\partial \widetilde\rho\left(\{x\},\> -\phi_1, -\phi_2, \cdots -\phi_D\right)}
{\partial x_{k',q'}}
\right|_{\{x\}=\frac{1}{D}}
\cr
&&
\begin{array}{lll}
= & 
\frac{D^2}{2!} \>
\left[
\frac{1}{2} \>
\frac{Z_E(2\beta)}{Z_E^2(\beta)} \> 
(1- \frac{Z_S(2\beta)}{Z_S^2(\beta)})
\right.
&\\
 & \quad
- \frac{1}{2} \> 
\left(
\frac{Z_E(2\beta)}{Z_E^2(\beta)} 
-\frac{Z_S(2\beta)}{Z_S^2(\beta)}\frac{Z_E(3\beta)}{Z_E^3(\beta)} 
+\frac{Z_E(2\beta)}{Z_E^2(\beta)} 
-\frac{Z_S(2\beta)}{Z_S^2(\beta)}\frac{Z_E(3\beta)}{Z_E^3(\beta)} 
\right.
&\\
 & \quad
\left.
-\frac{Z_S(2\beta)}{Z_S^2(\beta)}\frac{Z_E(2\beta)}{Z_E^2(\beta)} 
+\frac{Z_S(3\beta)}{Z_S^3(\beta)}\frac{Z_E(3\beta)}{Z_E^3(\beta)} 
-\frac{Z_S(2\beta)}{Z_S^2(\beta)}\frac{Z_E(2\beta)}{Z_E^2(\beta)} 
+\frac{Z_S(3\beta)}{Z_S^3(\beta)}\frac{Z_E(3\beta)}{Z_E^3(\beta)} 
\right) \>
&\\
 & \quad
- \frac{1}{2} \> 
\left(
\frac{Z_E(2\beta)}{Z_E^2(\beta)} 
-\frac{Z_S(2\beta)}{Z_S^2(\beta)}\frac{Z_E(3\beta)}{Z_E^3(\beta)} 
+\frac{Z_E(2\beta)}{Z_E^2(\beta)} 
-\frac{Z_S(2\beta)}{Z_S^2(\beta)}\frac{Z_E(3\beta)}{Z_E^3(\beta)} 
\right.
&\\
 & \quad
\left.
-\frac{Z_S(2\beta)}{Z_S^2(\beta)}\frac{Z_E(2\beta)}{Z_E^2(\beta)} 
+\frac{Z_S(3\beta)}{Z_S^3(\beta)}\frac{Z_E(3\beta)}{Z_E^3(\beta)} 
-\frac{Z_S(2\beta)}{Z_S^2(\beta)}\frac{Z_E(2\beta)}{Z_E^2(\beta)} 
+\frac{Z_S(3\beta)}{Z_S^3(\beta)}\frac{Z_E(3\beta)}{Z_E^3(\beta)} 
\right) \>
&\\
 & \quad
\left.
+ 
\left(
1 - \frac{Z_S(2\beta)}{Z_S^2(\beta)}\frac{Z_E(2\beta)}{Z_E^2(\beta)}
-\frac{Z_S(2\beta)}{Z_S^2(\beta)} 
+ \frac{Z_S^2(2\beta)}{Z_S^4(\beta)} \frac{Z_E(2\beta)}{Z_E^2(\beta)}
\right)
\> \left(\frac{Z_E(2\beta)}{Z_E^2(\beta)}\right) 
\right]
&\\
\end{array}
\cr
&&
\begin{array}{lll}
 =  & 
\frac{D^2}{2!} \>
\left[
\frac{1}{2} \>
\frac{Z_E(2\beta)}{Z_E^2(\beta)} \> 
(1- \frac{Z_S(2\beta)}{Z_S^2(\beta)})
\right.
- \frac{Z_E(2\beta)}{Z_E^2(\beta)} \>
+2 \frac{Z_S(2\beta)}{Z_S^2(\beta)}\frac{Z_E(3\beta)}{Z_E^3(\beta)} \>
+ \frac{Z_S(2\beta)}{Z_S^2(\beta)}\frac{Z_E(2\beta)}{Z_E^2(\beta)} \>
&\\
  & \quad
+2 \frac{Z_S(3\beta)}{Z_S^3(\beta)}\frac{Z_E(3\beta)}{Z_E^3(\beta)} \>
\left.
- \frac{Z_S(2\beta)}{Z_S^2(\beta)}\frac{Z_E^2(2\beta)}{Z_E^4(\beta)} \>
+ \frac{Z_S^2(2\beta)}{Z_S^4(\beta)}\frac{Z_E^2(2\beta)}{Z_E^4(\beta)} \>
\right]
&\\
\end{array}
\cr
&&
\begin{array}{lll}
 = & 
-\frac{D^2}{4} \>
\frac{Z_E(2\beta)}{Z_E^2(\beta)} \>
\left(1-\frac{Z_S(2\beta)}{Z_S^2(\beta)}\right) \>
+ D^2 \>
\frac{Z_E(3\beta)}{Z_E^3(\beta)} \>
\left(\frac{Z_S(2\beta)}{Z_S^2(\beta)}+\frac{Z_S(3\beta)}{Z_S^3(\beta)}\right) \>
- \frac{D^2}{2} \>
\frac{Z_E^2(2\beta)}{Z_E^4(\beta)} \>
\frac{Z_S(2\beta)}{Z_S^2(\beta)} \>
\left(1 - \frac{Z_S(2\beta)}{Z_S^2(\beta)} \right)
&\\
\end{array}
\end{eqnarray}
which is also not a pretty expression.  

The last two terms give the same results as the first two, since they are 
complex conjugates of the first two terms.  For example, the fourth term is 
the complex conjugate of the first term, and the result after the averaging 
over the $\{\phi\}$ is real, so the final result for the fourth term equals the 
final result for the first term.  

Collecting the four terms gives the final result for the 
different-$x_{k,q}$ second derivatives to be 
\begin{eqnarray}
&&
\frac{1}{2!} \>
\sum_{k=1}^{D_S} \sum_{k'=1}^{D_S} 
\sum_{q=1}^{D_E} \sum_{q'=1}^{D_E} 
\left(1-\delta_{k,k'} \delta_{q,q'}\right)
\left.
\frac{\partial^2 f_{2\sigma^2}}{\partial x_{k,q} \> \partial x_{k',q'}} 
\right|_{\{x\}=\frac{1}{D}}
\cr
&&\begin{array}{lll}
 =  &  
\frac{D^2}{2}\> \frac{Z_E(2\beta)}{Z_E^2(\beta)} \>
\left( 1-\frac{Z_S(2\beta)}{Z_S^2(\beta)} \right)
\> + \> 
2 \> D^2 \> 
\frac{Z_E(3\beta)}{Z_E^3(\beta)} \> 
\left(\frac{Z_S(2\beta)}{Z_S^2(\beta)}-\frac{Z_S(3\beta)}{Z_S^3(\beta)}\right)
&\\
 & \quad
\> - \> 
2 \> D^2 \>
\frac{Z_E^2(2\beta)}{Z_E^4(\beta)} \>
\frac{Z_S(2\beta)}{Z_S^2(\beta)} \>
\left( 1 - \frac{Z_S(2\beta)}{Z_S^2(\beta)} \right) 
-\frac{D^2}{2} \>
\frac{Z_E(2\beta)}{Z_E^2(\beta)} \>
\left(1-\frac{Z_S(2\beta)}{Z_S^2(\beta)}\right) \>
&\\
 & \quad
+ 2 \> D^2 \>
\frac{Z_E(3\beta)}{Z_E^3(\beta)} \>
\left(\frac{Z_S(2\beta)}{Z_S^2(\beta)}+\frac{Z_S(3\beta)}{Z_S^3(\beta)}\right) \>
- D^2 \>
\frac{Z_E^2(2\beta)}{Z_E^4(\beta)} \>
\frac{Z_S(2\beta)}{Z_S^2(\beta)} \>
\left(1 - \frac{Z_S(2\beta)}{Z_S^2(\beta)} \right)
&\\
\end{array}
\cr
&&
\begin{array}{lll}
 =  & 
4 \> D^2 \> 
\frac{Z_E(3\beta)}{Z_E^3(\beta)} \> 
\left(\frac{Z_S(2\beta)}{Z_S^2(\beta)}-\frac{Z_S(3\beta)}{Z_S^3(\beta)}\right)
\>
\> - \> 
3 \> D^2 \>
\frac{Z_E^2(2\beta)}{Z_E^4(\beta)} \>
\frac{Z_S(2\beta)}{Z_S^2(\beta)} \>
\left( 1 - \frac{Z_S(2\beta)}{Z_S^2(\beta)} \right) 
&\\
\end{array}
\end{eqnarray}
which is the same as the same-$x_{k,q}$ term except for a negative sign.

\subsubsection{\textbf{0$^{\rm \> th}$, 1$^{\rm st}$, and 2$^{\rm nd}$ terms of
${\cal E}(2\sigma^2)$}}

To second order one has the final expression for $2\sigma^2$, 
now that all $\phi_{k,q}$ have correctly been taken into account, 
\begin{eqnarray}
\label{EqS:E2sigma2CORRECT}
{\cal E}\left(f_{2\sigma^2}\right) 
& \> = \> &
{\cal E}\left(
\left.f_{2\sigma^2}\right|_{\{x\}=\frac{1}{D}} 
\right) 
+ \> 
\frac{1}{2!} \>
{\cal E}\left( \left(x-\frac{1}{D}\right)^2\right) 
\>
\sum_{k=1}^{D_S} \sum_{q=1}^{D_E} 
\left.
\frac{\partial^2 f_{2\sigma^2} }
{\partial x_{k,q}^2 }
\right|_{\{x\}=\frac{1}{D}} 
\cr
& \>  \> & \quad
+ \> 
\frac{1}{2!} \>
{\cal E}\left(\left(x-\frac{1}{D}\right) \left(x'-\frac{1}{D}\right) \right) 
\>
\sum_{k=1}^{D_S} \sum_{k'=1}^{D_S} \sum_{q=1}^{D_E} \sum_{q'=1}^{D_E} 
\left(1-\delta_{k,k'}\delta_{q,q'}\right)
\left.
\frac{\partial^2 f_{2\sigma^2} }
{\partial x_{k,q} x_{k',q'} }
\right|_{\{x\}=\frac{1}{D}} 
\cr
&&
\begin{array}{lll}
 =  &
\frac{Z_E(2\beta)}{Z_E^2(\beta)}
\> 
\left(
1-\frac{Z_S(2\beta)}{Z_S^2(\beta)}
\right)
&\\
  & \quad
+ \> 
\left(\frac{D-1}{D^2\left(D+1\right)}\right) 
\>
\left[
-  4  D^2 \>
\frac{Z_E(3\beta)}{Z_E^3(\beta)} \>
\left( \frac{Z_S(2\beta)}{Z_S^2(\beta)}  - \frac{Z_S(3\beta)}{Z_S^3(\beta)} \right)
+  3  D^2 \>
\frac{Z_E^2(2\beta)}{Z_E^4(\beta)} \>
\left( \frac{Z_S(2\beta)}{Z_S^2(\beta)}  
- \frac{Z_S^2(2\beta)}{Z_S^4(\beta)} \right)
\right]
&\\
 & \quad
+ \> 
\left(\> - \> \frac{1}{D^2\left(D+1\right)}\right) 
\>
\left[
4 \> D^2 \> 
\frac{Z_E(3\beta)}{Z_E^3(\beta)} \> 
\left(\frac{Z_S(2\beta)}{Z_S^2(\beta)}-\frac{Z_S(3\beta)}{Z_S^3(\beta)}\right)
\>
\> - \> 
3 \> D^2 \>
\frac{Z_E^2(2\beta)}{Z_E^4(\beta)} \>
\frac{Z_S(2\beta)}{Z_S^2(\beta)} \>
\left( 1 - \frac{Z_S(2\beta)}{Z_S^2(\beta)} \right) 
\right]
&\\
\end{array}
\cr
&&
\begin{array}{lll}
 =  &
\frac{Z_E(2\beta)}{Z_E^2(\beta)}
\> 
\left(
1-\frac{Z_S(2\beta)}{Z_S^2(\beta)}
\right)
&\\
  & \quad
+ \> 
\left(\frac{D-1}{\left(D+1\right)}\right) 
\>
\left[
-  4  \>
\frac{Z_E(3\beta)}{Z_E^3(\beta)} \>
\left( \frac{Z_S(2\beta)}{Z_S^2(\beta)}  - \frac{Z_S(3\beta)}{Z_S^3(\beta)} \right)
+  3  \>
\frac{Z_E^2(2\beta)}{Z_E^4(\beta)} \>
\left( \frac{Z_S(2\beta)}{Z_S^2(\beta)}  
- \frac{Z_S^2(2\beta)}{Z_S^4(\beta)} \right)
\right]
&\\
 & \quad
+ \> 
\left(\> - \> \frac{1}{\left(D+1\right)}\right) 
\>
\left[
4 \> 
\frac{Z_E(3\beta)}{Z_E^3(\beta)} \> 
\left(\frac{Z_S(2\beta)}{Z_S^2(\beta)}-\frac{Z_S(3\beta)}{Z_S^3(\beta)}\right)
\>
\> - \> 
3 \> 
\frac{Z_E^2(2\beta)}{Z_E^4(\beta)} \>
\frac{Z_S(2\beta)}{Z_S^2(\beta)} \>
\left( 1 - \frac{Z_S(2\beta)}{Z_S^2(\beta)} \right) 
\right]
&\\
\end{array}
\cr
&&
\begin{array}{lll}
 = &
\frac{Z_E(2\beta)}{Z_E^2(\beta)}
\> 
\left(
1-\frac{Z_S(2\beta)}{Z_S^2(\beta)}
\right)
- \> 4 \> 
\frac{D}{\left(D+1\right)} \> 
\frac{Z_E(3\beta)}{Z_E^3(\beta)} \>
\left( \frac{Z_S(2\beta)}{Z_S^2(\beta)}  - \frac{Z_S(3\beta)}{Z_S^3(\beta)} \right)
+ \> 3 \> 
\frac{D}{\left(D+1\right)} \> 
\frac{Z_E^2(2\beta)}{Z_E^4(\beta)} \>
\left( \frac{Z_S(2\beta)}{Z_S^2(\beta)}  
- \frac{Z_S^2(2\beta)}{Z_S^4(\beta)} \right)
\> .
&\\
\end{array}
\end{eqnarray}
Equation~(\ref{EqS:E2sigma2CORRECT}) is written as Eq.~(\ref{sigma2}) in the main text, but is 
written in terms of free energies rather than partition functions.

In the limit of high temperature ($\beta\rightarrow 0$), one has 
that $Z_E(0)=D_E$ and $Z_S(0)=D_S$ to give 
\begin{equation}
\begin{array}{lcl}
\lim_{\beta\rightarrow 0} \> {\cal E}\left(f_{2\sigma^2}\right) 
& \> = \> &
\frac{D_E}{D_E^2} \> \left( 1 - \frac{D_S}{D_S^2}\right) \>
- 4 \frac{D}{D+1} 
\frac{D_E}{D_E^3}\left(\frac{D_S}{D_S^2}-\frac{D_S}{D_S^3}\right) \>
+ 3 \frac{D}{D+1} 
\frac{D_E^2}{D_E^4}\left(\frac{D_S}{D_S^2}-\frac{D_S^2}{D_S^4}\right)
\\
& \> = \> &
\frac{1}{D_E} \> \left( 1 - \frac{1}{D_S}\right) \>
- 4 \frac{D_E D_S}{D+1} 
\frac{1}{D_E^2}\left(\frac{1}{D_S}-\frac{1}{D_S^2}\right) \>
+ 3 \frac{D_E D_S}{D+1} 
\frac{1}{D_E^2}\left(\frac{1}{D_S}-\frac{1}{D_S^2}\right)
\\
& \> = \> &
\frac{1}{D_E} \> \frac{\left(D_S-1\right)}{D_S} \>
- \frac{1}{D+1} 
\frac{1}{D_E}\left(1-\frac{1}{D_S}\right) \>
\\
& \> = \> &
\frac{D}{D+1} \> \frac{1}{D_E} \> \frac{\left(D_S-1\right)}{D_S} \>
\\
& \> = \> &
\frac{D_S-1}{D+1} 
 = 
\frac{D_S-1}{D_E D_S+1} 
\> .
\\
\end{array}
\end{equation}

One can perform an expansion about $\beta=0$ (temperature $T$$=$$\infty$).  In particular, 
use that the average internal energy for the environment is given by
\begin{equation}
\left\langle E\left(n\beta\right)\right\rangle_E
\> = \> 
- \> \frac{\partial {\rm ln}\left(Z_E(n\beta)\right)}{\partial \left(n\beta\right)}
\> = \> 
- \> \frac{1}{Z_E(n\beta)} \> \frac{1}{n} \> 
\frac{\partial Z_E(n\beta)}{\partial \beta}
\end{equation}
so
\begin{equation}
\frac{\partial Z_E(n\beta)}{\partial \beta}
\> = \> 
-n \> \left\langle E\left(n\beta\right)\right\rangle_E \> Z_E(n\beta)
\> .
\end{equation}
Similarly for the derivatives of $Z_S(n\beta)$ for the system, 
\begin{equation}
\frac{\partial Z_S(n\beta)}{\partial \beta}
\> = \> 
-n \> \left\langle E\left(n\beta\right)\right\rangle_S \> Z_S(n\beta)
\> .
\end{equation}
Taking the limit $\beta=0$ gives the average internal energy at infinite temperature, 
$U_\infty^{(E)}$ and $U_\infty^{(S)}$, for the environment and system, respectively.  Thus 
\begin{equation}
\left. \frac{\partial Z_S(n\beta)}{\partial \beta} \right|_{\beta=0}
\> = \> 
-n \> U_\infty^{(S)} \> D_S
\quad\quad
{\rm and}
\quad\quad
\left. \frac{\partial Z_E(n\beta)}{\partial \beta} \right|_{\beta=0}
\> = \> 
-n \> U_\infty^{(E)} \> D_E
\> .
\end{equation}
Note that 
\begin{equation}
\left.
\frac{\partial\>\>}{\partial\beta} \left(\frac{Z_E^m(n\beta)}{Z_E^{mn}(\beta)}\right)
\right|_{\beta=0}
\> = \> 
-\frac{n m D_E^{m-1} D_E}{D_E^{mn}} U_\infty^{(E)} \> + \> 
\frac{n m D_E^{m} D_E}{D_E^{mn+1}} U_\infty^{(E)} 
\> = \> 
0
\end{equation}
and similarly for the system $Z_S$.  Thus, the first order term 
in the expansion about $\beta=0$ vanishes.  
This gives that for small $\beta$ the Taylor expansion is
\begin{equation}
{\cal E}\left(f_{2\sigma^2}\right) 
 \> \approx \> 
\frac{D_S-1}{D_E D_S+1} 
+ {\cal O}\left(\beta^2\right) 
\> .
\end{equation}

The second order terms should be in terms of the heat capacities 
at constant volume, $C_{E,v}$ and $C_{S,v}$, since 
\begin{equation}
\begin{array}{lcl}
C_{S,v} 
& \> = \> &
\frac{\partial\left\langle E\right\rangle_S}{\partial T}
 =  
k_{\rm B} \beta^2 
\frac{\partial\left\langle E\right\rangle_S}{\partial \beta}
 = 
- k_{\rm B} \beta^2 
\frac{\partial^2 {\rm ln}\left(Z_S(\beta)\right)}{\partial \beta^2}
\\
& \> = \> &
k_{\rm B} \beta^2 \> 
\left[
\frac{1}{Z_S(\beta)}\frac{\partial^2 Z_S(\beta)}{\partial \beta^2}
-\left(\frac{1}{Z_S(\beta)}\frac{\partial Z_S(\beta)}{\partial \beta}\right)^2
\right]
\>.
\\
\end{array}
\end{equation}
In order to calculate more easily the second-order term, define 
\begin{equation}
R_E(n_E\beta) = \frac{Z_E(n_E\beta)}{Z_E^{n_E}(\beta)}
\quad{\rm and}\quad
R_S(n_S\beta) = \frac{Z_S(n_S\beta)}{Z_S^{n_S}(\beta)}
\>
\end{equation}
and evaluated at $\beta$$=$$0$ gives
\begin{equation}
\begin{array}{lcl}
\left.R_E(n_E\beta)\right|_{\beta=0} 
& \>  = \> & 
\left. \frac{Z_E(n_E\beta)}{Z_E^{n_E}(\beta)} \right|_{\beta=0} \\
& \>  = \> & 
\frac{D_E}{D_E^{n_E}}
 = 
\frac{1}{D_E^{n_E-1}}. \\
\end{array}
\end{equation}
The first derivative is 
\begin{equation}
\begin{array}{lcl}
\frac{\partial R_E(n_E\beta)}{\partial\beta} 
& \> = \> & 
\frac{\partial\>\>\>}{\partial\beta}\left(
\frac{Z_E(n_E\beta)}{Z_E^{n_E}(\beta)}
\right) \\
& = & 
\frac{1}{Z_E^{n_E}(\beta)}\frac{\partial Z_E(n_E\beta)}{\partial \beta}
-\frac{n_E Z_E(n_E\beta)}{Z_E^{n_E+1}(\beta)}\frac{\partial Z_E(\beta)}{\partial\beta} 
\\
& = & 
-\frac{n_E Z_E(n_E\beta)}{Z_E^{n_E}(\beta)} \left\langle E(n_E\beta)\right\rangle_E
+\frac{n_E Z_E(n_E\beta)}{Z_E^{n_E}(\beta)} \left\langle E(\beta)\right\rangle_E
\>
\\
\end{array}
\end{equation}
and evaluated at $\beta=0$ gives
\begin{equation}
\begin{array}{lcl}
\left. \frac{\partial R_E(n_E\beta)}{\partial\beta} \right |_{\beta=0}
& \> = \> & 
\left. \frac{\partial\>\>\>}{\partial\beta}\left(
\frac{Z_E(n_E\beta)}{Z_E^{n_E}(\beta)}
\right) \right |_{\beta=0} 
\\
& \> = \> & 
\left.
-\frac{n_E Z_E(n_E\beta)}{Z_E^{n_E}(\beta)} \left\langle E(n_E\beta)\right\rangle_E
\right |_{\beta=0}
+
\left.
\frac{n_E Z_E(n_E\beta)}{Z_E^{n_E}(\beta)} \left\langle E(\beta)\right\rangle_E
\right |_{\beta=0}
\\
& \> = \> & 
-\frac{n_E D_E}{D_E^{n_E}} U_\infty^{(E)}
+
\frac{n_E D_E}{D_E^{n_E}} U_\infty^{(E)}
\\
& \> = \> & 
0
\> .
\end{array}
\end{equation}
The second order derivative is
\begin{equation}
\begin{array}{lcl}
\frac{\partial^2 R_E(n_E\beta)}{\partial\beta^2} 
& \> = \> & 
\frac{\partial^2\>\>\>}{\partial\beta^2}\left(
\frac{Z_E(n_E\beta)}{Z_E^{n_E}(\beta)}
\right) \\
& = & 
\frac{1}{Z_E^{n_E}(\beta)}\frac{\partial^2 Z_E(n_E\beta)}{\partial \beta^2}
- n_E 
\frac{1}{Z_E^{n_E+1}(\beta)}\frac{\partial Z_E(n_E\beta)}{\partial \beta}
\frac{\partial Z_E(\beta)}{\partial \beta}
\\
& & \quad 
-\frac{n_E Z_E(n_E\beta)}{Z_E^{n_E+1}(\beta)}
\frac{\partial^2 Z_E(\beta)}{\partial\beta^2} 
-\frac{n_E}{Z_E^{n_E+1}(\beta)}\frac{\partial Z_E(\beta)}{\partial\beta}
\frac{\partial Z_E(n_E \beta)}{\partial\beta} 
+\frac{n_E \left(n_E+1\right)Z_E(n_E\beta)}{Z_E^{n_E+2}(\beta)}
\left(\frac{\partial Z_E(\beta)}{\partial\beta} \right)^2
\> 
\end{array}
\end{equation}
or using the definition of the specific heat as
\begin{equation}
\frac{\partial^2 Z_E(n_E\beta)}{\partial \beta^2} \> = \> 
- \frac{1}{k_{\rm B} \beta^2} \> Z_E(n_E\beta) \> C_{E,v}(n_E \beta)
\end{equation}
with the limiting result
\begin{equation}
\begin{array}{lcl}
\left.
\frac{\partial^2 Z_E(n_E\beta)}{\partial \beta^2} 
\right |_{\beta=0} 
& \> = \> &
\left. 
-\frac{n_E}{k_{\rm B} \beta^2} \> Z_E(n_E\beta) \> C_{E,v}(n_E\beta)
\right |_{\beta=0} 
\\
& \> = \> &
- \frac{n_E}{k_{\rm B} \beta^2} \> D_E \> C_{E,v}(\infty)
\\
\end{array} 
\end{equation}
gives
\begin{equation}
\begin{array}{lcl}
\left.
\frac{\partial^2 R_E(n_E\beta)}{\partial\beta^2} 
\right |_{\beta=0} 
& \> = \> & 
\left.
\frac{\partial^2\>\>\>}{\partial\beta^2}\left(
\frac{Z_E(n_E\beta)}{Z_E^{n_E}(\beta)}
\right) 
\right |_{\beta=0} 
\\
& = & 
\frac{n_E}{D_E^{n_E}}
\left(- \frac{1}{k_{\rm B} \beta^2}\right) D_E C_{E,v}(\infty)
\\
& & \quad 
-\left( \frac{n_E D_E}{D_E^{n_E+1}} \right) 
\left(- \frac{n_E}{k_{\rm B} \beta^2}\right) D_E C_{E,v}(\infty)
\\
& = & 
\frac{n_E\> C_{E,v}(\infty)}{k_{\rm B}\>\beta^2}\>
\left(\frac{n_E}{D_E^{n_E-1}}-\frac{1}{D_E^{n_E-1}}\right)
\\
& = & 
\frac{n_E\left(n_E-1\right) \> C_{E,v}(\infty)}{k_{\rm B} \> \beta^2 \> D_E^{n_E-1} } 
\>.
\\
\end{array}
\end{equation}
Note that both 
\begin{equation}
\left.\frac{\partial R_E(n_E\beta)}{\partial\beta}\right|_{\beta=0} 
=0
\quad\quad
{\rm and} \>\> {\rm if} \>\> n_E=1 \>\>
\left.\frac{\partial R_E(n_E\beta)}{\partial\beta}\right|_{\beta=0, n_E=1} 
=0
\>.
\end{equation}
These greatly cut down on the number of non-zero terms from 
Eq.~(\ref{EqS:E2sigma2CORRECT}).  
One has that
\begin{eqnarray}
&&\begin{array}{lcl}
\frac{\partial^2\>\>\>}{\partial\beta^2}
\left[
\frac{Z_E(2\beta)}{Z_E^2(\beta)}
\> 
\left(
1-\frac{Z_S(2\beta)}{Z_S^2(\beta)}
\right)
- \> 4 \> 
\frac{D}{\left(D+1\right)} \> 
\frac{Z_E(3\beta)}{Z_E^3(\beta)} \>
\left( \frac{Z_S(2\beta)}{Z_S^2(\beta)}  - \frac{Z_S(3\beta)}{Z_S^3(\beta)} \right)
\right.
& \quad & \quad\quad\quad\quad\quad\quad
\\
\quad\quad\quad
\left.\left.
+ \> 3 \> 
\frac{D}{\left(D+1\right)} \> 
\frac{Z_E^2(2\beta)}{Z_E^4(\beta)} \>
\left( \frac{Z_S(2\beta)}{Z_S^2(\beta)}  
- \frac{Z_S^2(2\beta)}{Z_S^4(\beta)} \right)
\right]
\right|_{\beta=0}
\\
\end{array}
\cr
&&\begin{array}{lll}
 =  &
\frac{2 C_{E,v}(\infty)}{k_{\rm B}\beta^2 D_E}
\left(1-\frac{1}{D_S}\right)
-\frac{1}{D_E}\frac{2 C_{S,v}(\infty)}{k_{\rm B}\beta^2 D_S}
&\\
& \quad 
-4\frac{D}{D+1}\left[
\frac{6 C_{E,v}(\infty)}{k_{\rm B}\beta^2 D_E^2}\left(\frac{1}{D_S}-\frac{1}{D_S^2}\right)
+\frac{1}{D_E^2}\frac{C_{S,v}(\infty)}{k_{\rm B}\beta^2}\left(
\frac{2}{D_S}-\frac{6}{D_S^2}
\right)
\right]
&\\
&\quad  
+3\frac{D}{D+1}\left[
\frac{4 C_{E,v}(\infty)}{k_{\rm B}\beta^2 D_E^3}\left(\frac{1}{D_S}-\frac{1}{D_S^2}\right)
+\frac{1}{D_E^2}\frac{C_{S,v}(\infty)}{k_{\rm B}\beta^2}
\left(\frac{2}{D_S}-\frac{8}{D_S^3}\right)
\right]
&\\
\end{array}
\cr
&&\begin{array}{lll}
 = &
\frac{2 C_{E,v}(\infty)\left(D_S-1\right)}{k_{\rm B}\beta^2 D}
-\frac{1}{D}\frac{2 C_{S,v}(\infty)}{k_{\rm B}\beta^2}
&\\
&\quad  
-4\frac{1}{D(D+1)}\frac{2}{k_{\rm B}\beta^2}\left[
3 C_{E,v}(\infty)\left(D_S-1\right)
+C_{S,v}(\infty)\left(D_S-3\right)
\right]
&\\
&\quad  
+3\frac{1}{D(D+1)}\frac{2}{k_{\rm B}\beta^2}
\left[
\frac{2 C_{E,v}(\infty)}{D_E}\left(D_S-1\right)
+\frac{C_{S,v}(\infty)}{D_S}
\left(D_S^2-4\right)
\right]
&\\
\end{array}
\cr
&&
\begin{array}{lll}
 =  &
\frac{C_{E,v}(\infty)}{D\>k_{\rm B}\beta^2}
\left[
2 D_S -1 
-24 \frac{D_S-1}{D+1}
+ 12 \frac{D_S-1}{D_E\left(D+1\right)}
\right]
+
\frac{C_{S,v}(\infty)}{D\>k_{\rm B}\beta^2}
\left[
-2 
-8 \frac{\left(D_S-1\right)}{D+1}
+6 \frac{D_S^2-4}{D+1} 
\right]
\> .
&\\
\end{array}
\end{eqnarray}
Therefore the final result to second order about $\beta=0$ is
\begin{equation}
{\cal E}\left(f_{2\sigma^2}\right) 
 = 
\frac{D_S-1}{D+1}
+ \frac{1}{2!} \> \beta^2 \> 
\left\{
\frac{C_{E,v}(\infty)}{D\>k_{\rm B}\beta^2}
\left[
2 D_S -1 
-24 \frac{D_S-1}{D+1}
+ 12 \frac{D_S-1}{D_E\left(D+1\right)}
\right]
+
\frac{2 C_{S,v}(\infty)}{D\>k_{\rm B}\beta^2}
\left[
-1 
+2 \frac{\left(3 D_S^2-4D_S-8\right)}{D+1}
\right]
\right\}.
\end{equation}

One has to be cautious about the low-temperature ($\beta\rightarrow +\infty$) 
limit, since the analysis requires that $\beta \langle {  H}_{SE}\rangle$ 
be small.  
Then the partition function can be written as
\begin{equation}
Z_S(n \beta) = 
e^{-n \beta E_{0}^{(S)}} \left( g_{S} + 
\sum_{j=1}^{D_s-g_{S}} e^{- n \beta \left(E_j^{(S)}-E_0^{(S)}\right)}
\right)
\rightarrow_{\beta\rightarrow +\infty} 
g_{S} e^{-n \beta E_{0}^{(S)}}  
\> .
\end{equation}
Similarly for the partition function $Z_E(n\beta)$.  
Thus one has
\begin{equation}
\label{EqS:EsigmaLowT}
\begin{array}{lcl}
\lim_{\beta\rightarrow +\infty} \> {\cal E}\left(f_{2\sigma^2}\right) 
& \> = \> &
\frac{1}{g_{E}} \> 
\left( 1 - \frac{1}{g_{S}} \right) \>
- \frac{D}{D+1} \> \frac{1}{g_{E}^2} \> 
\left(\frac{1}{g_{S}} - \frac{1}{g_{S}^2}\right)
\\
& \> = \> &
\frac{g_{S}-1}{g_{E} g_{S}} \> 
\left( 1 - \frac{D}{\left(D+1\right) g_{E} g_{S} } \right) 
\> .
\\
\end{array}
\end{equation}
\end{widetext}
This expression goes to zero if the system ground state is 
non-degenerate.  For a highly degenerate system ground 
state ($g_{S}\gg 1$) the expression goes to $1/g_{E}$.  
Thus, in principle, one could use any system with $g_{S}>1$ and 
for a large bath $D\rightarrow +\infty$ at very low temperature measure 
${\cal E}(f_{2\sigma^2})$ in the system and from that deduce the 
degeneracy of the ground state of the bath.  

\subsection{Coupled entirety}

Our goal is to calculate in perturbation theory the expectation for 
$\sigma^2$, up to first order in the interaction Hamiltonian $\lambda H_I$ in Eq.~(\ref{EqS:Ham}).  
We then will show that for particular common symmetries this first order term is zero.  

Let us start with a formula 
from Wilcox, J. Math. Phys. 1967 (Eq.~4.1 of that paper)~\cite{WILC67} of
\begin{eqnarray}
\frac{\partial e^{{  H}(\lambda)}}{\partial\lambda} &=& 
\int_0^1 d\xi e^{\xi {  H}(\lambda)}
\frac{\partial {  H}(\lambda)}{\partial\lambda}
e^{-\xi {  H}(\lambda)}\>\>e^{{  H}(\lambda)} \\
&=& e^{{  H}(\lambda)}\int_0^1 d\xi e^{-\xi {  H}(\lambda)}
\frac{\partial {  H}(\lambda)}{\partial\lambda}
e^{\xi {  H}(\lambda)}\>\>.
\end{eqnarray}
Then one has
\begin{eqnarray}
e^{-\beta {  H}} & \approx & 
e^{-\beta {  H}_0}  
+ \left. \left\{
\frac{\partial e^{-\beta{  H}_0-\beta\delta{  H}_I}}
{\partial \lambda}\right\} \right|_{\lambda=0} \lambda \cr
& = &
e^{-\beta {  H}_0} 
+\left. \left\{
\int_0^1 d\xi e^{-\beta\xi{  H}}\frac{\partial (-\beta {  H})}{\partial\lambda}
e^{\beta\xi{  H}}\>\> e^{-\beta{  H}}
\right\}\right|_{\lambda=0} \lambda \cr
& \> = \> &
\left(1 - \left\{
\int_0^1 d\xi e^{-\beta\xi{  H}_0} {  H}_I
e^{\beta\xi{  H}_0}\>\> \right\}
\beta \lambda \right)e^{-\beta {  H}_0}    \\
&=&e^{-\beta {  H}_0}\left(1 - \left\{
\int_0^1 d\xi e^{\beta\xi{  H}_0} {  H}_I
e^{-\beta\xi{  H}_0}\>\> \right\}
\beta \lambda \right)
\>.
\end{eqnarray}

The wave function we start our dynamics with is given by Eq.~(\ref{EqS:PsiBeta}).
The first order perturbation comes from both the denominator and numerator of 
Eq.~(\ref{EqS:PsiBeta}).
First let us deal with the denominator.
Up to the first order, we have
\begin{widetext}
\begin{eqnarray}
\left <  \Psi_0\right | e^{-\beta {  H}}\left |  \Psi_0\right > &=&
\left <  \Psi_0\right | e^{-\beta {  H}_0} -\left\{
\int_0^1 d\xi e^{-\beta\xi{  H}_0} {  H}_I
e^{\beta\xi{  H}_0}\>\> \right\}
\beta \lambda e^{-\beta {  H}_0} + {\cal O}(\lambda^2)\left |  \Psi_0\right > \cr
&=&
\left <  \Psi_0\right | e^{-\beta {  H}_0}\left |  \Psi_0\right >-\beta \lambda\left \langle  
\Psi_0\right |  \int_0^1 d\xi e^{-\beta\xi{  H}_0} {  H}_I
e^{-\beta(1-\xi){  H}_0}\>\>  \left |  \Psi_0\right > +{\cal O}(\lambda^2)   \cr
&=&
\left <  \Psi_0\right | e^{-\beta {  H}_0}\left |  \Psi_0\right >-\beta \lambda 
\int_0^1 d\xi \left <  \Psi_0\right |   e^{-\beta\xi{  H}_0} {  H}_I
e^{-\beta(1-\xi){  H}_0}\>\>  \left |  \Psi_0\right > +{\cal O}(\lambda^2) .
\end{eqnarray}
\end{widetext}
According to the results in Ref.~\cite{HAMS00}, for large $D$ we have
\begin{equation}
{\bf Tr} A \approx D \left <  \Psi_0\right | A \left |  \Psi_0\right >
\end{equation}
where $A$ is an operator which is acting on a $D$-dimensional Hilbert space.
Then the denominator of Eq.~(\ref{EqS:PsiBeta}) reads
\begin{eqnarray}
D \left <  \Psi_0\right | e^{-\beta {  H}}\left |  \Psi_0\right >
&\approx &
{\bf Tr} e^{-\beta {  H}_0} - \beta \lambda \int_0^1 d\xi {\bf Tr} e^{-\beta\xi{  H}_0} {  H}_I
e^{-\beta(1-\xi){  H}_0} \cr
&=&
{\bf Tr} e^{-\beta {  H}_0} - \beta \lambda {\bf Tr} e^{-\beta{H}_0} {H}_I.
\end{eqnarray}
If we restrict the Hamiltonian into the Heisenberg type which is given by
\begin{eqnarray}
\label{Shamiltonian}
H_{S} &=&-\sum_{i=1}^{N_{S}-1}\sum_{j=i+1}^{N_{S}}\sum_{\alpha
=x.y,z}J_{i,j}^{\alpha }S_{i}^{\alpha }S_{j}^{\alpha } \label{SHAMS} \\ 
H_{E} &=&-\sum_{i=1}^{N_E-1}\sum_{j=i+1}^{N_E}\sum_{\alpha =x,y,z}\Omega
_{i,j}^{\alpha }I_{i}^{\alpha }I_{j}^{\alpha }  \label{SHAME}\\ 
H_{SE} &=&-\sum_{i=1}^{N_{S}}\sum_{j=1}^{N_E}\sum_{\alpha =x,y,z}\lambda
_{i,j}^{\alpha }S_{i}^{\alpha }I_{j}^{\alpha }.  \label{SHAMSE}
\end{eqnarray}
where $S$ and $I$ are referring to the spin-$1/2$ operator of 
the system and environment respectively,
then the first order term of the denominator of Eq.~(\ref{EqS:PsiBeta}) is zero.
To see this, we apply an unitary transformation $U$ which transforms $S\rightarrow -S$ 
and $I\rightarrow I$ or $S\rightarrow S$ and $I \rightarrow -I$
to the first order term. The transformation does not change the Hamiltonian 
$H_0=H_S+H_E$, but change the Hamiltonian $H_{I}$ into $-H_I$.
One has
\begin{equation}
{\bf Tr} e^{-\beta{H}_0} {H}_I
={\bf Tr} UU^+ e^{-\beta{H}_0} UU^+{H}_I 
=-{\bf Tr} e^{-\beta {H}_0} {H}_I.
\end{equation}
Therefore, the first order term has to be zero.

Now up to the first order, we have
\begin{equation}
\left <  \Psi_0\right | e^{-\beta {  H}}\left |  \Psi_0\right >\approx
{\bf Tr} e^{-\beta {  H}_0}/D=Z_0/D
\end{equation}
where $Z_0$ is the partition function of the unperturbed system.
Then the wave function is thus given approximately by
\begin{widetext}
\begin{eqnarray}
\left| \Psi_\beta\right\rangle 
&\>\approx \>&  
\sqrt{\frac{D}{Z_0}}\> 
e^{-\beta{  H}/2}\left| \Psi_0\right\rangle   \cr
&=&\sqrt{\frac{D}{Z_0}}\>    \left(1 - \left\{
\int_0^1 d\xi e^{-\beta\xi{  H}_0 /2} {  H}_I
e^{\beta\xi{  H}_0 /2}\>\> \right\}
\beta \lambda/2  +{\cal O}(\lambda^2)\right)e^{-\beta {  H}_0/2}     \left| \Psi_0\right\rangle
\>.
\end{eqnarray}
The corresponding bra is 
\begin{eqnarray}
\left\langle\Psi_\beta\right| 
& \> \approx\> & \sqrt{\frac{D}{Z_0}}\left\langle\Psi(0)\right|e^{-\beta {  H}/2}  \cr
&=& \sqrt{\frac{D}{Z_0}}\left\langle\Psi(0)\right| e^{-\beta {  H}_0/2}\left(1 - \left\{
\int_0^1 d\xi e^{\beta\xi{  H}_0/2} {  H}_I
e^{-\beta\xi{  H}_0/2}\>\> \right\}
\beta \lambda/2 +{\cal O}(\lambda^2)\right).
\end{eqnarray}
The density matrix of the entirety $S+E$ is given by
\begin{eqnarray}
\rho &=&\left| \Psi_\beta\right\rangle  \left\langle\Psi_\beta\right| \cr
&\approx&\frac{D}{Z_0} e^{-\beta {  H}/2} \left| \Psi_0\right\rangle 
\left\langle\Psi_0\right| e^{-\beta {  H}/2} \cr
&=&\frac{D}{Z_0} \left \{ e^{-\beta {  H}_0/2} \left| \Psi_0\right\rangle 
\left\langle\Psi_0\right| e^{-\beta {  H}_0/2} \right. \cr
&&-\frac{\beta}{2} \lambda e^{-\beta {  H}_0/2} \left| \Psi_0\right\rangle 
\left\langle\Psi_0\right| e^{-\beta {  H}_0/2}
\int_0^1 d\xi e^{\beta\xi{  H}_0/2} {  H}_I
e^{-\beta\xi{  H}_0/2} \cr
&& \left . -\frac{\beta}{2} \lambda
\int_0^1 d\xi e^{-\beta\xi{  H}_0 /2} {  H}_I
e^{\beta\xi{  H}_0 /2}\>\> e^{-\beta {  H}_0/2} \left| \Psi_0\right\rangle 
\left\langle\Psi_0\right| e^{-\beta {  H}_0/2} + {\cal O}(\lambda^2)
\right \}. \label{Sdm_w}
\end{eqnarray}
In the energy basis $\{\left |E_{ip}\right \rangle=\left | E_i \right \rangle 
\left | E_p \right \rangle \}$ of the unperturbed system,
the random wave function is given by
\begin{equation}
\left| \Psi_0\right\rangle =\sum_{i=1}^{D_S}\sum_{p=1}^{D_E} d_{ip} \left | E_{ip} \right \rangle
\end{equation}
where $d_{ip}$ is a Gaussian random number and $\sum_{ip} |d_{ip}|^2=1$. Hence, the density 
matrix of the random state is given by
\begin{equation}
\left| \Psi_0\right\rangle \left\langle\Psi_0\right| = 
\sum_{i=1}^{D_S}\sum_{j=1}^{D_S}\sum_{p=1}^{D_E}\sum_{q=1}^{D_E}
d_{ip} d_{jq}^* \left| E_{ip} \right \rangle \left \langle E_{jq} \right | .
\label{Sdm_r}
\end{equation}
Tracing out the degrees of freedom of the environment, one has
\begin{equation}
{\bf Tr}_E \left| \Psi_0\right\rangle \left\langle\Psi_0\right| = 
\sum_{i=1}^{D_S}\sum_{j=1}^{D_S}\sum_{p=1}^{D_E}
d_{ip} d_{jp}^* \left| E_{i} \right \rangle \left \langle E_{j} \right | .
\label{Srdm_r}
\end{equation}
Substituting Eq.~(\ref{Sdm_r}) into Eq.~(\ref{Sdm_w}), 
the density matrix of the entirety $S+E$ reads
\begin{eqnarray}
\rho &\approx& \frac{D}{Z_0} \sum_{i=1}^{D_S}\sum_{j=1}^{D_S}\sum_{p=1}^{D_E}\sum_{q=1}^{D_E}
d_{ip} d_{jq}^*  \left \{
e^{-\beta E_{ip}/2}\left| E_{ip} \right \rangle \left \langle E_{jq} 
\right |e^{-\beta E_{jq}/2} \right . \cr
&& - \frac{\beta}{2} \lambda e^{-\beta E_{ip}/2} \left| E_{ip} \right \rangle 
\left \langle E_{jq} \right | e^{-\beta E_{jq}/2} \int_0^1 d\xi e^{\beta\xi E_{jq}/2} {  H}_I
e^{-\beta\xi{  H}_0/2} \cr
&& \left . -\frac{\beta}{2} \lambda \int_0^1 d\xi e^{-\beta\xi{  H}_0 /2} {  H}_I
e^{\beta\xi E_{ip} /2} e^{-\beta E_{ip}/2} \left| E_{ip} \right \rangle 
\left \langle E_{jq} \right | 
e^{-\beta E_{jq}/2} + \cdots
\right \}.
\end{eqnarray}
Tracing out the degrees of freedom of the environment, 
we obtain the reduced density matrix of the system $S$,
\begin{eqnarray}
\widetilde\rho&=&{\bf Tr}_E \rho \cr
&\approx & \frac{D}{Z_0} \sum_{i=1}^{D_S}\sum_{j=1}^{D_S}\sum_{p=1}^{D_E}
\sum_{q=1}^{D_E}\sum_{l=1}^{D_E}
d_{ip} d_{jq}^*  \left \{
e^{-\beta E_{ip}/2}\left \langle E_l \right . \left| E_{ip} \right \rangle 
\left \langle E_{jq} \right | 
\left. E_l \right \rangle e^{-\beta E_{jq}/2} \right . \cr
&&- \frac{\beta}{2} \lambda e^{-\beta E_{ip}/2} \left \langle E_l \right . 
\left| E_{ip} \right \rangle \left \langle E_{jq} \right | e^{-\beta E_{jq}/2} 
\int_0^1 d\xi e^{\beta\xi E_{jq}/2} {  H}_I
e^{-\beta\xi{  H}_0/2}\left | E_l \right \rangle \cr
&& \left . -\frac{\beta}{2} \left \langle E_l \right |\lambda \int_0^1 d\xi 
e^{-\beta\xi{  H}_0 /2} {  H}_I
e^{\beta\xi E_{ip} /2} e^{-\beta E_{ip}/2} \left| E_{ip} \right \rangle 
\left \langle E_{jq} \right | \left. E_l \right \rangle
e^{-\beta E_{jq}/2} + \cdots
\right \} \cr
&=& \frac{D}{Z_0} \sum_{i=1}^{D_S}\sum_{j=1}^{D_S}\sum_{p=1}^{D_E}\sum_{q=1}^{D_E}\sum_{l=1}^{D_E}
d_{ip} d_{jq}^*  \left \{
e^{-\beta E_{ip}/2} \delta_{lp} \left| E_{i} \right \rangle \left \langle E_{j} \right | 
\delta_{lq} e^{-\beta E_{jq}/2} \right . \cr
&& - \frac{\beta}{2} \lambda e^{-\beta E_{ip}/2} \delta_{lp}\left| E_{i} \right \rangle 
\left \langle E_{jq} \right | e^{-\beta E_{jq}/2} \int_0^1 d\xi e^{\beta\xi E_{jq}/2} {  H}_I
e^{-\beta\xi{  H}_0/2}\left | E_l \right \rangle \cr
&&\left . -\frac{\beta}{2} \left \langle E_l \right |\lambda 
\int_0^1 d\xi e^{-\beta\xi{  H}_0 /2} {  H}_I
e^{\beta\xi E_{ip} /2} e^{-\beta E_{ip}/2} \left| E_{ip} \right \rangle 
\left \langle E_{j} \right | \delta_{lq}
e^{-\beta E_{jq}/2} + \cdots
\right \}.
\end{eqnarray}
Then the elements of the reduced density matrix of the system $S$, in the 
basis that diagonalizes $H_S$, reads
\begin{eqnarray}
{\widetilde\rho}_{i^\prime j^\prime}&=& 
\left \langle E_{i^\prime} \right | {\widetilde\rho} \left | E_{j^\prime}\right\rangle \cr
&\approx& 
\frac{D}{Z_0} \sum_{i=1}^{D_S}\sum_{j=1}^{D_S}\sum_{p=1}^{D_E}\sum_{q=1}^{D_E}\sum_{l=1}^{D_E}
d_{ip} d_{jq}^*  \left \{
e^{-\beta E_{ip}/2} \delta_{lp} \left\langle E_{i^\prime}\right .\left| E_{i} \right \rangle 
\left \langle E_{j} \right | \left . E_{j^\prime} \right \rangle 
\delta_{lq} e^{-\beta E_{jq}/2} \right . \cr
&& - \frac{\beta}{2} \lambda e^{-\beta E_{ip}/2} \delta_{lp} \left\langle E_{i^\prime}\right . 
\left| E_{i} \right \rangle \left \langle E_{jq} \right | e^{-\beta E_{jq}/2} 
\int_0^1 d\xi e^{\beta\xi E_{jq}/2} {  H}_I
e^{-\beta\xi{  H}_0/2}\left | E_l \right \rangle \left | E_{j^\prime}\right \rangle \cr
&&\left . -\frac{\beta}{2} \lambda \left\langle E_{i^\prime}\right|\left \langle E_l \right | 
\int_0^1 d\xi e^{-\beta\xi{  H}_0 /2} {  H}_I
e^{\beta\xi E_{ip} /2} e^{-\beta E_{ip}/2} \left| E_{ip} \right \rangle 
\left \langle E_{j} \right |\left . E_{j^\prime}\right\rangle \delta_{lq}
e^{-\beta E_{jq}/2} + \cdots
\right \} \cr
&=&
\frac{D}{Z_0} \sum_{i=1}^{D_S}\sum_{j=1}^{D_S}\sum_{p=1}^{D_E}\sum_{q=1}^{D_E}\sum_{l=1}^{D_E}
d_{ip} d_{jq}^*  \left \{
e^{-\beta E_{ip}/2} \delta_{lp} \delta_{i^\prime i} \delta_{j^\prime j} 
\delta_{lq} e^{-\beta E_{jq}/2} \right . \cr
&& - \frac{\beta}{2} \lambda e^{-\beta E_{ip}/2} \delta_{lp} \delta_{i^\prime i}  
e^{-\beta E_{jq}/2} \int_0^1 d\xi e^{\beta\xi E_{jq}/2} \left \langle E_{jq} \right | 
{  H}_I\left | E_{j^\prime l} \right \rangle 
e^{-\beta\xi E_{j^\prime l}/2} \cr
&&\left . -\frac{\beta}{2} \lambda  \int_0^1 d\xi e^{-\beta\xi E_{i^\prime l} /2} 
\left\langle E_{i^\prime l}\right|{  H}_I \left| E_{ip} \right \rangle
e^{\beta\xi E_{ip} /2} e^{-\beta E_{ip}/2}  \delta_{j^\prime j} \delta_{lq}
e^{-\beta E_{jq}/2} + \cdots
\right \}.
\end{eqnarray}
Let us look at the different orders of terms $\lambda$ of the reduced density matrix.
The zero oder is
\begin{equation}
{\cal O}({\widetilde\rho}_{i^\prime j^\prime})_{\lambda^0}
=\frac{D}{Z_0} \sum_{l=1}^{D_E}
d_{i^\prime l} d_{j^\prime l}^*  
e^{-\beta E_{i^\prime l}/2} e^{-\beta E_{j^\prime l}/2}
\end{equation}
which is the term we have analyzed for the uncoupled entirety.
The first order is
\begin{eqnarray}
{\cal O}({\widetilde\rho}_{i^\prime j^\prime})_{\lambda^1}
&=&
-\frac{\beta}{2} \lambda\frac{D}{Z_0} 
\sum_{i=1}^{D_S}\sum_{j=1}^{D_S}\sum_{p=1}^{D_E}\sum_{q=1}^{D_E}\sum_{l=1}^{D_E}
d_{ip} d_{jq}^*  \left \{
 e^{-\beta E_{ip}/2} \delta_{lp} \delta_{i^\prime i}  e^{-\beta E_{jq}/2} 
\int_0^1 d\xi e^{\beta\xi E_{jq}/2} \left \langle E_{jq} \right |{  H}_I
\left | E_{j^\prime l} \right \rangle 
e^{-\beta\xi E_{j^\prime l}/2} \right. \cr
&&\left . + \int_0^1 d\xi e^{-\beta\xi E_{i^\prime l} /2} 
\left\langle E_{i^\prime l}\right|{  H}_I \left| E_{ip} \right \rangle
e^{\beta\xi E_{ip} /2} e^{-\beta E_{ip}/2}  \delta_{j^\prime j} \delta_{lq}
e^{-\beta E_{jq}/2}
\right \} \cr
&=& 
-\frac{\beta}{2} \lambda\frac{D}{Z_0} 
\sum_{j=1}^{D_S}\sum_{q=1}^{D_E}\sum_{l=1}^{D_E}
d_{i^\prime l} d_{jq}^* 
e^{-\beta E_{i^\prime l}/2}  e^{-\beta E_{jq}/2} 
\int_0^1 d\xi e^{\beta\xi E_{jq}/2} \left \langle E_{jq} \right |
{  H}_I\left | E_{j^\prime l} \right \rangle 
e^{-\beta\xi E_{j^\prime l}/2}     \cr
&&
-\frac{\beta}{2} \lambda\frac{D}{Z_0} 
\sum_{i=1}^{D_S}\sum_{p=1}^{D_E}\sum_{l=1}^{D_E}
d_{ip} d_{j^\prime l}^*
\int_0^1 d\xi e^{-\beta\xi E_{i^\prime l} /2} 
\left\langle E_{i^\prime l}\right|{  H}_I \left| E_{ip} \right \rangle
e^{\beta\xi E_{ip} /2} e^{-\beta E_{ip}/2}  
e^{-\beta E_{j^\prime l}/2} \cr
\left ( j\rightarrow i, q\rightarrow p
\right ) \>\>\>\>\>\>
&=&
-\frac{\beta}{2} \lambda\frac{D}{Z_0} 
\sum_{i=1}^{D_S}\sum_{p=1}^{D_E}\sum_{l=1}^{D_E}
e^{-\beta E_{ip}/2} \left \{
d_{i^\prime l} d_{ip}^* 
e^{-\beta E_{i^\prime l}/2}  \int_0^1 d\xi e^{\beta\xi E_{ip}/2} 
\left \langle E_{ip} \right |{  H}_I\left | E_{j^\prime l} \right \rangle 
e^{-\beta\xi E_{j^\prime l}/2}  \right.   \cr
&& + \left .
d_{ip} d_{j^\prime l}^*
\int_0^1 d\xi e^{-\beta\xi E_{i^\prime l} /2} 
\left\langle E_{i^\prime l}\right|{  H}_I \left| E_{ip} \right \rangle
e^{\beta\xi E_{ip} /2}   
e^{-\beta E_{j^\prime l}/2}
\right \}.
\end{eqnarray}

We also need the complex conjugate of the reduced density matrix.
The zero order is
\begin{equation}
{\cal O}({\widetilde\rho}_{i^\prime j^\prime}^*)_{\lambda^0}
=\frac{D}{Z_0} \sum_{l^{\prime\prime}=1}^{D_E}
d_{i^\prime l^{\prime\prime}}^* d_{j^\prime l^{\prime\prime}}  
e^{-\beta E_{i^\prime l^{\prime\prime}}/2} e^{-\beta E_{j^\prime l^{\prime\prime}}/2}.
\end{equation}
The first order is ($\langle E_{ip}|H_I|E_{jq}\rangle$ is 
real for the Hamiltonian we are interested in.)
\begin{eqnarray}
{\cal O}({\widetilde\rho}_{i^\prime j^\prime}^*)_{\lambda^1}
&=&
-\frac{\beta}{2} \lambda\frac{D}{Z_0} 
\sum_{i^{\prime\prime\prime}=1}^{D_S}\sum_{p^{\prime\prime\prime}=1}^{D_E}
\sum_{l^{\prime\prime\prime}=1}^{D_E}
e^{-\beta E_{i^{\prime\prime\prime}p^{\prime\prime\prime}}/2} \left \{
d_{i^\prime l^{\prime\prime\prime}}^* d_{i^{\prime\prime\prime}p^{\prime\prime\prime}} 
e^{-\beta E_{i^\prime l^{\prime\prime\prime}}/2}  
\int_0^1 d\xi e^{\beta\xi E_{i^{\prime\prime\prime}p^{\prime\prime\prime}}/2} 
\left \langle E_{i^{\prime\prime\prime}p^{\prime\prime\prime}} \right |{  H}_I 
\left | E_{j^\prime l^{\prime\prime\prime}} \right \rangle 
e^{-\beta\xi E_{j^\prime l^{\prime\prime\prime}}/2}  \right.   \cr
&& + \left .
d_{i^{\prime\prime\prime}p^{\prime\prime\prime}}^* d_{j^\prime l^{\prime\prime\prime}}
\int_0^1 d\xi e^{-\beta\xi E_{i^\prime l^{\prime\prime\prime}} /2} 
\left\langle E_{i^\prime l^{\prime\prime\prime}}\right|{  H}_I \left| 
E_{i^{\prime\prime\prime}p^{\prime\prime\prime}} \right \rangle
e^{\beta\xi E_{i^{\prime\prime\prime}p^{\prime\prime\prime}} /2}   
e^{-\beta E_{j^\prime l^{\prime\prime\prime}}/2}
\right \}.
\end{eqnarray}

The expectation value for $\sigma^2$ that we want to calculate is
\begin{equation}
{\cal E}\left(2\sigma^2\right) 
 = 
{\cal E}\left(\sum_{i^\prime\ne j^\prime} \left|
{\widetilde\rho}_{i^\prime j^\prime}
\right|^2\right) 
 = 
\sum_{i^\prime\ne j^\prime}^{D_S} {\cal E}\left(\left|
{\widetilde\rho}_{i^\prime j^\prime}
\right|^2\right) 
 = 
\sum_{i^\prime\ne j^\prime}^{D_S} {\cal E}\left(
{\widetilde\rho}_{i^\prime j^\prime} {\widetilde\rho}_{i^\prime j^\prime}^*
\right) 
\>.
\end{equation}

The order $\lambda^0$ term for $\sigma^2$ is
\begin{eqnarray}
{\cal O}\left({\cal E}\left(2\sigma^2\right)\right)_{\lambda^0} 
&=&
\sum_{i^\prime\ne j^\prime}^{D_S} {\cal E}\left( {\cal O}\left (
{\widetilde\rho}_{i^\prime j^\prime} {\widetilde\rho}_{i^\prime j^\prime}^*
\right)_{{\cal O}(\lambda^0)} \right)     \cr
&=&
\left(\frac{D}{Z_0}\right)^2 \sum_{i^\prime\ne j^\prime}^{D_S} 
\sum_{l=1}^{D_E}\sum_{l^{\prime\prime}=1}^{D_E}
{\cal E}\left( d_{i^\prime l} d_{j^\prime l}^*  
d_{i^\prime l^{\prime\prime}}^* d_{j^\prime l^{\prime\prime}} \right)
e^{-\beta E_{i^\prime l}/2} e^{-\beta E_{j^\prime l}/2} 
e^{-\beta E_{i^\prime l^{\prime\prime}}/2} e^{-\beta E_{j^\prime l^{\prime\prime}}/2}
\end{eqnarray}
which is the term being analyzed for the uncoupled entirety with the approximation in the main text.

The order $\lambda^1$ term for $\sigma^2$ is (in the following, 
$\bf a$ and $\bf b$ are symbols for the calculation terms)
\begin{eqnarray}
{\cal O}\left({\cal E}\left(2\sigma^2\right)\right)_{\lambda^1} 
&=&
\sum_{i^\prime\ne j^\prime}^{D_S} {\cal E}\left( {\cal O}\left (
{\widetilde\rho}_{i^\prime j^\prime} {\widetilde\rho}_{i^\prime j^\prime}^*
\right)_{\lambda^1} \right)     \cr
&=&
\sum_{i^\prime\ne j^\prime}^{D_S} {\cal E}\left( {\cal O}\left (
{\widetilde\rho}_{i^\prime j^\prime} \right)_{\lambda^0}
{\cal O}\left( {\widetilde\rho}_{i^\prime j^\prime}^*\right)_{\lambda^1}
+
{\cal O}\left (
{\widetilde\rho}_{i^\prime j^\prime} \right)_{\lambda^1}
{\cal O}\left( {\widetilde\rho}_{i^\prime j^\prime}^*\right)_{\lambda^0}
 \right)   \cr
={\bf ab^*+a^*b}&=&
-\left(\frac{D}{Z_0}\right)^2 \frac{\beta}{2}\lambda 
\sum_{i^\prime\ne j^\prime}^{D_S} {\cal E}\Bigg ( \cr
{\bf Put \>\>a}\>\>\>\>&&
\sum_{l=1}^{D_E}
d_{i^\prime l} d_{j^\prime l}^*  
e^{-\beta E_{i^\prime l}/2} e^{-\beta E_{j^\prime l}/2}  \times  \cr
{\bf Put \>b^*}\>|^{i^{\prime\prime\prime}\rightarrow i}_{p^{\prime\prime\prime}\rightarrow p}
|^{l^{\prime\prime\prime}\rightarrow l^{\prime\prime}}
\>&&
\sum_{i=1}^{D_S}\sum_{p=1}^{D_E}\sum_{l^{\prime\prime}=1}^{D_E}
e^{-\beta E_{ip}/2} \left \{
d_{i^\prime l^{\prime\prime}}^* d_{ip} 
e^{-\beta E_{i^\prime l^{\prime\prime}}/2}  \int_0^1 d\xi e^{\beta\xi E_{ip}/2} 
\left \langle E_{ip} \right |{  H}_I\left | E_{j^\prime l^{\prime\prime}} \right \rangle 
e^{-\beta\xi E_{j^\prime l^{\prime\prime}}/2}  \right.   \cr
&& + \left .
d_{ip}^* d_{j^\prime l^{\prime\prime}}
\int_0^1 d\xi e^{-\beta\xi E_{i^\prime l^{\prime\prime}} /2} 
\left\langle E_{i^\prime l^{\prime\prime}}\right|{  H}_I \left| E_{ip} \right \rangle
e^{\beta\xi E_{ip} /2}   
e^{-\beta E_{j^\prime l^{\prime\prime}}/2}
\right \}     \cr
&& +   \cr
{\bf Put \>\>a^*}\>\>\>\>&&
\sum_{l^{\prime\prime}=1}^{D_E}
d_{i^\prime l^{\prime\prime}}^* d_{j^\prime l^{\prime\prime}}  
e^{-\beta E_{i^\prime l^{\prime\prime}}/2} e^{-\beta E_{j^\prime l^{\prime\prime}}/2} \times   \cr
{\bf Put \>\>b}\>\>\>\>&&
\sum_{i=1}^{D_S}\sum_{p=1}^{D_E}\sum_{l=1}^{D_E}
e^{-\beta E_{ip}/2} \left \{
d_{i^\prime l} d_{ip}^* 
e^{-\beta E_{i^\prime l}/2}  \int_0^1 d\xi e^{\beta\xi E_{ip}/2} \left \langle E_{ip} \right |
{  H}_I\left | E_{j^\prime l} \right \rangle 
e^{-\beta\xi E_{j^\prime l}/2}  \right.   \cr
&& + \left . \left .
d_{ip} d_{j^\prime l}^*
\int_0^1 d\xi e^{-\beta\xi E_{i^\prime l} /2} \left\langle E_{i^\prime l}
\right|{  H}_I \left| E_{ip} \right \rangle
e^{\beta\xi E_{ip} /2}   
e^{-\beta E_{j^\prime l}/2}
\right \} \right ).
\end{eqnarray}
The summation indices are all the same, so we pull them out to the from of the sum
\begin{eqnarray}
{\cal O}\left({\cal E}\left(2\sigma^2\right)\right)_{\lambda^1} 
&=&
-\left(\frac{D}{Z_0}\right)^2 \frac{\beta}{2}\lambda \sum_{i^\prime\ne j^\prime}^{D_S} 
{\cal E}\left ( \sum_{i=1}^{D_S}\sum_{p=1}^{D_E}\sum_{l^{\prime\prime}=1}^{D_E} \sum_{l=1}^{D_E}
\Bigg [ \right. \cr
{\bf Put \>\>a}\>\>\>\>&&
d_{i^\prime l} d_{j^\prime l}^*  
e^{-\beta E_{i^\prime l}/2} e^{-\beta E_{j^\prime l}/2}  \times  \cr
{\bf Put \>b^*}\>
\>&&
e^{-\beta E_{ip}/2} \left \{
d_{i^\prime l^{\prime\prime}}^* d_{ip} 
e^{-\beta E_{i^\prime l^{\prime\prime}}/2}  
\int_0^1 d\xi e^{\beta\xi E_{ip}/2} \left \langle E_{ip} \right |{  H}_I\left | 
E_{j^\prime l^{\prime\prime}} \right \rangle 
e^{-\beta\xi E_{j^\prime l^{\prime\prime}}/2}  \right.   \cr
&& + \left .
d_{ip}^* d_{j^\prime l^{\prime\prime}}
\int_0^1 d\xi e^{-\beta\xi E_{i^\prime l^{\prime\prime}} /2} 
\left\langle E_{i^\prime l^{\prime\prime}}\right|{  H}_I \left| E_{ip} \right \rangle
e^{\beta\xi E_{ip} /2}   
e^{-\beta E_{j^\prime l^{\prime\prime}}/2}
\right \}     \cr
&& +   \cr
{\bf Put \>\>a^*}\>\>\>\>&&
d_{i^\prime l^{\prime\prime}}^* d_{j^\prime l^{\prime\prime}}  
e^{-\beta E_{i^\prime l^{\prime\prime}}/2} e^{-\beta E_{j^\prime l^{\prime\prime}}/2} \times   \cr
{\bf Put \>\>b}\>\>\>\>&&
e^{-\beta E_{ip}/2} \left \{
d_{i^\prime l} d_{ip}^* 
e^{-\beta E_{i^\prime l}/2}  \int_0^1 d\xi e^{\beta\xi E_{ip}/2} \left \langle E_{ip} 
\right |{  H}_I\left | E_{j^\prime l} \right \rangle 
e^{-\beta\xi E_{j^\prime l}/2}  \right.   \cr
&& + \left .\left.\left.
d_{ip} d_{j^\prime l}^*
\int_0^1 d\xi e^{-\beta\xi E_{i^\prime l} /2} 
\left\langle E_{i^\prime l}\right|{  H}_I \left| E_{ip} \right \rangle
e^{\beta\xi E_{ip} /2}   
e^{-\beta E_{j^\prime l}/2}
\right \} \right ]\right ).
\end{eqnarray}
Rearranging the terms, one has
\begin{eqnarray}
{\cal O}\left({\cal E}\left(2\sigma^2\right)\right)_{\lambda^1} 
&=&
-\left(\frac{D}{Z_0}\right)^2 \frac{\beta}{2}\lambda \sum_{i^\prime\ne j^\prime}^{D_S} 
 \sum_{i=1}^{D_S}\sum_{p=1}^{D_E}\sum_{l^{\prime\prime}=1}^{D_E} \sum_{l=1}^{D_E}
\Bigg [  \cr
&&
e^{-\beta E_{i^\prime l}/2} e^{-\beta E_{j^\prime l}/2}  \times  \cr
{\bf Put \>ab^*}\>
\>&&
e^{-\beta E_{ip}/2} \left \{
{\cal E}\left(d_{i^\prime l} d_{j^\prime l}^* d_{i^\prime l^{\prime\prime}}^* d_{ip} \right)
e^{-\beta E_{i^\prime l^{\prime\prime}}/2}  \int_0^1 d\xi e^{\beta\xi E_{ip}/2} \left \langle 
E_{ip} \right |{  H}_I\left | E_{j^\prime l^{\prime\prime}} \right \rangle 
e^{-\beta\xi E_{j^\prime l^{\prime\prime}}/2}  \right.   \cr
&& + \left .
{\cal E}\left(d_{i^\prime l} d_{j^\prime l}^* d_{ip}^* d_{j^\prime l^{\prime\prime}}\right)
\int_0^1 d\xi e^{-\beta\xi E_{i^\prime l^{\prime\prime}} /2} \left\langle 
E_{i^\prime l^{\prime\prime}}\right|{  H}_I \left| E_{ip} \right \rangle
e^{\beta\xi E_{ip} /2}   
e^{-\beta E_{j^\prime l^{\prime\prime}}/2}
\right \}     \cr
&& +   \cr
&& 
e^{-\beta E_{i^\prime l^{\prime\prime}}/2} e^{-\beta E_{j^\prime l^{\prime\prime}}/2} \times   \cr
{\bf Put \>\>a^*b}\>\>\>\>&&
e^{-\beta E_{ip}/2} \left \{
{\cal E}\left(d_{i^\prime l^{\prime\prime}}^* d_{j^\prime l^{\prime\prime}} d_{i^\prime l} d_{ip}^* \right)
e^{-\beta E_{i^\prime l}/2}  \int_0^1 d\xi e^{\beta\xi E_{ip}/2} \left \langle E_{ip} \right | 
{  H}_I\left | E_{j^\prime l} \right \rangle 
e^{-\beta\xi E_{j^\prime l}/2}  \right.   \cr
&& + \left .\left.
{\cal E}\left(
d_{i^\prime l^{\prime\prime}}^* d_{j^\prime l^{\prime\prime}} d_{ip} d_{j^\prime l}^*\right)
\int_0^1 d\xi e^{-\beta\xi E_{i^\prime l} /2} 
\left\langle E_{i^\prime l}\right|{  H}_I \left| E_{ip} \right \rangle
e^{\beta\xi E_{ip} /2}   
e^{-\beta E_{j^\prime l}/2}
\right \} \right ].
\end{eqnarray}
We want to use the expectation value identities
\begin{equation}
{\cal E}\left( d_\alpha d_\beta d_\gamma^* d_\delta^* \right)
\>=\> 
{\cal E}\left( \left|d\right|^2\left|d\right|^2\right)
\left(\delta_{\alpha\gamma} \delta_{\beta\delta} +
\delta_{\alpha\delta} \delta_{\beta\gamma}\right)
\>+\>
{\cal E}\left(\left|d\right|^4\right)
\delta_{\alpha\beta}\delta_{\alpha\gamma}\delta_{\alpha\delta}.
\end{equation}
Notice that we do not have the term ${\cal E}(|d|^4)$ as the indices $i^\prime \neq j^\prime$.
We check the terms ${\cal E}(|d|^2|d|^2)$,
\begin{eqnarray}
{\cal E}\left(d_{i^\prime l} d_{j^\prime l}^* d_{i^\prime l^{\prime\prime}}^* d_{ip} \right)
&=&{\cal E}\left(|d|^2|d|^2\right) 
\delta_{i^\prime l,i^\prime l^{\prime\prime}}\delta_{j^\prime l,ip}
\\
{\cal E}\left(d_{i^\prime l} d_{j^\prime l}^* d_{ip}^* d_{j^\prime l^{\prime\prime}}\right)
&=&{\cal E}\left(|d|^2|d|^2\right)
\delta_{i^\prime l,ip}\delta_{j^\prime l,j^\prime l^{\prime\prime}}
\\
{\cal E}
\left(d_{i^\prime l^{\prime\prime}}^* d_{j^\prime l^{\prime\prime}} d_{i^\prime l} d_{ip}^* \right)
&=&{\cal E}\left(|d|^2|d|^2\right)
\delta_{i^\prime l^{\prime\prime},i^\prime l}\delta_{j^\prime l^{\prime\prime},ip}
\\
{\cal E}
\left(d_{i^\prime l^{\prime\prime}}^* d_{j^\prime l^{\prime\prime}} d_{ip} d_{j^\prime l}^*\right)
&=&{\cal E}\left(|d|^2|d|^2\right)
\delta_{i^\prime l^{\prime\prime},ip}\delta_{j^\prime l^{\prime\prime},j^\prime l}.
\end{eqnarray}
Then we have
\begin{eqnarray}
{\cal O}\left({\cal E}\left(2\sigma^2\right)\right)_{\lambda^1} 
&=&
-\left(\frac{D}{Z_0}\right)^2 \frac{\beta}{2}\lambda 
{\cal E}\left(|d|^2|d|^2\right)\sum_{i^\prime\ne j^\prime}^{D_S} 
 \sum_{i=1}^{D_S}\sum_{p=1}^{D_E}\sum_{l^{\prime\prime}=1}^{D_E} \sum_{l=1}^{D_E}
\Bigg [  \cr
&&
e^{-\beta E_{i^\prime l}/2} e^{-\beta E_{j^\prime l}/2}  \times  \cr
{\bf Put \>ab^*}\>
\>&&
e^{-\beta E_{ip}/2} \left \{
\delta_{i^\prime l,i^\prime l^{\prime\prime}}\delta_{j^\prime l,ip}
e^{-\beta E_{i^\prime l^{\prime\prime}}/2}  \int_0^1 d\xi e^{\beta\xi E_{ip}/2} 
\left \langle E_{ip} \right |{  H}_I\left | E_{j^\prime l^{\prime\prime}} \right \rangle 
e^{-\beta\xi E_{j^\prime l^{\prime\prime}}/2}  \right.   \cr
&& + \left .
\delta_{i^\prime l,ip}\delta_{j^\prime l,j^\prime l^{\prime\prime}}
\int_0^1 d\xi e^{-\beta\xi E_{i^\prime l^{\prime\prime}} /2} 
\left\langle E_{i^\prime l^{\prime\prime}}\right|{  H}_I \left| E_{ip} \right \rangle
e^{\beta\xi E_{ip} /2}   
e^{-\beta E_{j^\prime l^{\prime\prime}}/2}
\right \}     \cr
&& +   \cr
&& 
e^{-\beta E_{i^\prime l^{\prime\prime}}/2} e^{-\beta E_{j^\prime l^{\prime\prime}}/2} \times   \cr
{\bf Put \>\>a^*b}\>\>\>\>&&
e^{-\beta E_{ip}/2} \left \{
\delta_{i^\prime l^{\prime\prime},i^\prime l}\delta_{j^\prime l^{\prime\prime},ip}
e^{-\beta E_{i^\prime l}/2}  \int_0^1 d\xi e^{\beta\xi E_{ip}/2} \left \langle E_{ip} \right | 
{  H}_I\left | E_{j^\prime l} \right \rangle 
e^{-\beta\xi E_{j^\prime l}/2}  \right.   \cr
&& + \left .\left.
\delta_{i^\prime l^{\prime\prime},ip}\delta_{j^\prime l^{\prime\prime},j^\prime l}
\int_0^1 d\xi e^{-\beta\xi E_{i^\prime l} /2} \left\langle E_{i^\prime l}\right|{  H}_I 
\left| E_{ip} \right \rangle
e^{\beta\xi E_{ip} /2}   
e^{-\beta E_{j^\prime l}/2}
\right \} \right ]    \\
&=&
-\left(\frac{D}{Z_0}\right)^2 \frac{\beta}{2}\lambda 
{\cal E}\left(|d|^2|d|^2\right)\sum_{i^\prime\ne j^\prime}^{D_S}
\Bigg [     \cr
{\bf Put \>ab^*}\>\> &&
\left \{ 
\sum_{l=1}^{D_E} 
e^{-\beta E_{i^\prime l}/2} e^{-\beta E_{j^\prime l}/2}  
e^{-\beta E_{j^\prime l}/2} 
e^{-\beta E_{i^\prime l}/2}  \int_0^1 d\xi e^{\beta\xi E_{j^\prime l}/2} 
\left \langle E_{j^\prime l} \right |{  H}_I\left | E_{j^\prime l} \right \rangle 
e^{-\beta\xi E_{j^\prime l}/2}  \right.   \cr
&& + \left .
\sum_{l=1}^{D_E}
e^{-\beta E_{i^\prime l}/2} e^{-\beta E_{j^\prime l}/2}
e^{-\beta E_{i^\prime l}/2}
\int_0^1 d\xi e^{-\beta\xi E_{i^\prime l} /2} 
\left\langle E_{i^\prime l}\right|{  H}_I \left| E_{i^\prime l} \right \rangle
e^{\beta\xi E_{i^\prime l} /2}   
e^{-\beta E_{j^\prime l}/2}
\right \}   \cr
&&  +   \cr
{\bf Put \>\>a^*b}\>\>\>\>&&
\left \{ \sum_{l=1}^{D_E}
e^{-\beta E_{i^\prime l}/2} e^{-\beta E_{j^\prime l}/2}
e^{-\beta E_{j^\prime l}/2} 
e^{-\beta E_{i^\prime l}/2}  
\int_0^1 d\xi e^{\beta\xi E_{j^\prime l}/2} 
\left \langle E_{j^\prime l} \right |{  H}_I\left | E_{j^\prime l} \right \rangle 
e^{-\beta\xi E_{j^\prime l}/2}  \right.   \cr
&& + \left .\left.
\sum_{l=1}^{D_E}
e^{-\beta E_{i^\prime l}/2} e^{-\beta E_{j^\prime l}/2}
e^{-\beta E_{i^\prime l}/2}
\int_0^1 d\xi e^{-\beta\xi E_{i^\prime l} /2} 
\left\langle E_{i^\prime l}\right|{  H}_I \left| E_{i^\prime l} \right \rangle
e^{\beta\xi E_{i^\prime l} /2}   
e^{-\beta E_{j^\prime l}/2}
\right \} \right ]   \cr
&=&
-\left(\frac{D}{Z_0}\right)^2 \frac{\beta}{2}\lambda 
{\cal E}\left(|d|^2|d|^2\right)\sum_{i^\prime\ne j^\prime}^{D_S}
\sum_{l=1}^{D_E}
\Bigg [    \cr
{\bf Put \>ab^*}\>\> &&
\left \{  
e^{-\beta E_{i^\prime l}} e^{-\beta E_{j^\prime l}}  
\left \langle E_{j^\prime l} \right |{  H}_I\left | E_{j^\prime l} \right \rangle 
  +
e^{-\beta E_{i^\prime l}} e^{-\beta E_{j^\prime l}}
 \left\langle E_{i^\prime l}\right|{  H}_I \left| E_{i^\prime l} \right \rangle
\right \}   \cr
&&  +   \cr
{\bf Put \>\>a^*b}\>\>\>\>&&
\left \{ 
e^{-\beta E_{i^\prime l}} e^{-\beta E_{j^\prime l}}
\left \langle E_{j^\prime l} \right |{  H}_I\left | E_{j^\prime l} \right \rangle 
+\left.
e^{-\beta E_{i^\prime l}} e^{-\beta E_{j^\prime l}}
\left\langle E_{i^\prime l}\right|{  H}_I \left| E_{i^\prime l} \right \rangle
\right \} \right ].
\end{eqnarray}
The final results for the first order term of $\sigma^2$ is
\begin{equation}
{\cal O}\left({\cal E}\left(2\sigma^2\right)\right)_{\lambda^1}=
-\left(\frac{D}{Z_0}\right)^2 \beta\delta {\cal E}(|d|^2|d|^2)\sum_{i^\prime\ne j^\prime}^{D_S}
\sum_{l=1}^{D_E} e^{-\beta E_{i^\prime l}} e^{-\beta E_{j^\prime l}} 
\left (   
\left \langle E_{j^\prime l} \right |{  H}_I\left | E_{j^\prime l} \right \rangle 
  +
 \left\langle E_{i^\prime l}\right|{  H}_I \left| E_{i^\prime l} \right \rangle
 \right ).
\end{equation}
Changing the indices 
$i^\prime \rightarrow i$, $j^\prime \rightarrow j$ and $l\rightarrow p$, we have
\begin{equation}
{\cal O}\left({\cal E}\left(2\sigma^2\right)\right)_{\lambda^1}=
-\left(\frac{D}{Z_0}\right)^2 \beta\delta {\cal E}\left(|d|^2|d|^2\right)\sum_{i\ne j}^{D_S}
\sum_{p=1}^{D_E} e^{-\beta E_{ip}} e^{-\beta E_{jp}} 
\left (   
\left \langle E_{ip} \right |{  H}_I\left | E_{ip} \right \rangle 
  +
 \left\langle E_{jp}\right|{  H}_I \left| E_{jp} \right \rangle
 \right ).
\end{equation}
Note that if one set $\beta=0$, the first order is 
zero and the results for the ``$X$" state from \cite{JIN13a} are retrieved.

Changing the sum 
\begin{equation}
\sum_{i\ne j}^{D_S} \Rightarrow 
\sum_i^{D_S} \sum_j^{D_S} \left(1-\delta_{ij}\right) 
\end{equation}
gives
\begin{eqnarray}
{\cal O}\left({\cal E}\left(2\sigma^2\right)\right)_{\lambda^1}&=&
-\left(\frac{D}{Z_0}\right)^2 \beta\delta {\cal E}\left(|d|^2|d|^2\right)\sum_i^{D_S} \sum_j^{D_S} 
\left(1-\delta_{ij}\right)
\sum_{p=1}^{D_E} e^{-\beta E_{ip}} e^{-\beta E_{jp}} 
\left (   
\left \langle E_{ip} \right |{  H}_I\left | E_{ip} \right \rangle 
  +
 \left\langle E_{jp}\right|{  H}_I \left| E_{jp} \right \rangle
 \right )   \cr
 &=&
 -\left(\frac{D}{Z_0}\right)^2 \beta\delta {\cal E}\left(|d|^2|d|^2\right)
\left[ \sum_i^{D_S} \sum_j^{D_S}\sum_{p=1}^{D_E} e^{-\beta E_{ip}} e^{-\beta E_{jp}} 
\left (   
\left \langle E_{ip} \right |{  H}_I\left | E_{ip} \right \rangle 
  +
 \left\langle E_{jp}\right|{  H}_I \left| E_{jp} \right \rangle
 \right )\right . \cr
 && \left . -
 2\sum_i^{D_S} \sum_{p=1}^{D_E} e^{-2 \beta E_{ip}} 
\left \langle E_{ip} \right |{  H}_I\left | E_{ip} \right \rangle 
 \right ]  \cr
 &=&
 -2\left(\frac{D}{Z_0}\right)^2 \beta\delta {\cal E}\left(|d|^2|d|^2\right)
\left[ \sum_i^{D_S} \sum_j^{D_S}\sum_{p=1}^{D_E} e^{-\beta E_{ip}} e^{-\beta E_{jp}}  
\left \langle E_{ip} \right |{  H}_I\left | E_{ip} \right \rangle 
\right . \cr
 && \left . -
 \sum_i^{D_S} \sum_{p=1}^{D_E} e^{-2 \beta E_{ip}} 
\left \langle E_{ip} \right |{  H}_I\left | E_{ip} \right \rangle 
 \right ] \cr
&=&
-2\left(\frac{D}{Z_0}\right)^2 \beta\delta {\cal E}\left(|d|^2|d|^2\right)\left[ \sum_i^{D_S} 
\sum_j^{D_S}\sum_{p=1}^{D_E}
 e^{-\beta E_{i}} e^{-\beta E_{j}} e^{-2\beta E_p}
\left \langle E_{ip} \right |{  H}_I\left | E_{ip} \right \rangle 
\right . \cr
 && \left . -
 \sum_i^{D_S} \sum_{p=1}^{D_E} e^{-2 \beta E_{ip}} 
\left \langle E_{ip} \right |{  H}_I\left | E_{ip} \right \rangle 
 \right ] \cr
 &=&
 -2\left(\frac{D}{Z_0}\right)^2 \beta\delta {\cal E}\left(|d|^2|d|^2\right)\left[
 \sum_j^{D_S} e^{-\beta E_j} \sum_i^{D_S}\sum_{p=1}^{D_E}
 e^{-\beta E_{i}}e^{-2\beta E_p}
 \left \langle E_{ip} \right |{  H}_I\left | E_{ip} \right \rangle 
\right . \cr
 && \left . -
 \sum_i^{D_S} \sum_{p=1}^{D_E} e^{-2 \beta E_{ip}} 
\left \langle E_{ip} \right |{  H}_I\left | E_{ip} \right \rangle 
 \right ] \cr
 &=&
 -2\left(\frac{D}{Z_0}\right)^2 \beta\delta {\cal E}\left(|d|^2|d|^2\right)\left[
 Z_S {\bf Tr} e^{-\beta H_S}e^{-2\beta H_E} H_I  -
 {\bf Tr} e^{-2\beta (H_S+H_E)}H_I 
 \right ].
\end{eqnarray}
By applying the same symmetry argument as above, transform $S\rightarrow -S$ and $I\rightarrow I$ 
or alternatively transform 
$S\rightarrow S$ and $I\rightarrow -I$, one has
\begin{eqnarray}
{\bf Tr} e^{-\beta H_S}e^{-2\beta H_E} H_I &=&
{\bf Tr} e^{-\beta H_S}e^{-2\beta H_E} U^+ H_I U 
=-{\bf Tr} e^{-\beta H_S}e^{-2\beta H_E} H_I \\
{\bf Tr} e^{-2\beta (H_S+H_E)}H_I &=&
{\bf Tr} e^{-2\beta (H_S+H_E)}U^+ H_I U 
=-{\bf Tr} e^{-2\beta (H_S+H_E)}H_I .   
\end{eqnarray}
\end{widetext}
The terms of traces have to be zero.
Therefore, if there exists such symmetry in the entirety $S+E$, 
such as the system with the Hamiltonian described in Eqs.~(\ref{SHAMS}-\ref{SHAMSE}), 
the first order of $\sigma^2$ is 
\begin{equation}
{\cal O}\left({\cal E}\left(2\sigma^2\right)\right)_{\lambda^1}=0 .
\end{equation}

Calculating the second order term of $\sigma^2$ is much more complicated as the 
perturbation term comes from both the denominator and numerator of Eq.~(\ref{EqS:PsiBeta}).
We are not going to calculate the second order term of $\sigma^2$.
We may conjecture that the second order term is zero from the simulation results, and 
the $\sigma$ of the uncoupled entirety is a lower bond for the $\sigma$ of the coupled entirety. 

We have not calculated the first-order term for ${\cal E}\left(\delta^2\right)$.  However, the 
numerical results from Appendix~A can be used to form an ansatz that the first order 
term either vanishes or is small for Hamiltonians with the symmetry that makes the 
first-order term of ${\cal E}\left(\sigma^2\right)$ be zero.  

\bibliographystyle{apsrev4-1}
\bibliography{decoherence}

\begin{thebibliography}{61}
\bibitem [{30s()}]{30spins}%
  \BibitemOpen
  \href@noop {} {}\bibinfo {note} {The coefficients of the $7$th-order
  polynomial fit to the data (see Fig.~8) of the 30-spin system are
  $a_0=0.34\times10^{-7}$, $a_1=-0.59\times10^{-7}$, $a_2=0.56\times10^{-6}$,
  $a_3=-0.18\times10^{-5}$, $a_4=0.41\times10^{-5}$, $a_5=-0.28\times10^{-5}$,
  $a_6=0.12\times10^{-5}$, and $a_7=-0.23\times10^{-6}$}\BibitemShut {NoStop}%
\bibitem [{40s()}]{40spins}%
  \BibitemOpen
  \href@noop {} {}\bibinfo {note} {The coefficients of the $7$th-order
  polynomial fit to the data (see Fig.~8) of the 40-spin system are
  $a_0=0.14\times10^{-9}$, $a_1=-0.65\times10^{-9}$, $a_2=0.27\times10^{-8}$,
  $a_3=-0.12\times10^{-7}$, $a_4=0.11\times10^{-5}$, $a_5=0.46\times10^{-8}$,
  $a_6=-0.80\times10^{-7}$, and $a_7=0.12\times10^{-7}$}\BibitemShut {NoStop}%
\end{thebibliography}
\end{document}